\documentclass[preprint]{emulateapj}
\usepackage{apjfonts}
\usepackage{graphicx}

\newcommand{\oi}{[O\,{\sc i}]}
\newcommand{\oii}{[O\,{\sc ii}]}
\newcommand{\oiii}{[O\,{\sc iii}]}
\newcommand{\oiv}{[O\,{\sc iv}]}

\newcommand{\Ni}{[N\,{\sc i}]}
\newcommand{\nii}{[N\,{\sc ii}]}

\newcommand{\sii}{[S\,{\sc ii}]}
\newcommand{\siii}{[S\,{\sc iii}]}
\newcommand{\siv}{[S\,{\sc iv}]}

\newcommand{\hei}{He\,{\sc i}}
\newcommand{\heii}{He\,{\sc ii}}

\newcommand{\neii}{[Ne\,{\sc ii}]}
\newcommand{\neiii}{[Ne\,{\sc iii}]}
\newcommand{\neiv}{[Ne\,{\sc iv}]}
\newcommand{\nev}{[Ne\,{\sc v}]}

\newcommand{\ariii}{[Ar\,{\sc iii}]}
\newcommand{\ariv}{[Ar\,{\sc iv}]}

\newcommand{\cliii}{[Cl\,{\sc iii}]}
\newcommand{\cliv}{[Cl\,{\sc iv}]}
\newcommand{\kriv}{[Kr\,{\sc iv}]}
\newcommand{\krv}{[Kr\,{\sc v}]}
\newcommand{\xeiii}{[Xe\,{\sc iii}]}
\newcommand{\baii}{Ba\,{\sc ii}}

\newcommand{\feiii}{[Fe\,{\sc iii}]}
\newcommand{\feiv}{[Fe\,{\sc iv}]}

\newcommand{\ha}{H$\alpha$}
\newcommand{\hb}{H$\beta$}

\newcommand{\fii}{[F\,{\sc ii}]}
\newcommand{\fiii}{[F\,{\sc iii}]}
\newcommand{\fiv}{[F\,{\sc iv}]}

\newcommand{\kms}{km s$^{-1}$}

\slugcomment{Draft version August 18, 2010}
\shorttitle{
The Origin and Evolution of the Halo PN BoBn 1
}
\shortauthors{Otsuka et al.}

\begin{document}
\title{
The Origin and Evolution of the Halo PN BoBn 1: \\From a Viewpoint of
Chemical Abundances Based on Multiwavelength Spectra
}
\author{Masaaki Otsuka\altaffilmark{1},
Akito Tajitsu\altaffilmark{2}, 
Siek Hyung\altaffilmark{3}, 
Hideyuki Izumiura\altaffilmark{4}
}
\altaffiltext{1}{Space Telescope Science Institute, 3700 San Martin Dr., Baltimore,
MD 21218, USA; otsuka@stsci.edu}
\altaffiltext{2}{Subaru Telescope, NAOJ, 650 North A'ohoku Place, Hilo, HI 96720, USA}
\altaffiltext{3}{School of Science Education (Astronomy), Chungbuk National
University, 12 Gaeshin-dong Heungduk-gu, CheongJu, \\~~Chungbuk 361-763, Korea}
\altaffiltext{4}{Okayama Astrophysical Observatory, NAOJ, 3037-5, Honjo,
Kamogata-cho, Asakuchi-shi, Okayama, 
719-0232, Japan}

\begin{abstract}
We have performed a comprehensive chemical abundance analysis of 
the extremely metal-poor ([Ar/H]$<$--2) halo planetary nebula (PN) 
BoBn 1 based on {\it IUE} archive data, Subaru/HDS spectra, VLT/UVES 
archive data, and {\it Spitzer}/IRS spectra. We have detected over 600 
lines in total and calculated ionic and elemental abundances 
of 13 elements using detected optical recombination lines (ORLs) 
and collisionally excited lines (CELs). The estimations of C, N, O, 
and Ne abundances from the ORLs and Kr, Xe, and Ba from the CELs 
are done the first for this nebula, empirically and theoretically. 
The C, N, O, and Ne abundances from ORLs are systematically larger 
than those from CELs. 
The abundance discrepancies apart from O could be explained by a 
temperature fluctuation model, and that of O might be by a hydrogen deficient cold component model. 
We have detected 5 fluorine and several slow neutron capture elements (the {\it s}-process). The amounts 
of [F/H], [Kr/H], and [Xe/H] suggest that BoBn 1 is the most F-rich 
among F detected PNe and is a heavy {\it s}-process element rich PN. 
We have confirmed dust in the nebula that 
is composed of amorphous carbon and PAHs with 
a total mass of 5.8$\times$10$^{-6}$ $M_{\odot}$. The photo-ionization models built with non-LTE theoretical stellar 
atmospheres indicate that the progenitor was a 1-1.5 $M_{\odot}$ 
star that would evolve into a white dwarf with an $\sim$0.62 $M_{\odot}$ 
core mass and $\sim$0.09 $M_{\odot}$ ionized nebula. We have measured 
a heliocentric radial velocity of +191.6 $\pm$1.3 km s$^{-1}$ 
and expansion velocity 2$V_{\rm exp}$ of 40.5 $\pm$ 3.3 km s$^{-1}$ 
from an average over 300 lines. The derived elemental 
abundances have been reviewed from the standpoint of theoretical 
nucleosynthesis models. It is likely that the elemental abundances 
except N could be explained either by a 1.5 $M_{\odot}$ single star 
model or by a binary model composed of 0.75 $M_{\odot}$ + 1.5 $M_{\odot}$ 
stars. Careful examination implies that BoBn 1 has evolved from a 0.75 $M_{\odot}$ + 1.5 $M_{\odot}$ 
binary and experienced coalescence during the evolution to become 
a visible PN, similar to 
the other extremely metal-poor halo PN, K 648 in M 15.

\end{abstract}

\section{Introduction}
Planetary Nebulae (PNe) represent a stage in the evolution of 
low- to intermediate-mass stars with initial masses of 1-8 $M_{\odot}$. At the end
of their life, a star of such mass evolves first into a red giant branch
(RGB) star, then an asymptotic giant branch (AGB) star, next a PN, and finally a white
dwarf.  During their evolution, such stars eject a large amount of their mass.
The investigation of chemical abundances in PNe enables the determination of
how much of a progenitor's mass becomes a PN, when and how elements
synthesized in the progenitor were brought to the surface, and how
chemically rich the Galaxy was when the progenitors were born.

Currently, over 1,000 objects are regarded as PNe in the Galaxy (Acker
et al. 1991). Of these, about 14 objects 
have been identified as halo members from their location and kinematics since the PN 
K 648 was discovered in M 15 (Pease 1928). Halo PNe are interesting objects as they provide direct 
insight into the final evolution of old, low-mass halo stars, and they are able to 
convey important information for the study of low-mass star evolution and the early 
chemical conditions of the Galaxy. However, in extremely metal-poor and 
C- and N-rich ([C,N/O]$\ga$0, [Ar/H]$<$--2) halo PNe, 
there are unresolved issues on chemical abundances and 
evolution time scales. BoBn 1 (PN G108.4--76.1) is one of the C- and N-rich and extremely
metal-poor halo PNe ([C, N/O]$>$1, [Ar/H]=--2.22$\pm$0.09, [Fe/H]=--2.39$\pm$0.14; this work), 
which composes a class of PN together with K 648 (Otsuka 2007, see Table 20) and H 4-1 (Otsuka et
al. 2003).

The progenitors of halo PNe are generally thought to be $\sim$0.8 $M_{\odot}$ 
stars, which is the typical mass of a halo star. Above mentioned three metal-poor C- and N-rich halo
PNe, however, show signatures that they have evolved from massive
progenitors. For example, they would become N-rich, but would not
C-rich if they have evolved from $\sim$0.8 $M_{\odot}$ single stars with
[Fe/H]$\sim$--2.3 ($Z$=10$^{-4}$), according to the current stellar
evolution models (e.g., Fujimoto et al. 2000). To become C-rich PNe, the third dredge-up (TDU) must
take place in the late AGB phase. The efficiency of the TDU depends on the initial mass and composition, with 
increasing efficiency in models of increasing mass, or decreasing metallicity. 
At halo metallicities, it is predicted that the TDU is efficient in stars with 
initial masses greater than $\sim$1 $M_{\odot}$ (Karakas 2010; Stancliffe 2010).
Also, current stellar evolutionary models predict that the post-AGB evolution 
of a star with an initial mass $\sim$0.8 $M_{\odot}$ proceeds too slowly 
for a visible PN to be formed. The origin and evolution of halo PNe are 
still one of the unresolved big problems in this research field.

How did these progenitor stars become visible C- and N-rich halo PNe? 
To answer this key question would deepen understanding of low-mass star 
evolution, in particular, extremely metal-poor C-rich stars found in the 
Galactic halo, and Galactic chemical evolution at early phases. 
If we can accurately estimate elemental abundances and ejected masses, 
then we can directly estimate elemental yields synthesized by PNe 
progenitors which might provide a constraint to the growth-rate of core
mass, the number of thermal pulses and dredge-up mass. Hence we can 
build realistic stellar evolution models and Galactic chemical evolution models. 
We have observed these extremely metal-poor C- and N-rich 
halo PNe using the Subaru/High-Dispersion Spectrograph (HDS) and we also utilized
collecting archival data carefully in order to revise our picture of these objects. 
In this paper, we focus on BoBn 1.

The known nebular and stellar parameters of BoBn 1 are listed in Table \ref{bb1.natures}.
Zijlstra et al. (2006) have associated BoBn 1 with the leading 
tail of the Sagittarius (Sgr) Dwarf Spheroidal Galaxy, which traces 
several halo globular clusters. The heliocentric distance to the Sgr dwarf Galaxy is 
$\sim$24.8 kpc (Kunder \& Chaboyer 2009), while the distance to this object is between 
16.5 (Henry et al. 2004) and 29 kpc (Kingsburgh \& Barlow 1992). 

BoBn 1 is a unique PN in that it might possess information about the 
chemical building-up history of the Galactic halo. 
Otsuka et al. 
(2008a) found that the [C/Fe] and [N/Fe] abundances of BoBn 1 are
compatible with those of carbon-enhanced metal-poor (CEMP) stars. 
The C and N overabundances of CEMP can be explained by theoretical binary
interaction models (e.g., Komiya et al. 2007; Lau et al. 2007). 
Otsuka et al. (2008a) detected two fluorine (F) lines and found 
that BoBn 1 is the most F-enhanced and metal-poor PN among F-detected
PNe. They found that the C, N, and F overabundances of 
BoBn1 are comparable to those of the CEMP star HE1305+0132 (Schuler et
al. 2007). Through a comparison between the observed enhancements of C, N, 
and F with the theoretical binary nucleosynthesis model by Lugaro et
al. (2008), they concluded that BoBn 1 might share its origin and 
evolution with CEMP-{\it s} stars such as HE1305+0132, and if that is the
case the slow neutron capture process (the {\it s}-process) should be 
considered.

According to current evolutionary models of low- to intermediate-mass 
stars, the {\it s}-process elements are synthesized by slowly capturing
neutrons during the thermal pulse AGB phase. The {\it
s}-process elements together with carbon are brought to the
stellar surface by the TDU. If we could find signatures that BoBn 1 
has experienced binary evolutions such as mass transfer from a massive
companion and coalescence, the issues on chemical abundances and
evolutionary time scale would be simultaneously 
resolved. 
It would also be of great significance to reveal the origin of these elements
in the early Galaxy through the study of metal-poor objects such as BoBn 1.
We will search {\it s}-process elements and investigate their enhancement in BoBn 1.

In this paper, we present a chemical abundance analysis of BoBn 1 using
the newly obtained Subaru/HDS spectra, ESO VLT/UVES, {\it Spitzer}/IRS and
{\it IUE} archive data. We detect several candidate collisional
excited lines (CELs) of {\it s}-process elements and optical
recombination lines (ORLs) of N, O, and Ne. We determine ionic and
chemical abundances of 13 elements using ORLs and CELs. We construct 
a detailed photo-ionization model to derive the properties of the central star, 
ionized nebula, and dust. We also check consistency between our abundance
estimations and the model. Finally, we compare the empirically derived abundances
with the theoretical nucleosynthesis model values and discuss evolutionary
scenarios for BoBn 1.

\begin{table*}
\centering
\footnotesize
\caption{Nebular and Stellar Parameters of BoBn 1}
\begin{tabular}{@{}lll@{}}
\hline\hline
Quantity&Value &References\\
\hline
Name &BoBn 1 (PN G108.4--76.1)&discovered by Boeshaar \& Bond (1977)\\
Position (J2000.0)&$\alpha$=00:37:16.03 $\delta$=--13:42:58.48 &\\
Distance (kpc)&22.5;29 &Hawley \& Miller (1978);Kingsburgh \& Barlow (1992)\\
              &18.2;16.5&Mal'kov (1997);Henry et al. (2004)\\
              &24.8&Kunder \& Chaboyer (2009)\\
Size (arcsec) &$\sim$2 (diameter) &This work (see Fig. \ref{bb1.image})\\
$\log F(\rm H\beta)$ (erg cm$^{-2}$ s$^{-1}$)&--12.54;--12.43&Cuisinier et al.
 (1996);Wright et al. (2005)\\
&--12.38;--12.53(observed);--12.44(de-redden) &Kwitter et al. (2003);This work;This work\\
$c(\rm H\beta)$&0.18;0.0;0.09 &Cahn et al. (1992);Kwitter et al. (2003);This work\\
Rad. Velocity (km s$^{-1}$)&191.6 (heliocentric) &This work \\
Exp. Velocity ($2V_{\rm exp}$) &See Table \ref{exp}&This work\\
$T_{\epsilon}$ (K) &See Table \ref{diano_table}&This work\\
$n_{\epsilon}$ (cm$^{-3}$) &See Table \ref{diano_table}&This work\\
Abundances &See Table \ref{sam.abun} &therein Table \ref{sam.abun}\\
\\
$\log\,(L_{\star}/L_{\odot})$&3.57;3.72;3.07 &Mal'kov (1997);Zijlstra et al. (2006);This work\\
$\log\,g$ (cm s$^{-2}$)&5.52;6.5&Mal'kov (1997);This work \\
$M_{\star}$ (M$_{\odot}$)&0.575;0.62&Mal'kov (1997);This work\\
$T_{\star}$ (K)&125\,000;96\,300&Howard et al. (1997);Mal'kov (1997)\\
              &125\,260&This work\\
Magnitude &16(B),14.6(R),16.13(J),15.62(H),15.18(K)&Simbad data base\\
\hline
\end{tabular}
\label{bb1.natures}
\end{table*}

\section{Data \& Reductions}
\subsection{Subaru/HDS observations}

\begin{figure}
\centering
\includegraphics[width=80mm]{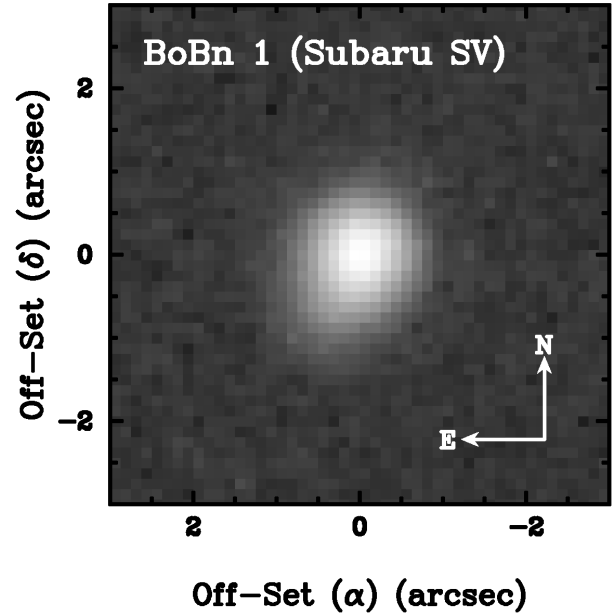}
\caption{The optical image of BoBn 1 taken by the Subaru/HDS 
silt viewer camera. The seeing was 0.4$''$ when we took this image and
 the nebular FWHM was measured to $\sim$1$''$.}
\label{bb1.image}
\end{figure}

The spectra of BoBn 1 were taken using the High-Dispersion Spectrograph
(HDS; Noguchi et al. 2002) attached to one of the two
Nasmyth foci of the 8.2-m Subaru telescope atop Mauna Kea, Hawaii on
October 6th 2008 (program ID: S08B-110, PI: M.Otsuka).
In Fig. \ref{bb1.image}, we present the optical image of BoBn 1 taken
by the HDS silt viewer camera ($\sim0.\hspace{-2pt}''12$
pixel$^{-1}$, no filters) during the HDS observation. The sky
condition was clear and stable, and the seeing was between $0.\hspace{-2pt}''4$
and $0.\hspace{-2pt}''6$. The FWHM of the image is $\sim$1$''$.
BoBn 1 shows a small protrusion toward the southeast.

Spectra were taken for two wavelength ranges, 3600-5400 {\AA}
(hereafter, the blue region spectra) and 4600-7500 {\AA} (the red region
spectra). An atmospheric dispersion corrector (ADC) was used to minimize the
differential atmospheric dispersion through the broad wavelength
region. In these spectra, there are many recombination lines of hydrogen,
helium, \& metals and collisionally excited lines (CELs). These numerous spectral lines allowed
us to derive reliable chemical compositions.
We used a slit width of $1.\hspace{-2pt}''2$ (0.6 mm) and a 2$\times$2 on-chip binning,
which enabled us to achieve a nominal spectral resolving power of
$R$=30\,000 with a 4.3 binned pixel sampling.
The slit length was set to avoid overlap of the echelle
diffraction orders at the shortest wavelength portion of the observing
wavelength range in each setup.
This corresponds to $8''$ (4.0mm), in which the nebula fits
well and can allow us to directly subtract sky background from the object frames.
The CCD sampling pitch along the slit length projected on the sky is
$\sim0.\hspace{-2pt}''276$ per binned pixel.
The achieved S/N is $>$40 at the nebular continuum
level even in both ends of each Echelle order.
The resulting resolving power is around
$R$ $>$33\,000, which derived from the mean of the full
width at half maximum (FWHM) of narrow Th-Ar and night sky lines.
All the data were taken as a series of 1800 sec exposure for weak emission-lines and 300 sec exposures for strong
emission-lines. The total exposure times were 16\,200 sec for red region spectra and 7200 sec for blue region spectra.
During the observation, we took several bias, instrumental flat lamp,
and Th-Ar comparison lamp frames, which were necessary for data reduction.
For the flux calibration, blaze function correction, and airmass
correction, we observed a standard star HR9087 at three different airmass.

\subsection{VLT/UVES archive data}
We also used archival high-dispersion spectra of BoBn 1, which are available from the
European Southern Observatory (ESO) archive. These spectra were observed 
on August 2002 (program ID: 069.D-0413, PI: M.Perinotto) and June 2007 
(program ID: 079.D-0788, PI: A.Zijlstra), using the Ultraviolet
Visual Echelle Spectrograph (UVES; Dekker et al. 2000) at the Nasmyth 
B focus of KUEYEN, the second of the four 8.2-m telescopes of the ESO 
Very Large Telescope (VLT) at Paranal, Chile. We call the August 2002 data
``UVES1'' and the June 2007 data ``UVES2'' hereafter. We used these data to
compensate for unobserved spectral regions and order gaps in the HDS
spectra. We normalized these data to the HDS spectra using the
intensities of detected lines in the overlapped regions between HDS and
UVES1 {\&} 2.

These archive spectra covered the wavelength range of 3300-6600 {\AA} in
UVES1 and 3300-9500 {\AA} in UVES2. The entrance slit
size in both observations was 11$''$ in length and 
1$.\hspace{-2pt}''5$ in width, giving $R$ $>$30\,000 derived from
Th-Ar and sky lines. The CCDs used in UVES have 15
$\mu$m pixel sizes. 
For UVES1 an 1$\times$1 binning CCD pattern was chosen. For UVES2 a 2$\times$2 on-chip binning pattern was chosen. 
The sampling pitch along the wavelength dispersion was $\sim$0.015-0.02
{\AA} pixel$^{-1}$ for UVES1 and $\sim$0.03-0.04 {\AA} pixel$^{-1}$ for
UVES2. The exposure time for UVES1 was 2700 sec $\times$ 4 frames, 10\,800 sec
in total. The exposure time for UVES2 was 1500 sec $\times$ 2 frames,
3000 sec in total. The standard star Feige 110 was observed for
flux calibration.

\begin{figure*}
\centering
\includegraphics[width=160mm]{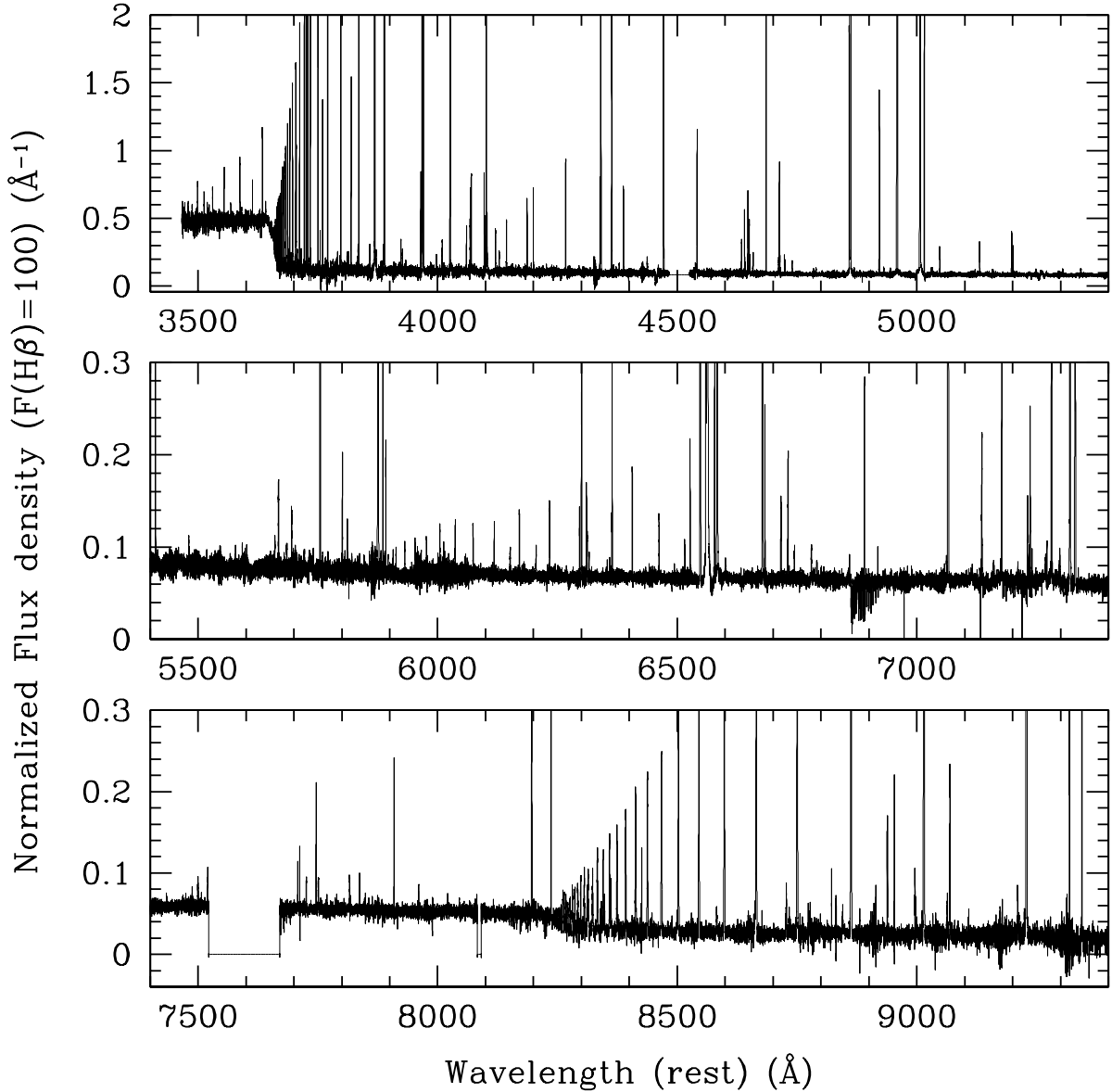}
\caption{The spectrum of BoBn 1. The flux density is normalized so that 
H$\beta$ flux $F(H\beta)$ = 100.}
\label{bb1.spec}
\end{figure*}

In Fig. \ref{bb1.spec}, we present the combined HDS and UVES spectrum of BoBn 1 normalized
to the {\hb} flux. The spectrum for the 
wavelength region of 3650-7500 {\AA} is from the HDS data and that of
3450-3650 {\AA} \& $>$7500 {\AA} is from the UVES data.

The observation logs are summarized in Table \ref{obslog}. The detected lines in
the Subaru/HDS and VLT/UVES spectra are listed in Appendix A.

\begin{table}
\centering
\caption{Journal of the HDS and UVES observations.\label{obslog}}
\begin{tabular}{@{}cccccr@{}}
\hline\hline
Instr.&Obs.Date&seeing&Range&binning&Exp.\\ 
      &        &($''$)      &(\AA)&&(sec)\\
\hline
HDS &2008/10/06&0.4-0.6&3650-5400&2$\times$2&1800$\times$4\\
    &         &0.4-0.6&3650-5400&2$\times$2&600$\times$3\\
    &         &0.4-0.6&4600-7500&2$\times$2&1800$\times$9\\
    &         &0.4-0.6&4600-7500&2$\times$2&600$\times$3\\
UVES&2002/08/04&0.8-1.5&3300-6600&1$\times$1&2700$\times$4\\
    &2007/06/30&0.5-0.7&3300-9500&2$\times$2&1500$\times$2\\
\hline
\end{tabular}
\end{table}

\subsection{{\it IUE} archive data}
We complemented optical spectra with UV spectra obtained by the 
International Ultraviolet Explorer ({\it IUE}) to derive C$^{+}$,
C$^{2+}$, N$^{2+}$, and N$^{3+}$ abundances from semi-forbidden lines C\,{\sc ii}], C\,{\sc
iii}], N\,{\sc iii}], and N\,{\sc iv}], since these emission lines cannot be observed in
the optical region. These {\it IUE} spectra were retrieved from 
the Multi-mission Archive at the STScI (MAST). 
We collected the high- and low-resolution {\it IUE} spectra taken by the Short
Wavelength Prime (SWP) and Long Wavelength Prime/Long Wavelength
Redundant (LWP/LWR) cameras. Our used data set is listed 
in Table \ref{iueobs}, and the wavelength dispersion mode is indicated in the column 3.
All the {\it IUE} observations were made using the large aperture (10.3$\times$23 arcsec$^{2}$). 
SWP and LWP/LWR spectra cover the wavelength range of 1150-1980 {\AA}
and 1850-3350 {\AA}, respectively. 
For each SWP and LWP/LWR spectra, we did median combine to improve the S/N. 
The combined short wavelength spectrum was used to measure fluxes of emission-lines 
in $\lesssim$1910 {\AA} because this allowed us to separate C~{\sc iii}$]$$\lambda$1906/08 and 
C~{\sc iv} $\lambda$1548/51 lines. C~{\sc iii}$]$$\lambda$1906/08 are important as a density diagnostic. 
The combined long wavelength spectrum was for measurements of emission-line fluxes in 
$\gtrsim$2000 {\AA}. The measured line fluxes were normalized to the {\hb} flux using theoretical ratios of He\,{\sc ii} 
$I$($\lambda$1640)/($\lambda$4686) for the short wavelength spectrum and
$I$($\lambda$2512)/($\lambda$4686) for the long wavelength spectrum, respectively, adopting an electron temperature $T_{\epsilon}$ =
8840 K and density $n_{\epsilon}$ = 10$^{4}$ cm$^{-3}$ as given by Storey \& Hummer (1995), then normalized to the {\hb} flux.  
The interstellar extinction correction was made using equation (\ref{redc}) (see section 3.1). 
The observed and normalized fluxes of detected lines are listed in the columns 4 and 5 of Table \ref{iue_list}, respectively.

\begin{table}
\centering
\caption{Journal of {\it IUE} observations. \label{iueobs}}
\begin{tabular}{@{}cccccr@{}}
\hline\hline
Camera&Data ID.&disp.&Range &Obs.Date&Exp.time\\
      &      &  &(\AA) &&(sec)\\ 
\hline
LWR&  16515 &low&1850-3350&    1983-08-03&  6780 \\
LWP&  23692 &low&1850-3350&    1992-08-13&  1500 \\
LWP&  23697 &low&1850-3350&    1992-08-14&  7200 \\
LWP&  23699 &low&1850-3350&    1992-08-15&  12\,000\\
SWP&  45367 &high&1150-1980&    1992-08-18&  7200 \\
LWP&  23713 &low&1850-3350&    1992-08-18&  1800 \\
SWP&  45369 &high&1150-1980&    1992-08-18 & 10\,500\\
SWP&  45371 &high&1150-1980&     1992-08-19&  19\,800\\
SWP&  45386 &high&1150-1980&    1992-08-21&  9000 \\
\hline
\end{tabular}
\end{table}

\subsection{{\it Spitzer} archive data}
\begin{figure*}
\centering
\includegraphics[width=160mm]{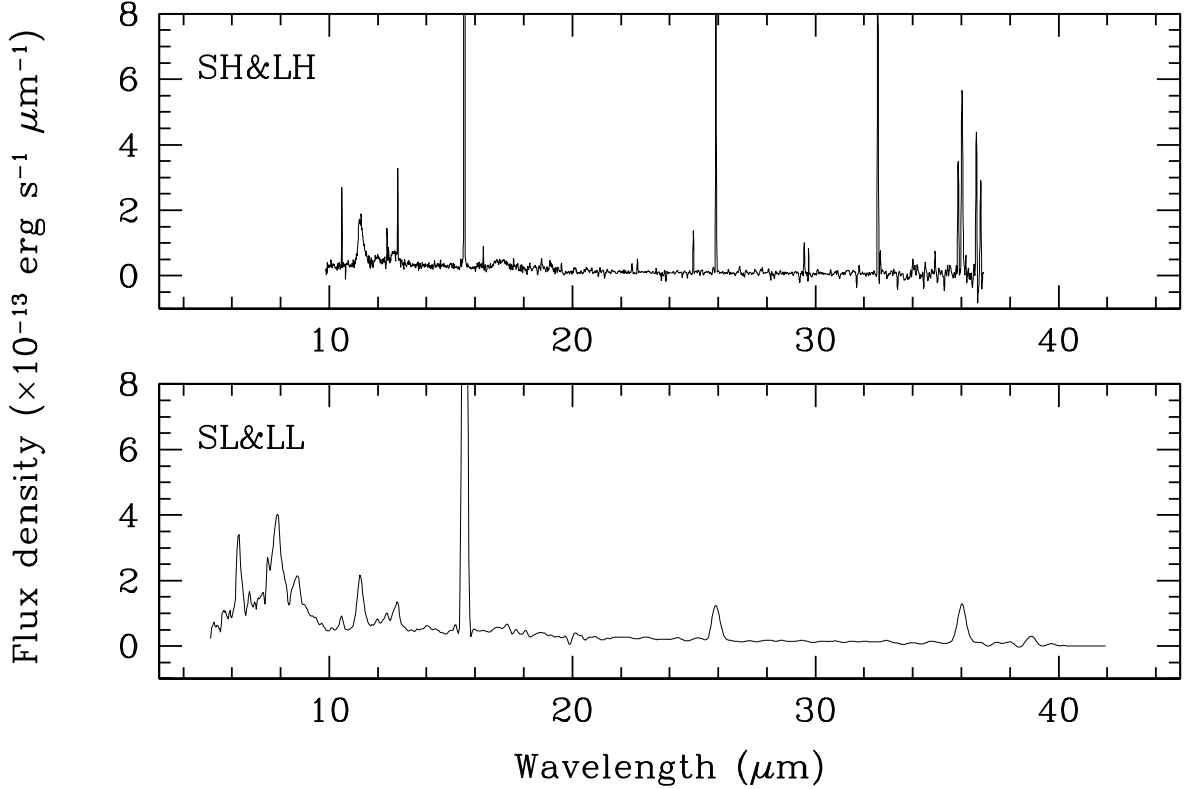}
\caption{The Spitzer spectra of BoBn 1 obtained by the SH and LH ({\it
 upper panel}) and the SL and LL modules ({\it lower panel}).}
\label{bb1.spec.spit}
\end{figure*}

We used two data sets (program IDs: P30333 PI: A.Zijlstra; P30652 PI:
J.Bernard-Salas) taken by the {\it Spitzer} space telescope in December 2006. 
The data were taken by the Infrared 
Spectrograph (IRS, Houck et al. 2004) with the SH (9.5-19.5 $\mu$m), 
LH (5.4-37$\mu$m), SL (5.2-14.5 $\mu$m) and LL (14-38$\mu$m) modules. 
In Fig. \ref{bb1.spec.spit} we present the {\it Spitzer} spectra of BoBn 1. 
We downloaded these data using {\it Leopard} provided by the Spitzer Science Center. The
one-dimensional spectra were extracted using {\it Spice} version c15.0A. 
We extracted a region within $\pm$1$''$ from the center of each spectral
order summed up along the spatial direction. For SH and LH spectra, 
we subtracted sky background using off-set spectra. We normalized the SL and LL 
data to the SH and LH using the measured fluxes of [S~{\sc iv}] $\lambda$10.5$\mu$m, 
H~{\sc i} $\lambda$12.4 $\mu$m, [Ne~{\sc ii}] $\lambda$12.8$\mu$m, 
[Ne~{\sc iii}] $\lambda$15.6$\mu$m, and [Ne~{\sc iii}] $\lambda$36.0$\mu$m. Finally, 
the measured line fluxes were normalized to the {\hb} flux. 
The observed line ratio 
H~{\sc i} $I$(11.2$\mu$m)/$I$($\lambda$4861) (3.1$\times$10$^{-3}$)
is consistent with the theoretical value (3.15$\times$10$^{-3}$) for $T_{\epsilon}$ =
8840 K and $n_{\epsilon}$ = 10$^{4}$ cm$^{-3}$ as given
by Storey \& Hummer (1995). We did not therefore perform interstellar extinction correction.

\begin{figure*}
\centering
\includegraphics[width=160mm]{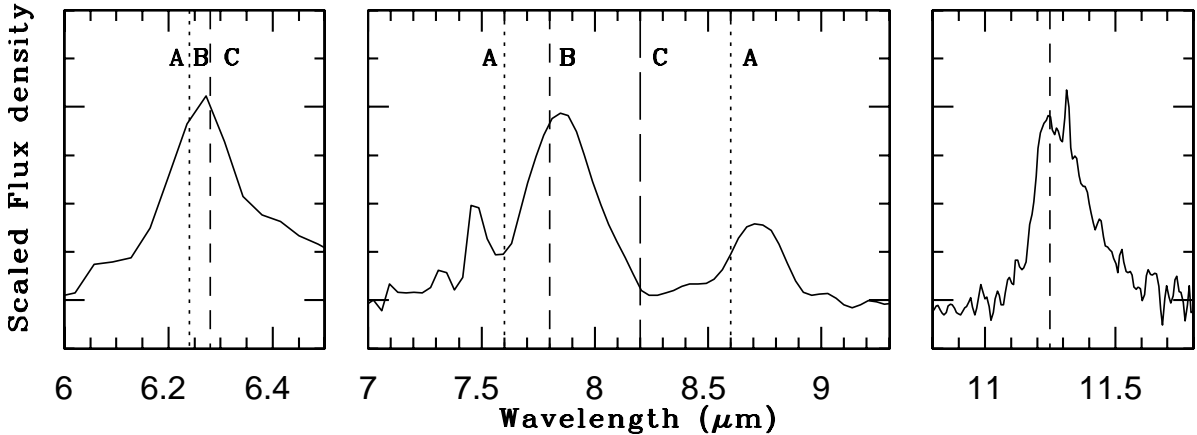}
\caption{The 6-11$\mu$m PAHs profiles. The PAHs 6.2, 7.9, and 8.7 $\mu$m 
profile classification (A, B, and C) of Peeters et al. (2002) and the 11.2 $\mu$m 
classification of van Diedenhoven et al. (2004) are given. }
\label{pah}
\end{figure*}

The observed and normalized fluxes of detected lines are listed in Table \ref{spitzer_list}.
In addition to the ionized gas emissions, the amorphous carbon dust continuum and 
the polycyclic aromatic hydrocarbons (PAHs) feature around 6.2, 7.7, 8.7, and 11.2 $\mu$m are found for 
the first time. In Fig. \ref{pah}, we present these PAH features. The 11.2 $\mu$m emission 
line is a complex of the narrow width H~{\sc i} 11.2 $\mu$m and the broad PAH 11.2 $\mu$m. 
The 6.2, 7.7, and 8.7 $\mu$m bands emit strongly in ionized PAHs, while the 11.2 $\mu$m does 
in neutral PAHs (Bernard-Salas et al. 2009). According to the PAH line profile classifications by 
Peeters et al. (2002) and van Diedenhoven et al. (2004), BoBn 1's PAH line-profiles 
belong to class B. Bernard-Salas et al. (2009) classified 10 of 14 Magellanic Clouds (MCs) PNe 
into class B based on {\it Spitzer} spectra. In measuring PAH band fluxes, we used local continuum 
subtracted spectrum by a spline function fitting. We followed Bernard-Salas et al. (2009) and measured 
integrated fluxes between 6.1 and 6.6 $\mu$m for the 6.2 $\mu$m PAH band, 7.2-8.3 $\mu$m for the 7.7 
$\mu$m PAH, 8.3-8.9 $\mu$m for the 8.6 $\mu$m PAH, and 11.1-11.7 $\mu$m for the 11.2 $\mu$m. 
The observed PAH flux ratios $I$(6.2$\mu$m)/$I$(11.2$\mu$m) and $I$(7.7$\mu$m)/$I$(11.2$\mu$m) follow a 
correlation among MCs PNe, shown in Fig. 2 of Bernard-Salas et al. (2009). BoBn 1 has a hot central 
star ($>$10$^{5}$ K), so that ionized PAH might be dominant. However, these line ratios of BoBn 1 
are somewhat lower than those of excited MCs PNe. One must take a look at a part of the neutral PAH 
emissions in a photodissociation region (PDR), too. 

We found a plateau between 10 and 14 $\mu$m, which are believed to be related to PAH clusters 
(Bernard-Salas et al. 2009). Meanwhile, MgS feature around 30 $\mu$m sometimes observed in C-rich PNe, 
was unseen in BoBn 1.

\subsection{Data reduction}
Data reduction and emission line analysis were performed mainly with 
a long-slit reduction package {\tt noao.twodspec} in
{\tt IRAF}\footnote{{\tt IRAF} is distributed by
the National Optical Astronomy Observatories, which are operated by 
the Association of Universities for Research in Astronomy (AURA), 
Inc., under a cooperative agreement with the National Science Foundation.}.
Data reduction was performed in a standard manner.

First, we made a zero-intensity level correction to all frames
including flat lamp, object, and Th-Ar comparison frames
using the overscan region of each frame and the mean bias frames.
We also removed cosmic ray events and hot pixels from the object frames.
Second, we trimmed the overscan region and removed scattered light 
from the flat lamp and object frames. Third, we made a CCD sensitivity
correction to the object frames using the median flat frames.
Fourth, we extracted a two-dimensional spectrum from each echelle
diffraction order of each object frame and made a wavelength calibration
using at least two Th-Ar frames taken before and after the object frame.
We referred to the Subaru/HDS comparison
atlas\footnote{http://www.naoj.org/Observing/Instruments/HDS/wavecal.html} and a Th-Ar atlas.
For the wavelength calibration, we fitted the wavelength dispersion against
the pixel number with a fourth- or fifth-order polynomial function. With this
order, any systematic trend did not show up in the residuals and the fitting appears to be satisfactory.
We also made a distortion correction along the slit length direction
using the mean Th-Ar spectrum as a reference. We fitted the slit image
in the Th-Ar spectrum with a two-dimensional function.
For the HDS spectra, we adopted fourth- and third-order polynomial functions
for the wavelength and space directions, respectively. The fitting
residual was of the order of 10$^{-4}$ {\AA}. For the UVES spectra, we adopted
third- and second-order polynomial functions for the wavelength and
space directions, respectively. The fitting residual was
of the order of 10$^{-3}$ {\AA}. 
Fifth, we determined a sensitivity function using sky-subtracted
standard star frames and obtained  sky-subtracted and flux-calibrated two-dimensional PN spectra.
The probable error in the flux calibration was estimated to be less than 5 ${\%}$.
Finally, we made a spatially integrated one-dimensional spectrum, and we
combined all the observed echelle orders using {\tt IRAF} task {\tt scombine}.

In measuring emission line fluxes, we assumed that the line profiles 
were all Gaussian and we applied multiple Gaussian fitting techniques.

\section{Results}


\begin{table}
\centering
\caption{The detected lines in the {\it IUE} spectra.\label{iue_list}}
\begin{tabular}{@{}clccr@{}}
\hline\hline
$\lambda_{\rm lab.}$&Ion&$f$($\lambda$)&$F$($\lambda$)&$I$($\lambda$)\\
(\AA)&&&(erg s$^{-1}$ cm$^{-2}$)&$[$$I$({\hb})=100$]$\\
\hline
1485&	N\,{\sc iv}]&	                   1.306&1.41(--13) $\pm$ 5.47(--14) &45.81 $\pm$ 17.84\\
1548&	C\,{\sc iv}&	                   1.239&3.27(--12) $\pm$ 5.04(--14) &1052.6 $\pm$ 20.22\\
1551&	C\,{\sc iv}&	                   1.237&1.62(--12) $\pm$ 4.27(--14) &519.3 $\pm$ 14.96\\	
1640&	He\,{\sc ii}&	                   1.177&5.13(--13) $\pm$ 7.40(--14) &162.97 $\pm$ 23.57\\	
1750&	N\,{\sc iii}]&	                   1.154&1.52(--13) $\pm$ 4.16(--14) &48.06 $\pm$ 13.16\\	
1906&	C\,{\sc iii}]&	                   1.255&2.60(--12) $\pm$ 2.50(--14) &838.88 $\pm$ 12.65\\
1908&	C\,{\sc iii}]&	                   1.258&1.87(--12) $\pm$ 2.33(--14) &602.72 $\pm$ 10.28\\
2324&	C\,{\sc ii}]&       1.388&2.87(--13) $\pm$ 3.60(--14) &36.64 $\pm$ 4.63\\	
    &+[O\,{\sc iii}]&            \\
2424&	[Ne\,{\sc iv}]&	                   1.134&1.41(--13) $\pm$ 1.46(--14) &17.12 $\pm$ 1.78\\	
2512&	He\,{\sc ii}&	                   0.969&2.66(--14) $\pm$ 1.29(--14) &3.13 $\pm$ 1.52\\
\hline
\end{tabular}
\tablecomments{$X$($-Y$) stands for $X$ $\times$ 10$^{-Y}$. 
$I$({\hb}) = 3.63(--13) $\pm$ 6.47(--14) (see section 3.1).}
\end{table}

\begin{table}
\centering
\caption{The detected lines in the {\it Spitzer} spectra.
\label{spitzer_list}}
\begin{tabular}{@{}clcr@{}}
\hline\hline
$\lambda_{\rm lab.}$&Ion&$F$($\lambda$)&$I$($\lambda$)\\
($\mu$m)&&(erg s$^{-1}$ cm$^{-2}$)&$[$$I$({\hb})=100$]$\\
\hline
6.2&	PAH   &2.62(--14) $\pm$ 1.52(--15)    &7.22 $\pm$ 1.35\\
7.7&	PAH   &8.13(--14) $\pm$ 2.31(--15)    &22.39 $\pm$ 4.04\\
8.6&	PAH   &2.02(--14) $\pm$ 1.23(--15)    &5.57 $\pm$ 1.05\\
10.5&	$[$S\,{\sc iv}$]$   &7.36(--15) $\pm$ 1.93(--16)    &1.92 $\pm$ 0.05\\
11.3&	H\,{\sc i}          &1.18(--15) $\pm$ 1.81(--16)    &0.31 $\pm$ 0.05\\
11.3&	PAH   &4.97(--14) $\pm$ 9.65(--15)    &13.68 $\pm$ 3.61\\
12.4&	H\,{\sc i}	    &3.89(--15) $\pm$ 3.01(--16)    &1.02 $\pm$ 0.08\\
12.5&	He\,{\sc i}	    &1.72(--15) $\pm$ 4.17(--16)     &0.45 $\pm$ 0.11\\
12.8&	$[$Ne {\sc ii}$]$   &9.53(--15) $\pm$ 3.13(--16)    &2.49 $\pm$ 0.08\\
14.6&	He\,{\sc i}	    &5.85(--16) $\pm$ 2.40(--16)    &0.15 $\pm$ 0.06\\
15.6&	$[$Ne\,{\sc iii}$]$ &6.17(--13) $\pm$ 4.98(--15)    &161.13 $\pm$ 1.30\\
16.4&	He\,{\sc i}	    &2.19(--15) $\pm$ 2.65(--16)    &0.57 $\pm$ 0.07\\
18.7&	$[$S\,{\sc iii}$]$  &2.65(--15) $\pm$ 1.79(--16)    &0.69 $\pm$ 0.05\\
19.1&	H\,{\sc i}	    &1.14(--15) $\pm$ 2.64(--16)    &0.30 $\pm$ 0.07\\
25.9&	$[$O\,{\sc iv}$]$   &4.77(--14) $\pm$ 5.20(--16)    &12.46 $\pm$ 0.14\\
36.0&	$[$Ne\,{\sc iii}$]$ &5.10(--14)	$\pm$ 7.04(--15)    &13.32 $\pm$ 1.84\\
\hline
\end{tabular}
\tablecomments{$X$($-Y$) stands for $X$ $\times$ 10$^{-Y}$. $I$({\hb}) = 3.63(--13) $\pm$ 6.47(--14) (see section 3.1).}
\end{table}

\subsection{Interstellar reddening correction \label{irc}}
We have detected over 600 emission lines in total. Before proceeding to the chemical abundance analysis, 
it is necessary to correct the spectra for the effects of absorption due to the Earth's 
atmosphere and interstellar reddening. The former was performed using 
experimental functions measured at the Keck observatories and the ESO/VLT.
The interstellar reddening correction was made by determining
the reddening coefficient at {\hb}, $c$({\hb}).
We fitted the observed intensity ratio of {\ha} to {\hb} 
with the theoretical ratios computed by Storey \& Hummer (1995). 
Two different situations are assumed for some lines: 
Case A assumes that the nebula is transparent to
the lines of all series of hydrogen; Case B assumes the nebula is 
partially opaque to the lines of the Lyman series but is transparent for the
Balmer series of hydrogen. Initially, we assumed that 
$T_{\epsilon}$ = 10$^{4}$ K and $n_{\epsilon}$ 
= 10$^{4}$ cm$^{-3}$ in Case B, and we estimated $c$({\hb}) = 0.087
$\pm$ 0.004 from the HDS spectra. That is an intermediate value between
Cahn et al. (1992) and Kwitter et al. (2003) (see Table 1). 
For the UVES spectra, we estimated $c$({\hb}) = 0.066 by the same manner. 
From Seaton's (1979) relation $c$({\hb}) 
= 1.47$E(B-V)$ one obtains $E(B-V)$ = 0.06 for the HDS spectra and 0.04 for 
UVES spectra, which are comparable to the Galactic value (0.02) to the direction to 
BoBn 1 measured by the Galactic extinction model of Schlegel et al. (1998).

All of the line
intensities were then de-reddened using the formula: 
\medskip

\begin{equation}
\label{redc}
\log_{10}\left[\frac{I(\lambda)}{I{\rm (H\beta)}}\right] = 
\log_{10}\left[\frac{F(\lambda)}{F{\rm (H\beta)}}\right] + c({\rm H\beta})f(\lambda),
\end{equation}
\medskip

\noindent
where $I(\lambda)$ is the de-reddened line flux;
$F(\lambda)$ is the observed line flux;
and $f(\lambda)$ is the interstellar extinction at $\lambda$. 
We adopted the reddening law of Cardelli et al. (1989) with the standard
value of $R_{V}$ = 3.1 for $f(\lambda)$. We
observed $F(H\beta)$ = 2.56$\times$10$^{-13}$ $\pm$
2.13$\times$10$^{-16}$ erg s$^{-1}$ cm$^{-2}$ ($X$($-Y$) stands for $X$ $\times$
10$^{-Y}$, hereafter) within the $1.\hspace{-2pt }''2$ slit 
in the HDS observation. We estimated captured light from BoBn 1 using the
image presented in Fig. \ref{bb1.image}, to be about 
86.8 $\%$ of the light from BoBn 1 in the HDS observation. 
The intrinsic observed {\hb} flux is 2.95(--13) $\pm$ 3.45(--16) erg
s$^{-1}$ cm$^{-2}$ and the de-reddened {\hb} flux is 3.63(--13) $\pm$
6.47(--14) erg s$^{-1}$ cm$^{-2}$ including the error of $c$({\hb}).

\subsection{Radial and expansion velocities }
\begin{figure*}
\centering
\includegraphics[width=140mm]{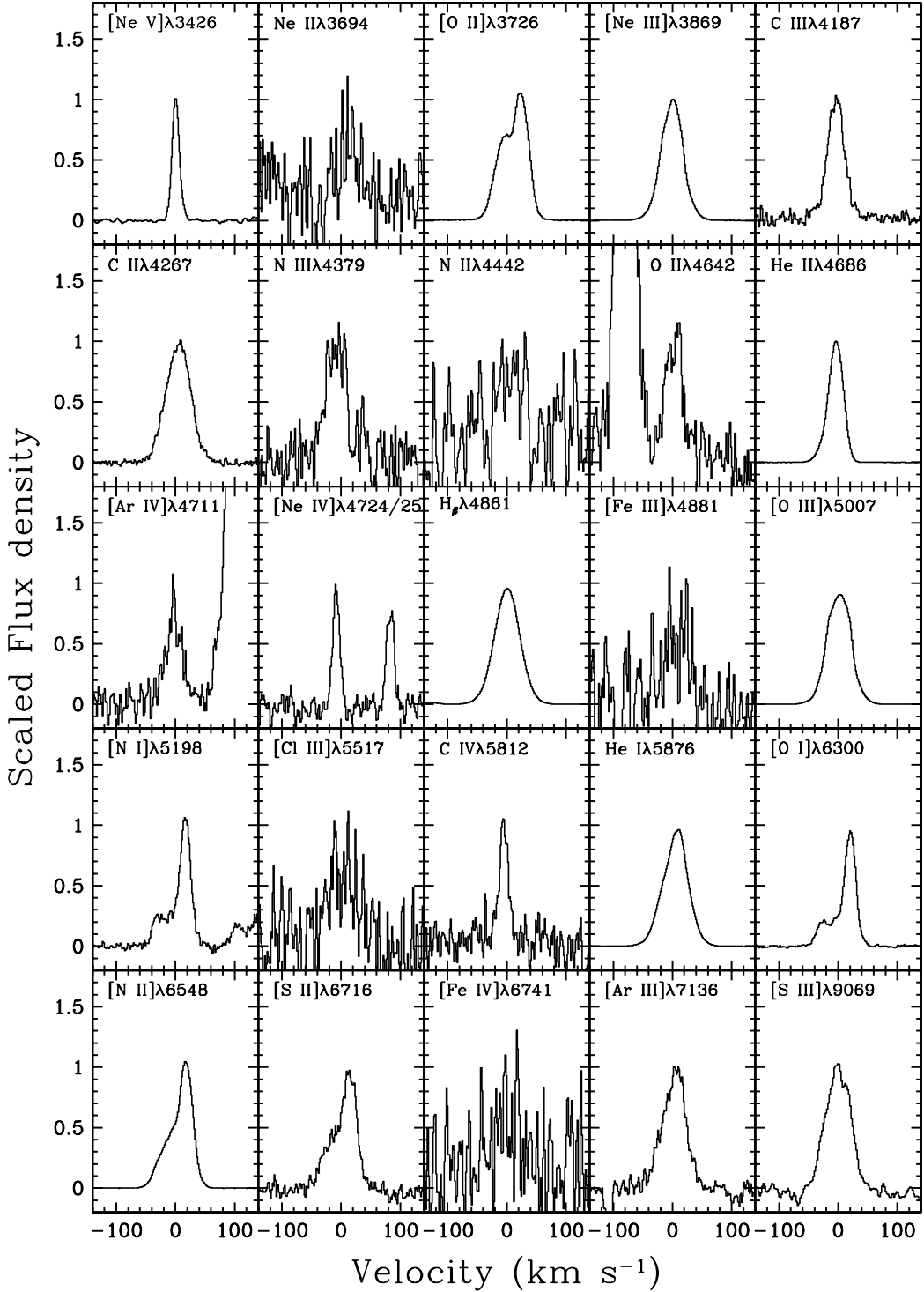}
\caption{
The line-profiles of selected ions. Vertical and horizontal axes are scaled flux
density and velocity with respect to the systemic radial velocity of 
+191.60 {\kms}, respectively. All are from the HDS
 spectra except [Ne\,{\sc v}]$\lambda$3426 and [S\,{\sc
 iii}]$\lambda$9069 which are from UVES1 and UVES2, respectively. 
 \label{lp}}
\end{figure*}

We present the line-profiles of selected ions in Fig. \ref{lp}. 
The observed wavelength at the time of observation was corrected 
to the averaged line-of-sight heliocentric radial velocity of +191.60 $\pm$ 1.25 
{\kms} among over 300 lines detected in the HDS spectra. The
line-profiles can be represented by a single Gaussian for weak forbidden
lines such as [Ne\,{\sc v}] $\lambda$3426 and metal recombination lines
such as O\,{\sc ii} $\lambda$4642. For the others, the profiles can be
represented by the sum of two or three Gaussian components.

Most of the
detected lines are asymmetric profile, in particular the profiles of 
low-ionization potential ions show strong asymmetry. The asymmetric line-profiles are
sometimes observed in bipolar PNe having an equatorial disk
structure. The similar line-profiles are also observed in the halo PN H4-1 (Otsuka et
al. 2006). H4-1 has an equatorial disk structure and multi-polar
nebulae. The elongated nebular shape of BoBn 1 (Fig. \ref{bb1.image})
might indicate the presence of such an equatorial disk. 
The receding ionized gas (especially, 
low-ionization potential ions) from the observers around the central star 
would be strongly weakened by the equatorial disk. In contrast, the relatively 
large extent bipolar flows perpendicular to the equatorial disk might be 
unaffected by the disk. Due to such a geometry, we observe asymmetric
line-profiles.

In Table \ref{exp}, we present twice the expansion velocity 2$V_{\rm
exp}$ measured from 
selected lines. When we fit line-profile with two or three Gaussian components, we define 
that 2$V_{\rm exp}$ corresponds to the difference between the positions of 
the red and blue shifted Gaussian peak components. We call expansion velocity measured 
by this method `2$V_{\rm exp}$(a)'. 

When we can fit line-profile with
single Gaussian, we determine 2$V_{\rm exp}$ from following equation,

\begin{equation}
2V_{\rm exp} = (V_{\rm FWHM}^2 - V_{\rm therm}^2 - V_{\rm instr}^2)^{1/2},
\label{expf}
\end{equation}

\noindent where $V_{\rm FWHM}$ is the velocity FWHM of each Gaussian component.
$V_{\rm therm}$ and $V_{\rm instr}$ ($\sim$9 {\kms}) are the thermal broadening and the 
instrumental broadening, respectively (Robinson et al. 1982). $V_{\rm therm}$ is represented by
21.4$(T_{4}/A)^{1/2}$, where $T_{4}$ is the electron temperature 
(in units of 10$^{4}$ K) and $A$ is the atomic weight of the target
ion. For CELs, we adopted $T_{4}$ listed in Table \ref{temp_ne_cles}. 
For ORLs, we adopted $T_{4}$ = 0.88. We call twice the expansion velocity measured 
by Equation (\ref{expf}) `2$V_{\rm exp}$(b)'.  We converted 2$V_{\rm exp}$(a)
into 2$V_{\rm exp}$(b) using the relation 2$V_{\rm exp}$(b) = (1.74
$\pm$ 0.12) $\times$ 2$V_{\rm exp}$(a) for 2$V_{\rm exp}$(b) $>$ 16
{\kms}, which can be applied only to BoBn 1. We adopted 2$V_{\rm
exp}$(b) as twice the expansion velocities for BoBn
1. The averaged $2V_{\rm exp}$(b) is 40.5 $\pm$ 3.28 {\kms} among
selected lines listed in Table \ref{exp} and 
33.04 $\pm$ 2.61 {\kms} among over 300 detected lines in the HDS
spectra.

\begin{figure}
\centering
\includegraphics[width=80mm]{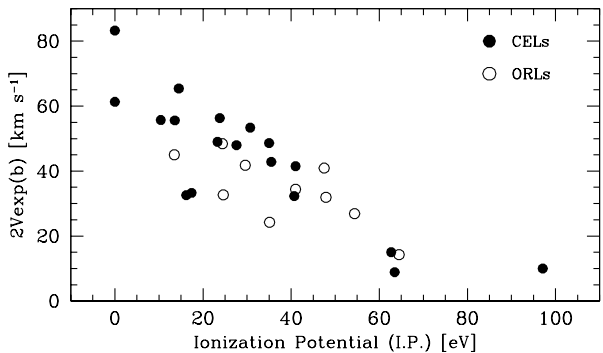}
\caption{Relation between 2$V_{\rm exp}$(b) and ionization potential
 (I.P.). The filled circles are the values from the CELs and the open
 circles are from the ORLs. \label{ip_exp}}
\end{figure}

In Fig. \ref{ip_exp} we present relation between 2$V_{\rm exp}$(b) and
ionization potential (I.P.). When we assume that BoBn 1 has a standard
ionized structure (i.e., high I.P. lines are emitted from close regions to the central
star and low I.P. lines are from far regions), expansion velocity of
BoBn 1 seems to be proportional to the distance from the central
star. BoBn 1 might have Hubble type flows. We found that 2$V_{\rm exp}$(b)
values from ORLs are slightly smaller than CELs with the same I.P., for example, O\,{\sc ii} \& {\oiii}
(35.5 eV) and Ne\,{\sc ii} \& [Ne\,{\sc iii}] (41.0 eV). O\,{\sc ii} and
Ne\,{\sc ii} might be emitted from colder regions than {\oiii} and [Ne\,{\sc
iii}] do.

\begin{table}
\centering
\caption{Twice the expansion velocities from selected lines. \label{exp}}
\begin{tabular}{lcccc}
\hline\hline
Ion	&$\lambda_{\rm lab}$&I.P.&$2V_{\rm exp}$(a)&$2V_{\rm exp}$(b)\\
	&({\AA}) &(eV)&({\kms})&({\kms})\\
\hline
$[$Ne\,{\sc v}$]$ &3425&97.1&$\cdots$	&10.0\\
Ne\,{\sc ii}	  &3694&41.0&$\cdots$	&34.4\\
$[$O\,{\sc ii}$]$  &3726&13.6&32.0	&55.6\\
$[$Ne\,{\sc iii}$]$&3868&41.0&$\cdots$	&41.5\\
$[$F\,{\sc iv}$]$  &4060&62.7    &$\cdots$	&15.0\\	
C\,{\sc iii}	  &4187&47.9&$\cdots$	&31.9\\		
C\,{\sc ii}	  &4267&24.4&$\cdots$	&48.4\\
N\,{\sc iii}	  &4379&47.5&$\cdots$	&40.9\\
N\,{\sc ii}	  &4442&29.6&$\cdots$	&41.8\\
O\,{\sc ii}	  &4642&35.5&$\cdots$	&24.2\\
He\,{\sc ii}	  &4686&54.4&$\cdots$	&26.9\\
$[$Ar\,{\sc iv}$]$ &4711&40.7&$\cdots$	&32.3\\
$[$Ne\,{\sc iv}$]$ &4724&63.5&$\cdots$	&8.9\\
$[$F\,{\sc ii}$]$  &4798&17.4&$\cdots$	&33.3\\
H$\beta		$ &4861&13.5&$\cdots$	&45.0\\
$[$Fe\,{\sc iii}$]$&4881&16.2&$\cdots$	&32.6\\
$[$O\,{\sc iii}$]$ &5007&35.5&$\cdots$	&42.8\\
$[$N\,{\sc i}$]$	  &5198&0   &35.3	&61.3\\
$[$Cl\,{\sc iii}$]$&5517&23.8&$\cdots$	&56.3\\	
C\,{\sc iv}	  &5812&64.5&$\cdots$	&14.3\\
$[$F\,{\sc iii}$]$  &5733&35.0&$\cdots$	&48.6\\
He\,{\sc i}	  &5876&24.6&$\cdots$	&32.7\\
$[$O\,{\sc i}$]$	  &6300&0   &48.0	&83.3\\		
$[$N\,{\sc ii}$]$  &6548&14.5&37.7	&65.4\\
$[$S\,{\sc ii}$]$  &6716&10.4&32.1	&55.7\\
$[$Fe\,{\sc iv}$]$ &6741&30.7&$\cdots$	&53.4\\
$[$Ar\,{\sc iii}$]$&7135&27.6&$\cdots$	&48.0\\
$[$S\,{\sc iii}$]$ &9069&23.3&$\cdots$	&49.7\\
\hline
\end{tabular}
\tablecomments{The probable error of $2V_{\rm exp}$(b) is within 5
 {\kms}. We assume 2$V_{\rm exp}$(b) as twice the expansion velocity of BoBn 1
 (see text).}
\end{table}

\begin{table*}
\centering
\footnotesize
\caption{Atomic data references for CELs. \label{atomf}}
\begin{tabular}{@{}lll@{}}
\hline\hline
Line &transition probabilities $A_{ji}$ &collisional strength $\Omega_{ij}$\\
\hline
$[$C\,{\sc i}$]$ &Froese-Fischer \& Saha (1985)&Johnson et al. (1987); P\'{e}quignot \& Aldrovandi (1976)\\
$[$C\,{\sc ii}$]$&Nussbaumer \& Storey (1981); Froese-Fischer (1994)  &Blum \& Pradhan (1992)\\
C\,{\sc iii}$]$&Wiese et al. (1996)
&Berrington et al. (1985)\\
C\,{\sc iv}&Wiese et al. (1996)                   &Badnell \& Pindzola (2000); Martin et al. (1993)            \\
{\Ni}&Wiese et al. (1996) &P\'{e}quignot \& Aldrovandi (1976); Dopita et al. (1976)\\
{\nii}&Wiese et al. (1996) &Lennon \& Burke (1994)\\
N\,{\sc iii}$]$&Brage et al. (1995); Froese-Fischer (1983) &Blum \& Pradhan (1992)\\
N\,{\sc iv}$]$&Wiese et al. (1996) &Ramsbottom et al. (1994)\\
{\oi} &Wiese et al. (1996) &Bhatia \& Kastner (1995)\\
{\oii} &Wiese et al. (1996) &McLaughlin \& Bell (1993); Pradhan (1976)\\
{\oiii} &Wiese et al. (1996) &Lennon \& Burke (1994)\\
{\oiv}&Wiese et al. (1996) &Blum \& Pradhan (1992)\\
$[$F\,{\sc ii}$]$&Storey \& Zeippen (2000); Baluja \& Zeippen (1988)& Butler \& Zeippen (1994)              \\
$[$F\,{\sc iii}$]$&Naqvi (1951)                    &See Text                 \\
$[$F\,{\sc iv}$]$&Garstang (1951); Storey \& Zeippen (2000)                     & Lennon \& Burke (1994)               \\
$[$Ne\,{\sc ii}$]$&Saraph \& Tully (1994)                    &Saraph \& Tully (1994)                \\
{\neiii} &Mendoza (1983); Kaufman \& Sugar (1986) &McLaughlin \& Bell(2000)\\
{\neiv} &Becker et al. (1989); Bhatia \& Kastner (1988)&Ramsbottom et al. (1998)\\
{\nev}&Kaufman \& Sugar (1986); Bhatia \& Doschek (1993)&Lennon \& Burke (1994)\\
{\sii} &Verner et al. (1996); Keenan et al. (1993)&Ramsbottom et al. (1996)\\
{\siii} &Tayal \& Gupta (1999) &Froese Fischer et al. (2006)\\
{\siv}   &Johnson et al. (1986); Dufton et al. (1982); Verner et al. (1996)&Dufton et al. (1982)             \\
{\cliii} &Mendoza \& Zippen (1982a); Kaufman \& Sugar (1986) &Ramsbottom et al. (2001)\\
{\cliv}  &Mendoza \& Zippen (1982b); Ellis \& Martinson (1984)&Galavis et al. (1995)              \\
         &Kaufman \& Sugar (1986)\\
{\ariii} &Mendoza (1983); Kaufman \& Sugar (1986)&Galavis et al. (1995) \\
{\ariv}  &Mendoza \& Zippen (1982a); Kaufman \& Sugar (1986) &Zeippen et al. (1987)\\
{\feiii} &Garstang (1957); Nahar \& Pradhan (1996)& Zhang (1996)              \\
{\feiv} &Froese-Fischer \& Rubin (1998); Garstang (1958)&Zhang \& Pradhan (1997)                 \\
{\kriv}  &Bi{\' e}mont \& Hansen (1986) & Sch{\" o}ning (1997)              \\
{\krv}   & Bi{\' e}mont \& Hansen (1986) & Sch{\" o}ning (1997)               \\
$[$Rb\,{\sc v}$]$&   Persson et al. (1984) & $\cdots$               \\
{\xeiii} & Bi{\' e}mont et al. (1995)  & Sch{\" o}ning \& Butler (1998)  \\
{\baii} &  Klose et al.  (2002)        & Sch{\" o}ning \& Butler (1998)             \\
\hline
\end{tabular}
\end{table*}

\subsection{Plasma diagnostics} 
\subsubsection{CEL diagnostics}
We have detected a large number of collisionally excited lines (CEL), 
useful for estimations of the temperatures ($T_{\epsilon}$)
and densities ($n_{\epsilon}$). The electron temperature and density diagnostic lines analyzed here 
arise from various ions, which have a wide variety of ionization
potentials ranging from 0 ({\Ni} \& {\oi}) to 63.5 eV ({\neiv}). 
We have examined the electron temperature and density structure within the nebula of
BoBn 1 using 17 diagnostic line ratios. The {\oi}, {\neiii}, {\neiv}, {\sii}, and {\siii} zone electron temperatures and {\Ni}, C\,{\sc
iii}] and {\neiii} zone electron densities are estimated for the first time. Electron
temperatures and densities were derived from each diagnostic
ratio for each line by solving level populations for a multi-level 
($\geqq$ 5 for almost all the ions) atomic model using the collision strengths $\Omega_{ij}$ ($j$$>$$i$) 
and spontaneous transition probabilities $A_{ji}$ for each ion from the
references given in Table \ref{atomf}.

\begin{figure*}
\centering
\includegraphics[scale=0.6]{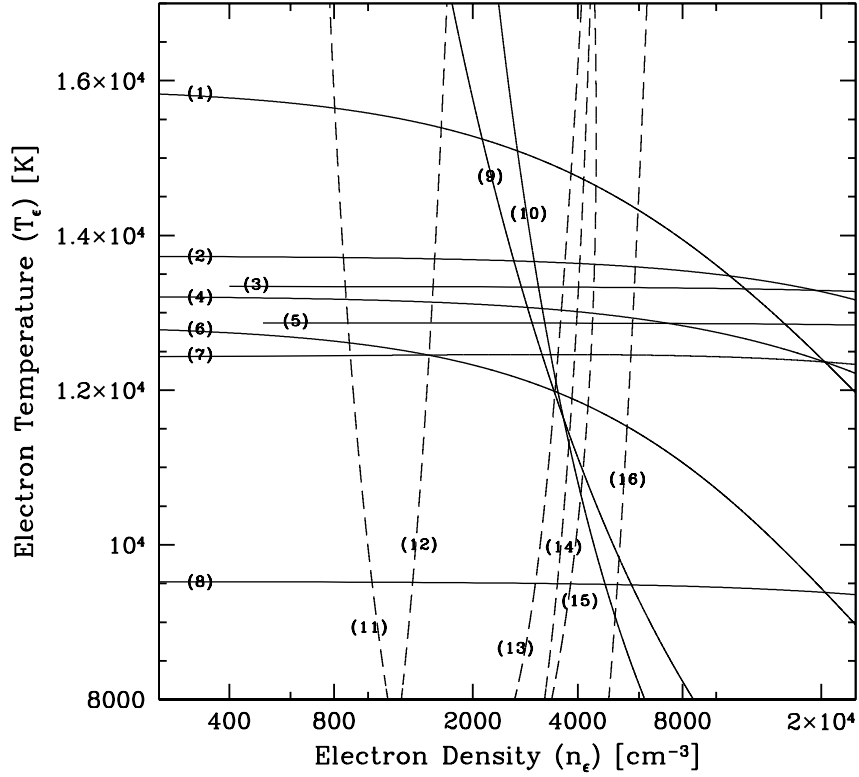}
\caption{Plasma diagnostic diagram. Each curve is labeled with an ID
 number given in Table \ref{diano_table}. For $T_{\epsilon}$({\nii}) and 
$T_{\epsilon}$({\oii}), we corrected for recombination contributions to
 {\nii} $\lambda$5755 and {\oii} $\lambda\lambda$7320/30, respectively
 (see text). \label{diagno_figure}}
\end{figure*}

The derived electron temperatures and densities are
listed in Table \ref{diano_table}. Fig. \ref{diagno_figure} is the diagnostic diagram that plots the loci 
of the observed diagnostic line ratios on the $n_{\epsilon}$--$T_{\epsilon}$ 
plane. This diagram shows that
most CELs in BoBn 1 are emitted from
$T_{\epsilon}$$\sim$12\,000--16\,000 K and $\log_{10}$\,$n_{\epsilon}$$\sim$3.5 cm$^{-3}$ ionized gas.

First, we calculated electron densities assuming a constant electron
temperature of 12\,800 K. Estimated electron densities range 
from 1030 ({\Ni}) to 5740 cm$^{-3}$ ({\sii}). Although {\sii} and {\oii}
have similar ionization potentials, a large discrepancy between their
electron densities is found (see Table 8).  Kniazev et al. (2008) and Kwitter et
al. (2003) estimated {\sii} electron densities as large as 9600 and 7100 cm$^{-3}$, 
respectively. Stanghellini \& Kaler (1989), Copetti \& Writzl (2002), and 
Wang et al. (2004) found that the {\sii}  density is systematically larger 
than the {\oii} density in a large number of samples.
The curve yielded by the {\sii}$\lambda$6716/31 ratio in the $n_{\epsilon}$
vs $T_{\epsilon}$ plane indicates higher electron density than critical 
density of these lines, 1600 $\&$ 4100 cm$^{-3}$ at $T_{\epsilon}$ = 
12\,800 K for $\lambda$6716 and $\lambda$6731, respectively (cf. 5740 cm$^{-3}$ in Fig.
\ref{diagno_figure}). This density discrepancy is 
not due to the errors in the {\oii} atomic data. Wang 
et al. (2004) also found the density discrepancy between {\sii} and 
{\oii} that might be likely caused by errors in the transition probabilities of {\oii} 
given by Wiese et al. (1996). In the case of BoBn 1, this possibility 
can be ruled out because we obtained similar {\oii} electron densities even when with the other 
transition probabilities. The high {\sii} density might be due to
high-density blobs in the outer nebula. This could have contributed 
to producing the small {\sii}$\lambda$6717/$\lambda$6731 ratio and 
give rise to an apparently high density. Zhang et al. (2005a) pointed 
out the possibility that a dynamical plow by the ionization front 
effects yields large density of {\sii} because the ionization potential 
of S$^{+}$ is close to the H$^{+}$ edge.
Since the estimated upper limit to the 
{\sii} density is close to the H$^{+}$ density derived from the Balmer
decrement (see below), this explanation might be plausible. In BoBn 1, caution is 
necessary when using the {\sii} electron density.

Next, we calculated the electron temperature. An 
average electron density of 3370 cm$^{-3}$, which excluded 
the $n_{\epsilon}$({\Ni}) and $n_{\epsilon}$({\sii}), was adopted 
when estimating electron temperatures except the $T_{\epsilon}$({\oi}). 
{\Ni} and {\oi} are representative of the very
outer part of the nebula and probably do not coexist with most of 
the other ions. $T_{\epsilon}$({\oi}) was, therefore, estimated 
using $n_{\epsilon}$({\Ni}).

To obtain the {\nii}, {\oii} and {\oiii} temperatures it is necessary
to subtract the recombination contamination to the 
{\nii} $\lambda$5755, {\oii} $\lambda\lambda$7320/30, and {\oiii} $\lambda$4363 lines,
respectively. For  {\nii} $\lambda$5755, Liu et al. (2000) 
estimated the contamination to {\nii} $\lambda$5755, 
$I_{R}$({\nii}$\lambda$5755) in the range 5000 $\leqq$ $T_{\epsilon}$ 
$\leqq$ 20\,000 K as
\medskip

\begin{equation}
\label{rni}
\frac{I_{R}(\rm [N\,{\sc II}]\lambda5755)}{I(\rm H\beta)} = 
3.19\left(\frac{T_{\epsilon}}{10^4}\right)^{0.33}\times\frac{\rm N^{2+}}{\rm H^{+}},
\end{equation}
\medskip

\noindent
where N$^{2+}$/H$^{+}$ is the doubly ionized nitrogen
abundance. Adopting the value derived from the ORL analysis (see Section \ref{ir})
and using Equation (\ref{rni}), we estimated $I_{R}$({\rm
{\nii}}$\lambda$5755) $\sim$0.1, which is approximately 7 $\%$ of the
observed value. Given the corrected {\nii} $\lambda$5755 intensity,
the {\nii} temperature is 12\,000 K, which is 400 K lower than that
obtained without taking into account the recombination effect.

The same effect also exists for the {\oii} $\lambda\lambda$7320/30 lines. We 
estimated the recombination contribution using the
doubly ionized oxygen abundance derived from O\,{\sc ii} lines and the 
equation of Liu et al. (2000) for these lines in the range 5000 $\leqq$
$T_{\epsilon}$ $\leqq$ 10\,000 K,   
\medskip

\begin{equation}
\label{roii}
\frac{I_{R}(\rm [O\,{\sc II}]\lambda\lambda7320/30)}{I(\rm H\beta)} = 
9.36\left(\frac{T_{\epsilon}}{10^4}\right)^{0.44}\times\frac{\rm O^{2+}}{\rm H^{+}}.
\end{equation}
\medskip

\noindent Using Equation (\ref{roii}), we estimated a contribution
of $\sim$7 $\%$ of the observed value and obtained 12\,100 K, lower by
700 K than that without the recombination contribution. 
For {\oiii} $\lambda$4363, we estimated the recombination contribution using the
O$^{3+}$ abundance derived from the fine-structure line 
[O {\sc iv}] $\lambda$25.9$\mu$m adopting $T_{\epsilon}$({\neiv}) and $n_{\epsilon}$({\ariv}) and 
the equation of Liu et al. (2000). We chose the value from this line because 
O\,{\sc iii} lines could be affected by star light
excitation and the abundance derived from them could be erroneous. Assuming the ratio of
O$^{3+}$(ORLs)/O$^{3+}$(CELs) = 10, we estimated the recombination 
contribution to {\oiii} $\lambda$4363 less than 1$\%$ of the
observed value, which has a negligible effect on the  
$T_{\epsilon}$({\oiii}) derivation.

The electron temperature in BoBn 1 ranges from 9520 ({\oi}) to 14\,920 K
({\neiv}). Our estimated electron temperatures except for {\oii} are
comparable to those of Kwitter et al. (2003) and Kniazev et al. (2008). Their estimated temperatures are 
12\,400-13\,720 K for {\oiii}, 11\,320-11\,700 K for {\nii}, and 13\,250 K for {\ariii}. 
Note that the {\oii} electron temperature of 8000 K of Kwitter et al. (2003) 
was estimated adopting the {\sii} density of 7100 cm$^{-3}$. Our 
$n_{\epsilon}$--$T_{\epsilon}$ plane predicts that the {\oii} 
temperature is $\sim$10\,000 K when adopting {\sii} density.

\begin{figure}
\centering
\includegraphics[width=80mm]{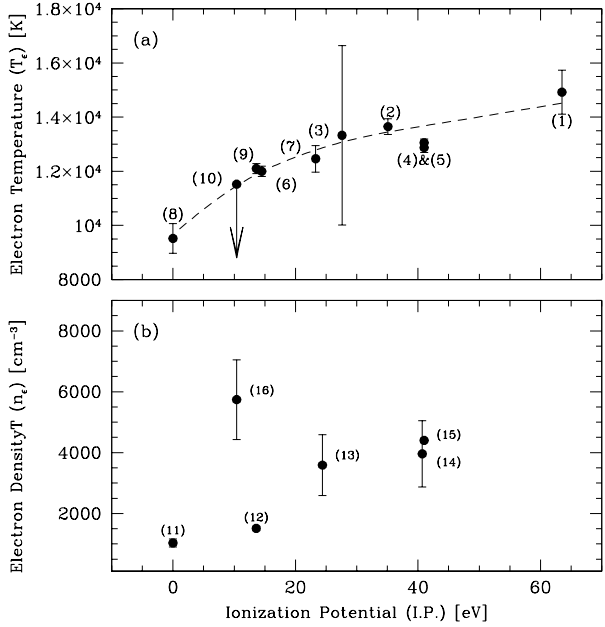}
\caption{Electron temperature ({\it upper}) and density ({\it lower}) versus
 ionization potential. Each value is labeled with an ID
 number given in Table \ref{diano_table}. 
\label{ip_te_ne}}
\end{figure}

The ionic abundances derived from the CELs depend strongly on the electron
temperature. In the case of {\oiii}$\lambda$5007, for example, only 500
K change makes a difference of over 10$\%$ for O$^{2+}$
abundance. It is therefore essential to find the proper electron temperature 
for each ionized stage of each ion. To that end, we examined
the behavior of the electron temperature and density as a function of
I.P. The upper panel of Fig. \ref{ip_te_ne} shows 
that $T_{\epsilon}$ is increasing proportional to I.P.
The observed behavior of $T_{\epsilon}$ is consistent with the schematic picture 
of stratified physical conditions in ionized nebula, where the electron 
temperature of ions in the inner part should be hotter than that in the outer part.
$n_{\epsilon}$ is simply monotonically
increasing up to $\sim$40 eV as I.P., except for {\sii}. 
The [S~{\sc ii}] might be emitted in high-density blobs in the outer nebula 
as we mentioned above.

To minimize the estimated error for ionic abundances due to electron temperature, 
we have assumed a 7-zone model for BoBn 1 by reference to Fig. \ref{ip_te_ne}. 
Adopted $T_{\epsilon}$ and $n_{\epsilon}$ for each ion are
presented in Table \ref{temp_ne_cles}. $T_{\epsilon}$({\oi}) and $n_{\epsilon}$([N\,{\sc i}]) 
are adopted for ions in zone 0, which have $<$10 eV. $T_{\epsilon}$({\nii}) 
and $n_{\epsilon}$({\sii}) are adopted for ions in zone 1 
(I.P. $<$ 11.3 eV). $T_{\epsilon}$({\nii}) and $n_{\epsilon}$({\oii}) 
are for zone 2 (11.3-20 eV). $T_{\epsilon}$({\siii}) and 
$n_{\epsilon}$(C\,{\sc iii}]) are for zone 3 (20-25 eV). For zone 4 
(25-41 eV), 5 (41-63.5 eV), and zone 6 ($>$ 63.5 eV), we adopted 
$n_{\epsilon}$({\ariv}) and $T_{\epsilon}$({\oiii}), the averaged value from 
$T_{\epsilon}$({\oiii}) and $T_{\epsilon}$({\neiv}), and
$T_{\epsilon}$({\neiv}), respectively.


\begin{table*}
\centering
\caption{Plasma diagnostics.\label{diano_table}}
\begin{tabular}{@{}lrlll@{}}
\hline\hline
&ID&Diagnostic&Ratio&Result\\
\hline  
$T_{\epsilon}$                &(1)&[Ne\,{\sc iv}] ($\lambda$2422+$\lambda$2425)/($\lambda$4715/16/25/26)&100.78$\pm$10.84&14\,920$\pm$810\\
(K)              &(2)&[O\,{\sc iii}] ($\lambda$4959+$\lambda$5007)/($\lambda$4363)&85.01$\pm$4.32&13\,650$\pm$290\\    
 &(3)&[Ar\,{\sc iii}] ($\lambda$7135)/($\lambda$5192)&85.70$\pm$35.63&13\,330$\pm$3\,310\\
 &(4)&[Ne\,{\sc iii}] ($\lambda$15.5$\mu$m)/($\lambda$3869+$\lambda$3967)&0.20$\pm$0.01&13\,050$\pm$140\\
 &(5)&[Ne\,{\sc iii}] ($\lambda$3869+$\lambda$3967)/($\lambda$3344)&333.67$\pm$14.63&12\,870$\pm$170\\
 &(6)&[N\,{\sc ii}] ($\lambda$6548+$\lambda$6583)/($\lambda$5755)&57.55$\pm$1.77&12\,000$\pm$190$^{a}$\\
 &(7)&[S\,{\sc iii}] ($\lambda$9069)/($\lambda$6312)&7.42$\pm$0.53&12\,460$\pm$490\\
 &(8)&[O\,{\sc i}] ($\lambda$6300+$\lambda$6363)/($\lambda$5577)&68.65$\pm$10.76&9520$\pm$550\\
 &(9)&[O\,{\sc ii}] ($\lambda$3726+$\lambda$3729)/($\lambda$7320+$\lambda$7330)&10.47$\pm$0.21&12\,100$\pm$180$^{b}$\\
 &(10)&[S\,{\sc ii}] ($\lambda$6716+$\lambda$6731)/($\lambda$4069+$\lambda$4076)&13.84$\pm$5.17&12\,420$_{-3590}$\\
&    &Average$^{\dagger}$ &&13\,050\\
\cline{2-5} 
 & &He\,{\sc i} ($\lambda$7281)/($\lambda$6678)&0.21$\pm$0.01&9430$\pm$310\\
 & &He\,{\sc i} ($\lambda$7281)/($\lambda$5876)&0.05$\pm$0.01&7340$\pm$110\\
 & &He\,{\sc i} ($\lambda$6678)/($\lambda$4471)&0.83$\pm$0.02&7400$^{+1070}$\\
 & &He\,{\sc i} ($\lambda$6678)/($\lambda$5876)&0.27$\pm$0.01&9920$\pm$310\\
 & &Average &&8520\\
\cline{2-5} 
 & &(Balmer Jump)/(H 11)&&8840$\pm$210\\
\hline
$n_{\epsilon}$ &(11)&[N {\sc i}] ($\lambda$5198)/($\lambda$5200)&1.43$\pm$0.03&1030$\pm$130\\
(cm$^{-3}$)&(12)&[O\,{\sc ii}] ($\lambda$3726)/($\lambda$3729)&1.65$\pm$0.03&1510$\pm$60\\
 &(13)&C\,{\sc iii}] ($\lambda$1906)/($\lambda$1909)&1.39$\pm$0.03&3590$\pm$1000\\
 &(14)&[Ar\,{\sc iv}] ($\lambda$4711)/($\lambda$4740)&1.05$\pm$0.07&3960$\pm$1090\\
 &(15)&[Ne\,{\sc iii}] ($\lambda$15.5$\mu$m)/($\lambda$36.0$\mu$m)&12.11$\pm$1.69&4400$^{+9010}$\\
 &(16)&[S\,{\sc ii}] ($\lambda$6716)/($\lambda$6731)&8.57$\pm$0.03&5740$\pm$1\,310\\
 &    &Average$^{\dagger}$ &&3370\\
\cline{2-5} 
 &    &Balmer decrement&&5000-10\,000\\
\hline
\end{tabular}
\tablenotetext{$^{\dagger}$}{From ions with I.P.$>$ 13.6 eV.}
\tablenotetext{a}{Corrected for recombination contribution to [N\,{\sc ii}]$\lambda$5755 (see text).}
\tablenotetext{b}{Corrected for recombination contribution to [O\,{\sc ii}] $\lambda\lambda$7320/30 (see text).}
\end{table*}

\begin{table}
\centering
\caption{Adopted $T_{\epsilon}$ and $n_{\epsilon}$ for CEL ionic abundance calculations.\label{temp_ne_cles}}
\begin{tabular}{@{}clcc@{}}
\hline\hline
Zone&Ions&$T_{\epsilon}$(K)&$n_{\epsilon}$(cm$^{-3}$)\\
\hline
0&C$^{0}$,N$^{0}$,O$^{0}$,Ba$^{+}$&9520&1030\\
1&C$^{+}$,S$^{+}$&12\,000&5740\\
2&O$^{+}$,N$^{+}$,F$^{+}$,Fe$^{2+}$&12\,000&1510\\
3&C$^{2+}$,Ne$^{+}$,S$^{2+}$,Cl$^{2+}$,Xe$^{2+}$&12\,460&3590\\
4&N$^{2+}$,O$^{2+}$,Ne$^{2+}$,Ar$^{2+}$,Ar$^{3+}$,S$^{3+}$&13\,650&3960\\
&Cl$^{3+}$,F$^{2+}$,Fe$^{3+}$,Kr$^{3+}$,Kr$^{4+}$&&\\
5&C$^{3+}$,N$^{3+}$,O$^{3+}$&14\,290$^{\dagger}$&3960\\
6&Ne$^{3+}$,Ne$^{4+}$,F$^{3+}$&14\,920&3960\\
\hline
\end{tabular}
\tablenotetext{$^{\dagger}$}{The averaged value from $T_{\epsilon}$({\oiii}) and $T_{\epsilon}$({\neiv}).}
\end{table}

\subsubsection{ORL diagnostics}

We detected a large number of optical recombination lines (ORL). 
C\,{\sc iii,iv}, O\,{\sc ii,iii,iv}, N\,{\sc ii,iii}, and Ne\,{\sc ii} are  for the first time detected. To
calculate ORL abundances, the electron temperature and density derived
from ORLs are needed. We estimated the electron temperature using the
Balmer discontinuity and {\hei} line ratios, and the electron density using the
Balmer decrement. The results are listed in Table \ref{diano_table}.

The ratio of the jump of continuum emission at the Balmer limit at 3646{\AA}
(BJ) to a given hydrogen emission line depends on the electron temperature.
Following Liu et al. (2001), we use this ratio to determine the
electron temperature. This temperature, $T_{\epsilon}$(BJ), is used to deduce ionic
abundances from ORLs. Defining 
BJ as $I(3646 {\rm {\AA}}) - I(3681 {\rm {\AA}})$, and taking the emissivities 
of H\,{\sc i} Balmer lines and H\,{\sc i}, He\,{\sc i}, 
and He\,{\sc ii} continuum emissivities, 
Liu et al. (2001) gave the following equation:
\medskip

\begin{eqnarray}
\label{balj}
&&T_{\epsilon}({\rm BJ}) = 368\left(1 + 0.259{\rm \frac{N(He^{+})}{N(H^{+})}} 
+ 3.409{\rm \frac{N(He^{2+})}{N(H^{+})}}\right) \nonumber\\
 &&\times \left(\frac{I(3646\,{\rm {\AA}}) 
- I(3681\,{\rm {\AA}})}{I{\rm (H\,11)}}\right)^{-1.5}
\end{eqnarray}
\medskip

\noindent
[$I({\rm 3646\,{\AA}})-I({\rm 3681\,{\AA}})]/I({\rm H\,11})$ is in units of {\AA}$^{-1}$. 
$T_{\epsilon}$(BJ) is valid over a range from 4000 to 20\,000 K. 
The process was repeated until self-consistent values for the
N(He$^{+}$)/N(H$^{+}$), N(He$^{2+}$)/N(H$^{+}$) and $T_{\epsilon}$(BJ)
were reached, we estimated $T_{\epsilon}$(BJ) of 8840 K.

We estimated the {\hei} electron temperature $T_{\epsilon}$({\hei}) 
from the ratios of He\,{\sc i}$\lambda$7281/$\lambda$6678, $\lambda$7281/$\lambda$5876,
$\lambda$6678/$\lambda$4471, and $\lambda$6678/$\lambda$5876 assuming a
constant electron density = 10$^{4}$ cm$^{-3}$, estimated from
the Balmer decrement as described below. All the {\hei} line ratios we chose 
here are insensitive to the electron density.
We adopted the emissivities of {\hei} from Benjamin et al. (1999). We estimated
$T_{\epsilon}$({\hei}) values as 7340--9920 K. The 
$T_{\epsilon}$({\hei}) from $\lambda$7281/$\lambda$6678 ratio appears to be the
most reliable value because (i) {\hei} $\lambda$6678 and $\lambda$7281 levels have 
the same spin as the ground state and the
Case B recombination coefficients for these lines by Benjamin et
al. (1999) are more reliable than the other He~{\sc i} $\lambda$4471 and
$\lambda$5876; (ii) the effect of interstellar extinction is less due to
the close wavelengths. We adopted 9430 K as $T_{\epsilon}$({\hei}). Note that in Fig. \ref{diagno_figure} 
the electron temperatures and densities derived from the H~{\sc i} and 
He~{\sc i} are not presented.

\begin{figure}
\centering
\includegraphics[width=80mm]{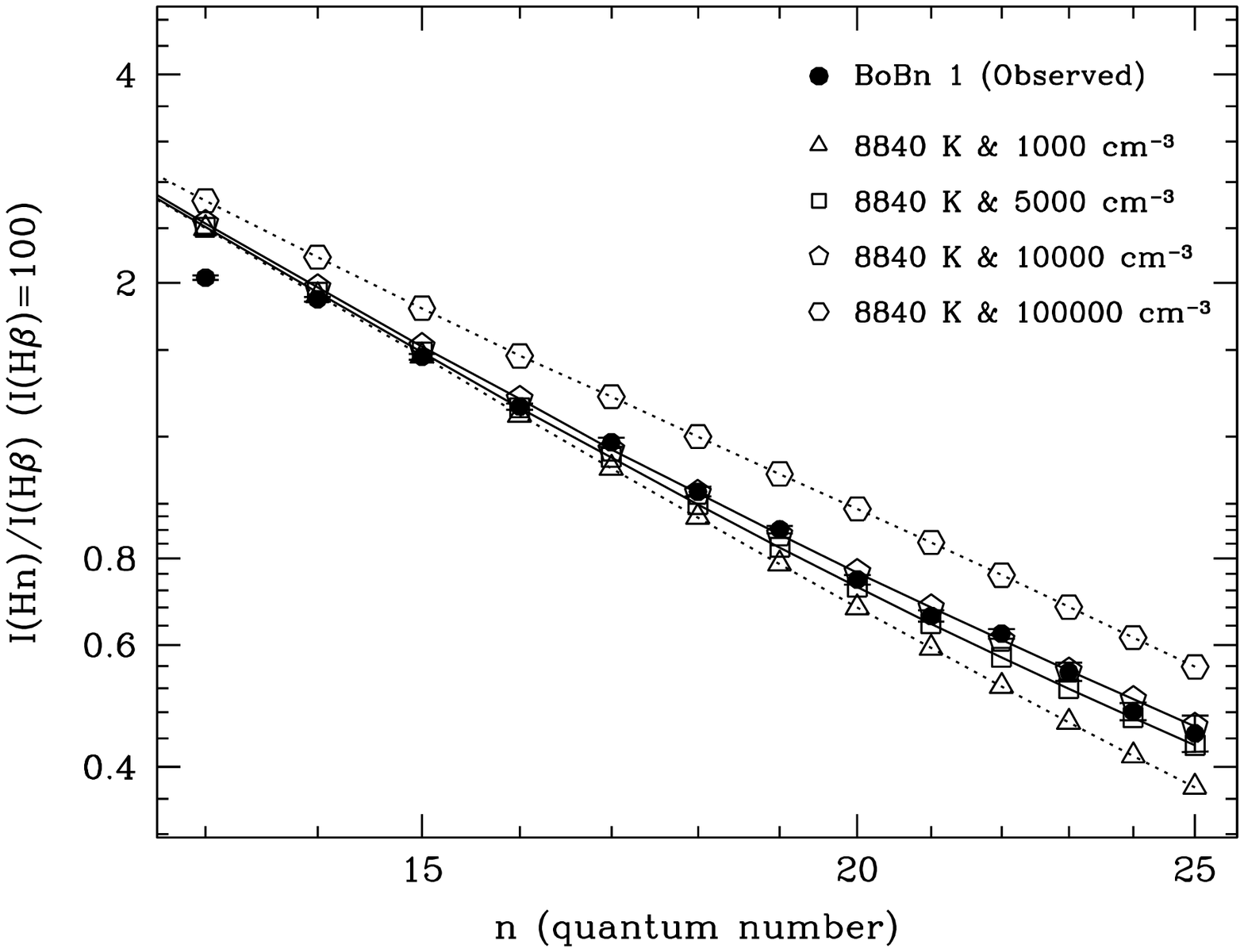}
\caption{Plot of the intensity ratio of the higher order Balmer lines (Hn, n= 11-25, n: quantum number of
 the upper level)
 to {\hb} (Case B assumption) with the theoretical intensity
 ratios for $T_{\epsilon}$ = 8840 K \& and different $n_{\epsilon}$'s. \label{highdens}}
\end{figure}

The intensity ratios of the high-order Balmer lines Hn (n $>$ 10, n: 
the principal quantum number of the upper level) to a lower Balmer line,
e.g., {\hb}, are also sensitive to the electron density.
In Fig. \ref{highdens}, we plot the ratio of higher-order Balmer lines to {\hb} with the theoretical
values by Storey \& Hummer (1995) for the cases of electron temperature of 8840 K (=
$T_{\epsilon}$(BJ)) \& electron densities of 1000, 5000, 10$^{4}$, and 10$^{5}$
cm$^{-3}$. This diagram indicates that the electron density in the ORL emitting
region is between 5000 and 10$^{4}$ cm$^{-3}$, which is fairly 
compatible with the CEL electron densities. 
Zhang et al. (2005b) estimated $T_{\epsilon}$({\hei}) from the ratio of He\,{\sc i}
$\lambda 7281/\lambda 6678$ for 48 PNe and found that high-density blobs
(10$^{5}$--10$^{6}$ cm$^{-3}$) might be present in nebula 
if $T_{\epsilon}$({\hei}) $\simeq$ $T_{\rm \epsilon}$(BJ). Fig.
\ref{highdens} indicates that such components do not coexist 
in BoBn 1.

\subsection{Ionic abundances from CELs \label{if}}
The derived CEL ionic abundances X$^{m+}$/H$^{+}$ are listed 
in Table \ref{cel_abund}.
X$^{m+}$ and H$^{+}$ are the number densities of an 
$m$ times ionized ion and ionized hydrogen, respectively. 
To estimate ionic abundances, we solved level populations for a multi-level 
atomic model. In the last one of the line series of each
ion, we present the adopted ionic abundances in bold face
characters. These values are estimated from the line intensity-weighted 
mean or average if there are two or more available lines. 
Over 10 ionic abundances of some elements are estimated for the first
time. These newly estimated ionic abundance would
reduce the uncertainty of estimation of each elemental abundance, in
particular, N, O, F, Ne, S, Fe, and some {\it s}-process elements, 
which are key elements to the 
nucleosynthesis in low-mass stars and chemical evolution in galaxies.

Ne$^{2+}$ (zone 4 ion) and S$^{2+}$ (zone 3) 
abundances are derived from CELs
 seen in both the UV-optical and mid-infrared regions. CELs in the mid-infrared,
 namely, fine-structure lines, have an advantage in derivations of ionic abundances. 
Since the excitation energy of the fine-structure lines is much lower than that of 
the other transition lines, ionic abundances derived from these lines are 
nearly independent of the electron temperature or temperature fluctuation in the
nebula. Note that these ionic abundances derived from fine-structure
 lines are almost consistent with those from other transition lines. This means 
 that the adopted electron temperature and density for the ions in zones 3 and 4 
are appropriate at least.

Followings are short comments on derivations of ionic abundances. 
We subtract the {\oiii} $\lambda$2322 contamination from C~{\sc ii}$]$ $\lambda$2324 
using the theoretical intensity ratio {\oiii}
$I$($\lambda$2322)/($\lambda$4363) of 0.24 and then estimate the C$^{+}$ abundance. 
The N$^{+}$ and O$^{+}$ abundances are derived only from {\nii} 
$\lambda\lambda$6548/83 and {\oii} $\lambda\lambda$3726/29 
to avoid recombination contamination, respectively, while The Ne$^{+}$ abundance is calculated 
from {\neii} $\lambda$12.8 $\mu$m by solving a two level atomic model.

We have detected two fluorine lines {\fiv} $\lambda\lambda$3996,4060 
and estimate a F$^{3+}$ abundance of 1.50(--8). Otsuka et al. (2008a) detected 
these fluorine lines and estimated a F$^{3+}$ abundance of 5.32(--8). 
The F$^{3+}$ abundance discrepancy 
between Otsuka et al. (2008a) and the present work is due to different adopted 
electron temperature and $c$({\hb}) values. We have detected candidates of 
{\fii} $\lambda\lambda$4790,4868 and {\fiii} $\lambda\lambda$5721/33. In
the previous section, we confirmed that BoBn 1 has no high-density components, 
larger than the critical density of these {\fiii} lines. 
The critical density of 
{\fii} $\lambda\lambda$4790/4868 is $\sim$2$\times$10$^{6}$ cm$^{-3}$, and that of 
{\fiii} $\lambda\lambda$5721/33 is $\sim$8$\times$10$^{6}$
cm$^{-3}$. Therefore, the effect of 
collisional de-excitation is negligibly small. Accordingly, the 
ratios of {\fii} $I$($\lambda$4790)/($\lambda$4868) and {\fiii} 
$I$($\lambda$5733)/($\lambda$5721) depend on their
transition probabilities. When adopting the transition probabilities by
Baluja \& Zeippen (1988) for {\fii} and Naqvi (1951) for {\fiii}, 
the theoretical intensity ratios of {\fii} 
$I$($\lambda$4790)/($\lambda$4868) and {\fiii} 
$I$($\lambda$5721)/($\lambda$5733) are $\sim$3.2 and $\sim$1, which are 
in agreement with our measurements (4.2$\pm$1.0 for {\fii} and 1.0$\pm$0.3
for {\fiii}). Hence, these four emission lines can be identified as 
{\fii} $\lambda\lambda$4790,4868 and {\fiii} $\lambda\lambda$5721/33. 
The F$^{2+}$ and F$^{3+}$ abundances are estimated from the each
detected line by solving the statistical equilibrium equations for the 
lowest five energy levels. For {\fiii} lines the relevant collision
strength has not been calculated. However, since F$^{2+}$ is
isoelectronic with Ne$^{3+}$, and collision strengths for the same levels
along an isoelectronic sequence tend to vary with effective nuclear
charge (Seaton 1958). We therefore assume that the collision strengths for {\fiii}
are $\sim$22$\%$ smaller than those for {\neiv} and estimate 
F$^{3+}$ abundances. Otsuka et al. (2008a) showed a correlation between [Ne/Ar] and [F/Ar] in PNe,
suggesting that Ne and F were synthesized in the same layer and 
carried to the surface by the third dredge-up. If this is the case, 
the ionic abundance ratios of F$^{2+}$ (I.P. = 35 eV) and F$^{3+}$ (62.7
eV) to F$^{+}$ (17.4 eV) should be comparable to those of Ne$^{2+}$ (41 eV) 
and Ne$^{3+}$ (63.5 eV) to Ne$^{+}$ (21.6 eV). Indeed, these ionic abundance ratios
follow our prediction; F$^{+}$:F$^{2+}$:F$^{3+}$ $\sim$1:34:1 and
Ne$^{+}$:Ne$^{2+}$:Ne$^{3+}$ $\sim$1:28:1. This means that our
identifications of two {\fii} and {\fiii} lines and the estimated ionic abundances are reliable. 
So far, fluorine has found in only a handful of PNe (see Zhang \& Liu 2005; 
Otsuka et al. 2008a). Among them BoBn 1 appears to be the most F-rich PN.

We subtract the contribution to {\cliii} $\lambda$8500.2 due to C\,{\sc 
iii} $\lambda$8500.32 using the C$^{3+}$ ORL abundance and give upper limit of Cl$^{2+}$ abundance from this 
line. The adopted Cl$^{2+}$ abundance is from {\cliii} $\lambda$5517
only. The Ar$^{3+}$ abundance is from {\ariv} 
$\lambda\lambda$4717/40. We adopt the S$^{+}$ abundance based on 
{\sii} lines except {\sii} $\lambda$4068, because {\sii} $\lambda$4068
could be partially blended with C\,{\sc iii} $\lambda$4068. 
To estimate Fe$^{2+}$ and Fe$^{3+}$ abundances, we solved a 33 level
model (from $^{5}D_{3}$ to b$^{3}P_{2}$) for [Fe\,{\sc iii}] and a 18 level
model (from $^{6}S_{5/2}$ to $^{2}F_{5/2}$) for [Fe\,{\sc iv}]. We adopt 
the transition probabilities of [Fe\,{\sc iv}] recommended by 
Froese-Fischer \& Rubin (1998). For those not considered by Froese-Fischer 
\& Rubin (1998), the values by Garstang (1958) were adopted.

\begin{figure*}
\centering
\includegraphics[width=160mm]{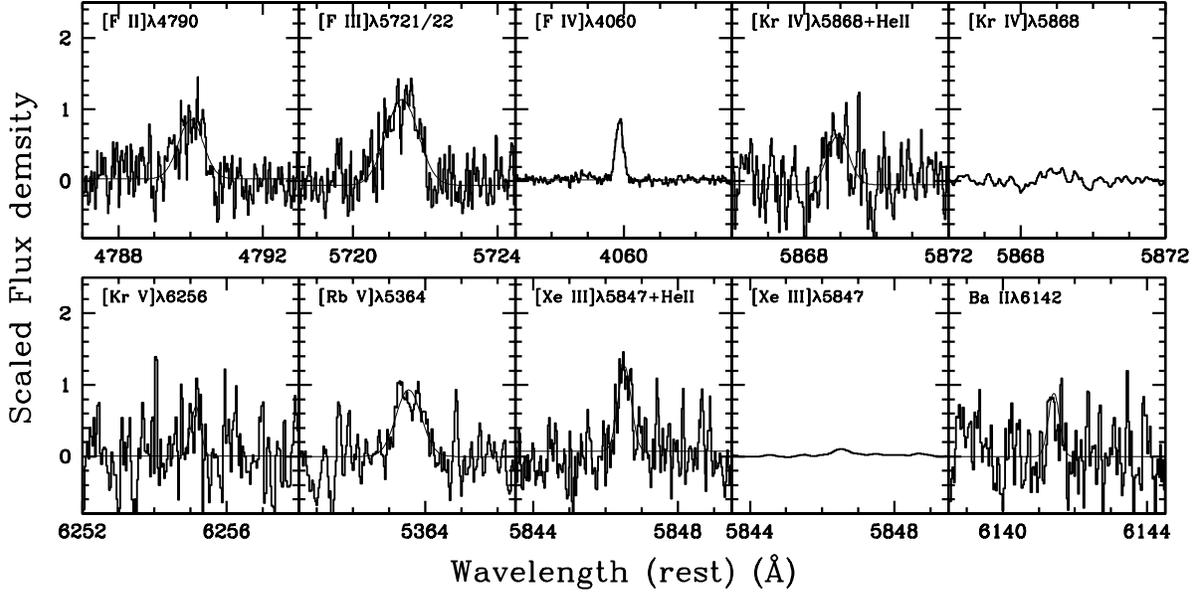}
\caption{Detected fluorine lines and candidates of {\it s}-process [Kr\,{\sc iv}]$\lambda$5867.7, [Kr\,{\sc v}]$\lambda$6256.1, 
[Rb\,{\sc v}]$\lambda$5363.6, [Xe\,{\sc iii}]$\lambda$5846.7, and Ba\,{\sc ii}$\lambda$6141.7. The thin lines indicate the fitted 
Gaussian profiles to each emission line. The isolated  [Kr\,{\sc iv}]$\lambda$5867.7 and 
[Xe\,{\sc iii}]$\lambda$5846.7 line-profiles after smoothing are also presented (see text for detail).}
\label{sprocess}
\end{figure*}

\subsection{Ionic abundances of heavy elements (Z $>$ 30)}
We have detected 10 emission line candidates 
of krypton (Kr), rubidium (Rb), xenon (Xe), and barium (Ba). Kr and Rb are light- {\it s}-process 
elements  (30 $\leqq$ Z $\leqq$40, 
Z: atomic number), Xe and Ba are heavy {\it s}-process (Z $\geqq$ 41). 
Kr has been detected in over 100 PNe, while the latter three
{\it s}-process elements have been detected in only a handful of PNe
(Sharpee et al. 2007). Selected line profiles of these candidates are presented in Fig. \ref{sprocess}.
The Kr$^{3+}$, Kr$^{4+}$, Xe$^{2+}$, and 
Ba$^{+}$ abundances in this object are estimated for the first time.

We have detected two nebular lines of [Kr\,{\sc iv}] $\lambda\lambda$5346.7,5867.7
($^{4}S^{o}_{3/2}{-}^{2}D^{o}_{5/2}$ and $^{4}S^{o}_{3/2}{-}^{2}D^{o}_{3/2}$, respectively). 
For [Kr\,{\sc iv}] $\lambda$5346.7, the possibility of blending with 
C\,{\sc iii} $\lambda$5345.85 (multiplet V13.01) is low, and the
contribution to this [Kr\,{\sc iv}] line is probably negligible 
because other V13.01 C\,{\sc iii} lines are not detected. For [Kr\,{\sc iv}]
$\lambda$5867.7, we estimated the contamination from He\,{\sc ii}
$\lambda$ 5867.7 using theoretical ratios of He\,{\sc ii} $I$($\lambda$5867.7) to $I$($\lambda$5828.4),
$I$($\lambda$5836.5), $I$($\lambda$5857.3), $I$($\lambda$5882.12), and 
$I$($\lambda$5896.8) given by Storey \& Hummer (1995) assuming
$T_{\epsilon}$ = 8840 K and $n_{\epsilon}$ = 10$^{4}$ cm$^{-3}$, from
which the contribution from He\,{\sc ii} $\lambda$5867.7 is estimated to be $\sim$64
$\%$. In Fig. \ref{sprocess}, we present the isolated 
[Kr\,{\sc iv}] $\lambda$5867.7 profile.
The observed ratio of [Kr\,{\sc iv}] $I$($\lambda$5867.7)/$I$($\lambda$5346.7) ($<$1.8) is comparable
to the theoretical value ($\sim$1.5) if the
population at each energy level does not exceed the critical densities
of $\sim$1.5(+6) cm$^{-3}$ ($^{2}D^{o}_{5/2}$) and
$\sim$2(+7) cm$^{-3}$ ($^{2}D^{o}_{3/2}$). 
We therefore identified these two lines as [Kr\,{\sc iv}] 
$\lambda\lambda$5346.7,5867.7.

[Kr\,{\sc v}] $\lambda$8243.4 ($^{3}P_{2}{-}^{1}D_{2}$) is likely to be 
blended with the blue wing of the Paschen series H {\sc i} $\lambda$8243.7 (n=3-43). 
Using theoretical ratios of H\,{\sc i} $I$($\lambda$8243.7) 
to $I$($\lambda$8247.7) and $I$($\lambda$8245.6) given by Storey \& Hummer (1995), 
we subtracted the contribution of H\,{\sc i} $\lambda$8243.7, then 
estimated the intensity of [Kr\,{\sc v}] $\lambda$8243.4. 
We found another nebular line [Kr\,{\sc v}] 
$\lambda$6256.1 ($^{3}P_{1}{-}^{1}D_{2}$). The theoretical intensity
ratio of $I$($\lambda$6256.1)/$I$($\lambda$8243.7) ($\sim$1.1) is in
good agreement with ours (1.2).

[Xe\,{\sc iii}] $\lambda$5846.8 ($^{3}P_{2}{-}^{1}D_{2}$) appears to be 
blended with He\,{\sc ii} $\lambda$5846.7. We subtracted the He~{\sc ii} 
$\lambda$5846.7 contribution from it using the theoretical 
ratios of  $I$($\lambda$5846.7) to $I$($\lambda$5828.4), 
$I$($\lambda$5836.5), $I$($\lambda$5857.3), $I$($\lambda$5882.12), 
and $I$($\lambda$5896.8) given by Storey \& Hummer (1995), then 
obtained an upper limit to the intensity of [Xe\,{\sc iii}] $\lambda$5846.8. 
In Fig. \ref{sprocess}, we present the isolated 
[Xe\,{\sc iii}] $\lambda$5846.8 profile.

Two Ba\,{\sc ii} recombination lines $\lambda\lambda$4934,6141.7 
($6s^{2}S_{1/2}{-}6p^{2}P^{o}_{1/2}$ and
$5d^{2}D_{5/2}{-}6p^{2}P^{o}_{3/2}$) are detected. Following Sharpee et 
al. (2007), we estimated the Ba$^{+}$ abundances adopting 
transition electron temperature and density (zone 0). 
The Ba$^{+}$ abundances from these lines are in good agreement each
other.

We have detected a candidate [Rb\,{\sc v}] $\lambda$5363.6 (auroral
line; $^{4}S^{o}_{3/2}{-}^{2}D^{o}_{3/2}$).  Rb is one of the important elements as tracers of the
neutron density. In the case of NGC 7027, 
Sharpee et al. (2007) argued the possibility that this line is 
O\,{\sc ii} $\lambda$5363.8 ($4f{F^{2}}[4]^{o}_{7/2}{-}4d^{'}{^{2}F}_{7/2}$). They also suggested 
that the intensity of O\,{\sc ii} $\lambda$5363.8 is comparable to 
O\,{\sc ii} $\lambda$4609.4 ($3d^{2}D_{5/2}{-}4f{F^{2}}[4]^{o}_{7/2}$),
arising from the lower level of O\,{\sc ii} $\lambda$5363.8. In BoBn 1, based on 
the ORL O$^{2+}$ abundance of 1.45(--4) from the 3d-4f O\,{\sc ii} lines (see next Section) 
the expected intensity of O\,{\sc ii} $\lambda$4609.4 is
$\sim$6.8(--5)$I$({\hb}), which is lower than the observed 
intensity of the [Rb\,{\sc v}] $\lambda$5363.6 candidate. No
$4f{F^{2}}[4]^{o}{-}4d^{'}{^{2}F}$ O\,{\sc ii} lines are detected in BoBn
1. Therefore, we consider that the detected line is [Rb\,{\sc v}]
$\lambda$5363.6. Since there are no available collision strengths for this line
at present, we do not estimate a Rb$^{4+}$ abundance.


\begin{table}
\caption{Ionic abundances from CELs. \label{cel_abund}}
\footnotesize
\begin{tabular}{@{}lccccc@{}}
\hline\hline
{X$^{m+}$}&
{$\lambda_{\rm lab}$}&
{$I$($\lambda_{\rm lab}$)}&
{$T_{\epsilon}$}&
{$n_{\epsilon}$}&
{X$^{m+}$/H$^{+}$}\\
{}        &
{({\AA}/$\mu$m)}&
{[$I$({\hb})=100]}&
{(K)}&
{(cm$^{-3}$)}&
{}\\
\hline
C$^{0}$ & 8727.12 & 7.93(--2) $\pm$ 4.73(--3) & 9520 & 1030 & {\bf 4.74(--7) $\pm$ 9.34(--8)} \\ 
C$^{+}$ & 2324 & 3.51(+1) $\pm$ 4.46(0) & 12000 & 5740 & {\bf 2.30(--5) $\pm$ 3.48(--6)} \\ 
C$^{2+}$ & 1906 & 8.39(+2) $\pm$ 1.27(+1) & 12460 & 3590 & 7.71(--4) $\pm$ 1.59(--4) \\ 
 & 1908 & 6.03(+2) $\pm$ 1.03(+1) &  &  & 7.70(--4) $\pm$ 1.59(--4) \\ 
 &  &  &  &  &   {\bf 7.71(--4) $\pm$ 1.59(--4)} \\ 
C$^{3+}$ & 1548 & 1.05(+3) $\pm$ 2.02(+1) & 14290 & 3960 & 2.42(--4) $\pm$ 4.47(--5) \\ 
 & 1551 & 5.19(+2) $\pm$ 1.50(+1) &  &  & 2.36(--4) $\pm$ 4.38(--5) \\ 
 &  &  &  &  &   {\bf 2.40(--4) $\pm$ 4.44(--5)} \\ 
N$^{0}$ & 5197.90 & 2.73(--1) $\pm$ 7.90(--3) & 9520 & 1030 & 4.81(--7) $\pm$ 1.01(--7) \\ 
  & 5200.26 & 1.91(--1) $\pm$ 5.47(--3) &  &  & 4.82(--7) $\pm$ 9.67(--8) \\ 
 &  &  &  &  &   {\bf 4.82(--7) $\pm$ 9.90(--8)} \\ 
N$^{+}$ & 5754.64 & 1.23(0) $\pm$ 1.40(--2) & 12000 & 1510 & 7.20(--6) $\pm$ 4.74(--7) \\ 
  & 6548.04 & 1.56(+1) $\pm$ 8.34(--1) &  &  & 5.84(--6) $\pm$ 3.75(--7) \\ 
  & 6583.46 & 5.04(+1) $\pm$ 1.28(0) &  &  & 6.40(--6) $\pm$ 2.79(--7) \\ 
 &  &  &  &  &   {\bf 6.27(--6) $\pm$ 3.02(--7)} \\ 
N$^{2+}$ & 1750 & 4.81(+1) $\pm$ 1.32(+1) & 13650 & 3960 & {\bf 6.24(--5) $\pm$ 1.88(--5)} \\ 
N$^{3+}$ & 1485 & 4.58(+1) $\pm$ 1.78(+1) & 14290 & 3960 & {\bf 3.83(--5) $\pm$ 1.65(--5)} \\ 
O$^{0}$ & 5577.34 & 1.70(--2) $\pm$ 2.64(--3) & 9520 & 1030 & 2.06(--6) $\pm$ 7.06(--7) \\ 
  & 6300.30 & 8.72(--1) $\pm$ 1.55(--2) &  &  & 2.04(--6) $\pm$ 3.97(--7) \\ 
  & 6363.78 & 2.91(--1) $\pm$ 9.91(--3) &  &  & 2.13(--6) $\pm$ 4.20(--7) \\ 
 &  &  &  &  &   {\bf 2.06(--6) $\pm$ 4.03(--7)} \\ 
O$^{+}$ & 3726.03 & 1.09(+1) $\pm$ 6.85(--2) & 12000 & 1510 & 4.00(--6) $\pm$ 2.23(--7) \\ 
  & 3728.81 & 6.61(0) $\pm$ 9.99(--2) &  &  & 4.03(--6) $\pm$ 2.42(--7) \\ 
  & 7319 & 1.02(0) $\pm$ 2.00(--2) &  &  & 7.13(--6) $\pm$ 5.67(--7) \\ 
  & 7330 & 7.81(--1) $\pm$ 1.38(--2) &  &  & 6.74(--6) $\pm$ 5.30(--7) \\ 
 &  &  &  &  &   {\bf 4.01(--6) $\pm$ 2.30(--7)} \\ 
O$^{2+}$ & 4363.21 & 5.57(0) $\pm$ 9.73(--2) & 13650 & 3960 & 4.76(--5) $\pm$ 4.87(--6) \\ 
  & 4931.23 & 4.36(--2) $\pm$ 3.88(--3) &  &  & 4.37(--5) $\pm$ 4.58(--6) \\ 
  & 4958.91 & 1.22(+2) $\pm$ 5.96(0) &  &  & 4.80(--5) $\pm$ 3.52(--6) \\ 
  & 5006.84 & 3.51(+2) $\pm$ 2.18(+1) &  &  & 4.76(--5) $\pm$ 3.94(--6) \\ 
 &  &  &  &  &   {\bf 4.77(--5) $\pm$ 3.83(--6)} \\ 
O$^{3+}$ & 25.9 & 1.25(+1) $\pm$ 1.36(--1) & 14290 & 3960 & {\bf 3.41(--6) $\pm$ 8.21(--8)} \\ 
F$^{+}$ & 4789.45 & 5.58(--2) $\pm$ 5.20(--3) & 12000 & 1510 & 2.16(--8) $\pm$ 2.23(--9) \\ 
  & 4868.99 & 1.34(--2) $\pm$ 3.00(--3) &  &  & 1.66(--8) $\pm$ 3.79(--9) \\ 
 &  &  &  &  &   {\bf 1.98(--8) $\pm$ 2.74(--9)} \\ 
F$^{2+}$ & 5721.20 & 2.70(--2) $\pm$ 2.93(--3) & 13650 & 3960 & 6.59(--7) $\pm$ 1.03(--7) \\ 
         & 5733.05 & 2.68(--2) $\pm$ 5.43(--3) &  &  & 6.70(--7) $\pm$ 1.55(--7) \\ 
&  &  &  &  &   {\bf 6.65(--7) $\pm$ 1.29(--7)} \\ 
F$^{3+}$ & 3996.92 & 4.09(--2) $\pm$ 2.53(--3) & 14920 & 3960 & 1.47(--8) $\pm$ 2.21(--9) \\ 
  & 4059.90 & 1.19(--1) $\pm$ 3.72(--3) &  &  & 1.51(--8) $\pm$ 2.12(--9) \\ 
 &  &  &  &  &   {\bf 1.50(--8) $\pm$ 2.14(--9)} \\ 
Ne$^{+}$ & 12.8 & 2.49(0) $\pm$ 8.17(--2) & 12460 & 3590 & {\bf 2.97(--6) $\pm$ 1.14(--7)} \\ 
Ne$^{2+}$ & 3342.42 & 8.47(--1) $\pm$ 2.75(--2) & 13650 & 3960 & 6.40(--5) $\pm$ 8.10(--6) \\ 
  & 3868.77 & 2.17(+2) $\pm$ 1.05(+1) &  &  & 8.41(--5) $\pm$ 6.75(--6) \\ 
  & 3967.46 & 6.39(+1) $\pm$ 4.13(--1) &  &  & 5.94(--5) $\pm$ 3.82(--6) \\ 
  & 4011.60 & 1.48(--2) $\pm$ 3.92(--3) &  &  & 9.71(--5) $\pm$ 2.65(--5) \\ 
  & 15.6 & 1.61(+2) $\pm$ 1.30(0) &  &  & 9.15(--5) $\pm$ 1.27(--6) \\ 
  & 36 & 1.33(+1) $\pm$ 1.84(0) &  &  & 9.07(--5) $\pm$ 1.26(--5) \\ 
 &  &  &  &  &   {\bf 8.32(--5) $\pm$ 4.33(--6)} \\ 
Ne$^{3+}$ & 2423.50 & 1.71(+1) $\pm$ 1.78(0) & 14920 & 3960 & 3.97(--6) $\pm$ 8.99(--7) \\ 
  & 4714.25 & 5.52(--2) $\pm$ 2.76(--3) &  &                & 4.35(--6) $\pm$ 1.23(--7) \\ 
  & 4715.80 & 1.87(--2) $\pm$ 2.12(--3) &  &                & 5.05(--6) $\pm$ 1.52(--6) \\ 
  & 4724.15 & 5.08(--2) $\pm$ 1.88(--3) &  &                & 3.60(--6) $\pm$ 1.01(--6) \\ 
  & 4725.62 & 4.52(--2) $\pm$ 2.58(--3) &  &                & 3.43(--6) $\pm$ 9.78(--7) \\ 
 &  &  &  &  &   {\bf 3.97(--6) $\pm$ 9.01(--7)} \\ 
Ne$^{4+}$ & 3345.83 & 3.22(--1) $\pm$ 1.98(--2) & 14920 & 3960 & 1.99(--7) $\pm$ 3.40(--8) \\ 
  & 3425.87 & 8.71(--1) $\pm$ 8.29(--3) &  &  & 1.97(--7) $\pm$ 3.15(--8) \\ 
 &  &  &  &  &   {\bf 1.98(--7) $\pm$ 3.22(--8)} \\ 
S$^{+}$ & 4068.60 & 3.95(--1) $\pm$ 8.37(--3) & 12000 & 5740 & 4.05(--8) $\pm$ 1.96(--9) \\ 
  & 4076.35 & 2.45(--2) $\pm$ 9.13(--3) &  &  & 7.44(--9) $\pm$ 2.79(--9) \\ 
  & 6716.44 & 1.23(--1) $\pm$ 4.55(--3) &  &  & 1.03(--8) $\pm$ 5.12(--10) \\ 
  & 6730.81 & 2.16(--1) $\pm$ 4.88(--3) &  &  & 1.03(--8) $\pm$ 4.01(--10) \\ 
 &  &  &  &  &   {\bf 1.03(--8) $\pm$ 4.41(--10)} \\ 
\hline
\end{tabular}
\end{table}

\setcounter{table}{9}
\begin{table}
\caption{Continued.}
\footnotesize
\begin{tabular}{@{}lccccc@{}}
\hline\hline
{X$^{m+}$}&
{$\lambda_{\rm lab}$}&
{$I$($\lambda_{\rm lab}$)}&
{$T_{\epsilon}$}&
{$n_{\epsilon}$}&
{X$^{m+}$/H$^{+}$}\\
{}        &
{({\AA}/$\mu$m)}&
{[$I$({\hb})=100]}&
{(K)}&
{(cm$^{-3}$)}&
{}\\
\hline
S$^{2+}$ & 6312.10 & 4.77(--2) $\pm$ 4.75(--3) & 12460 & 3590 & 6.87(--8) $\pm$ 1.07(--8) \\ 
  & 9068.60 & 3.78(--1) $\pm$ 9.97(--3) &  &  & 7.34(--8) $\pm$ 4.79(--9) \\ 
  & 18.7 & 6.92(--1) $\pm$ 4.67(--2) &  &  & 6.81(--8) $\pm$ 4.82(--9) \\ 
 &  &  &  &  &   {\bf 6.99(--8) $\pm$ 5.06(--9)} \\ 
S$^{3+}$ & 10.5 & 1.92(0) $\pm$ 5.03(--2) & 13650 & 3960 & {\bf 5.28(--8) $\pm$ 1.51(--9)} \\ 
Cl$^{2+}$           & 5517.66 & 1.81(--2) $\pm$ 2.82(--3) &12460 &3590 & {\bf 1.36(--9) $\pm$ 2.42(--10)} \\ 
  & 8500.20 & $<$1.81(--3) &  &  & $<$2.79(--9)\\ 
Cl$^{3+}$ & 8046.30 & 2.05(--2) $\pm$ 2.60(--3) & 13650 & 3960 & {\bf 7.82(--10) $\pm$ 1.04(--10)} \\ 
Ar$^{2+}$ & 5191.82 & 3.90(--3) $\pm$ 1.61(--3) & 13650 & 3960 & 1.23(--8) $\pm$ 5.17(--9) \\ 
  & 7135.80 & 2.73(--1) $\pm$ 1.14(--2) &  &  & 1.30(--8) $\pm$ 7.41(--10) \\ 
  & 7751.10 & 6.11(--2) $\pm$ 2.51(--3) &  &  & 1.22(--8) $\pm$ 6.83(--10) \\ 
 &  &  &  &  &   {\bf 1.29(--8) $\pm$ 7.30(--10)} \\ 
Ar$^{3+}$ & 4711.37 & 9.40(--2) $\pm$ 5.23(--3) & 13650 & 3960 & 7.53(--9) $\pm$ 5.46(--10) \\ 
  & 4740.17 & 8.99(--2) $\pm$ 3.71(--3) &  &  & 7.58(--9) $\pm$ 4.60(--10) \\ 
  & 7170.50 & 6.53(--3) $\pm$ 7.75(--4) &  &  & 4.32(--8) $\pm$ 6.13(--9) \\ 
  & 7262.70 & 4.58(--3) $\pm$ 3.99(--3) &  &  & 3.52(--8) $\pm$ 3.08(--8) \\ 
 &  &  &  &  &   {\bf 7.56(--9) $\pm$ 5.04(--10)} \\ 
Fe$^{2+}$ & 4881.00 & 2.14(--2) $\pm$ 4.84(--3) &12000  &1510  & 1.16(--8) $\pm$ 2.80(--9) \\ 
 & 5270.40 & 2.19(--2) $\pm$ 3.69(--3) &  &  & 1.04(--8) $\pm$ 1.79(--9) \\ 
 &  &  &  &  &   {\bf 1.10(--8) $\pm$ 2.29(--9)} \\ 
Fe$^{3+}$ & 6740.63 & 1.58(--2) $\pm$ 4.81(--3) & 13650 & 3960 & {\bf
 1.02(--7) $\pm$ 3.26(--8)} \\ 
Kr$^{3+}$ & 5346.02 & 4.56(--3) $\pm$ 2.02(--3)   & 13650 & 3960 & {\bf 1.41(--10) $\pm$ 6.31(--11)}    \\
          & 5867.70 & $<$8.20(--3) &  & &$<$1.89(--10) \\
Kr$^{4+}$ &6256.06 &  $<$6.27(--3))& 13650 & 3960 &  $<$4.09(--10)\\
          &8243.39 & $<$5.12(--3)&        &      &  $<$3.45(--10)\\
  &  &  &  &  &   {\bf $<$3.77(--10)} \\ 
Xe$^{2+}$ & 5846.66 & $<$1.56(--3) & 12460 & 3590 & {\bf $<$2.30(--11)}    \\ 
Ba$^{+}$ &4934.08 & 6.42(--3) $\pm$ 1.80(--3) &9550 &1030 &1.90(--10)
 $\pm$ 6.45(--11)\\  
          &6141.70 & 3.43(--3) $\pm$ 7.66(--4) &     &     &2.12(--10)
 $\pm$ 6.42(--11)\\  
 &  &  &  &  &   {\bf 1.98(--10) $\pm$ 6.44(--11)} \\ 
\hline
\end{tabular}
\end{table}

\subsection{Ionic abundances from ORLs \label{ir}}
We have detected many optical recombination lines (ORLs) of helium, carbon, 
nitrogen, oxygen, and neon. To our knowledge, the nitrogen, oxygen, and
neon ORLs are detected for the first time from this PN. These lines provide us with 
a new independent method to derive chemical abundances for BoBn 1. 
The recombination coefficient depends weakly on the electron temperature 
($\propto$ $T_{\epsilon}^{-1/2}$). The ionic abundances are, therefore, 
insignificantly affected by small-scale fluctuations of electron temperature.
This is the most important advantage of this determination method.
The ionic abundances from recombination lines are robust against
uncertainty in electron temperature estimation. 

The ORL ionic abundances 
X$^{m+}$/H$^{+}$ are derived
from 
\medskip

\begin{equation}
\label{abunr}
\frac{{\rm X^{m+}}}{{\rm H^{+}}} = \frac{\alpha({\rm
H\beta})}{\alpha({\rm X^{m+}})}
\frac{\lambda({\rm X^{m+}})}{\lambda({\rm H\beta})}\frac{I({\rm
X^{m+}})}{I({\rm H\beta})},
\end{equation}
\medskip

\noindent
where $\alpha({\rm X^{m+}})$ is the recombination coefficient for the
ion ${\rm X^{m+}}$. For calculating ORL ionic abundances, we
adopted $T_{\epsilon}$ of 8800 K and $n_{\epsilon}$ of 10$^{4}$
cm$^{-3}$ from the hydrogen recombination spectrum.

Effective recombination coefficients for the lines' parent multiplets
were taken from the references listed in Table \ref{atomr}. The 
recombination coefficients for each multiplet at a given electron density 
were calculated by fitting the polynomial functions of $T_{\epsilon}$. 
The recombination coefficient of each line was obtained by a branching 
ratio, $B(\lambda_{i})$, which is the ratio of the recombination 
coefficient of the target line, 
$\alpha(\lambda_{i})$ to the total recombination coefficient, 
$\sum_{i} \alpha(\lambda_{i})$ in a multiplet line. 
To calculate the branching ratio, we referred to Wiese et al. (1996)
except for O\,{\sc ii} 3p-3d and 3d-4f transitions and
Ne\,{\sc ii}. For O\,{\sc ii} 3p-3d and 3d-4f transition lines,
the branching ratios were provided by Liu et al. (1995) based on 
intermediate coupling. For Ne\,{\sc ii}, 
Kisielius et al. (1998) provided the branching ratios based on $LS$-coupling.

The estimated ORL ionic abundances are listed in Tables \ref{rec_hec}
and \ref{rec_none}. 
In general, a Case B assumption applies to the 
lines from levels having the same spin as the ground state, and a Case A
assumption applies to lines of other multiplicities. In the last one of 
the line series of each ion, we present the adopted ionic abundance and the
error, which are estimated from the line intensity-weighted mean.

\begin{table}
\centering
\caption{ Effective recombination coefficient References.}
\label{atomr}
\begin{tabular}{@{}l@{\hspace{25pt}}l@{\hspace{25pt}}c}
\hline\hline
Line&Transition&References\\
\hline
H\,{\sc i}  &All&(1),(2)\\
He\,{\sc i} &Singlet &(3)\\
           &Triplet &(4)\\
He\,{\sc ii}&3-4, 4-6 &(4)\\
C\,{\sc ii}&3d-4f, 3s-3p, 3p-3d &(5)\\
          &4f-7g, 4d-6f, 4f-6g&\\
C\,{\sc iii}&3s-3p&(6)\\
           &4f-5g&(4)\\
C\,{\sc iv} &2p-2s, 5fg-6gh&(4)\\
N\,{\sc ii} &3s-3p, 3p-3d &(7)\\
           &3d-4f &(8)\\
N\,{\sc iii}&3s-3p, 3p-3d, 4f-5g &(4)\\
O\,{\sc ii} &3s-3p&(9)\\
           &3p-3d, 3d-4f&(10)\\
O\,{\sc iii}&3s-3p &(4)\\
Ne\,{\sc ii} &3s-3p, 3p$^{'}$-3d$^{'}$, 3s$^{'}$-3p$^{'}$ &(11)\\
\hline
\end{tabular}
\tablerefs{
(1) Aller (1984). (2) Storey \&
	    Hummer (1995). (3) Benjamin et al. (1999). 
	    (4) P\'{e}quignot et al. (1991). (5) Davey et al.
	    (2000). (6) Nussbaumer \& Storey (1984). (7) Kisielius \&
	    Storey (2002). (8) Escalate \& Victor (1990). (9) Storey
	    (1994). (10) Liu et al. (1995). (11) Kisielius et al. (1998).
}
\end{table}

\subsubsection{Helium}
The He$^{+}$ abundances are estimated using electron density insensitive
five {\hei} lines to reduce intensity enhancement by collisional 
excitation from the He$^{0}$ 2$s$ $^{3}S$ level. The collisional excitation 
from the He$^{0}$ 2$s$ $^{3}S$ level enhances mainly the intensity of the triplet {\hei}
lines. We removed this contributions (1.4$\%$ for {\hei}
$\lambda$4387; up to 7.4$\%$ for {\hei} $\lambda$5876) from the observed line
intensities using the formulae given by Kingdon \& Ferland
(1995).

The He$^{2+}$ abundance is estimated from {\heii}~$\lambda$4686. 
Kniazev et al. (2008) estimated He$^{+}$ = 8.52(--2) and He$^{2+}$ =
1.53(--2), which are close to or slightly smaller than our values.

\subsubsection{Carbon}
We observed C\,{\sc ii} lines which arose from different transitions. The
ground state of C\,{\sc ii} line is a doublet (2$p$ $^{2}$$P_{0}$). 
The 3d-4f (multiplet V6), 
4d-6f (V16.04), 4f-6g (V17.04), and 4f-7g (V17.06)
lines, which have higher angular momentum as upper levels, 
are unaffected by both resonance fluorescence by starlight and 
recombination from excited $^{2}S$ and $^{2}D$ terms. Among these high angular momentum
lines, the V6 lines are the most case-insensitive and reliable. Comparison of the C$^{2+}$ abundance 
derived from C\,{\sc ii}~$\lambda$4267 with that of the other C\,{\sc ii} 
lines indicates that the observed C\,{\sc ii} lines are not populated by the intensity enhancement
mechanisms discussed above. Therefore we can safely use all the C\,{\sc ii} lines
for the estimation of C$^{2+}$ abundance.

All the observed C\,{\sc iii} lines are triplets. Since the ground state
of C\,{\sc iii} is singlet (2$s^{2}$ $^{1}S$), we adopted Case A assumption.
Unlike the case of C\,{\sc ii}, C\,{\sc iii} lines are relatively
case insensitive. Our estimated C$^{2+}$ and C$^{3+}$ abundances (Table \ref{rec_hec}) 
are in good agreement with Kanizev et al. (2008); their C$^{2+}$ and 
C$^{3+}$ are 7.78(--4) and 5.62(--4), respectively.

We estimate the C$^{4+}$ abundance using multiplet V8 and V8.01 lines. 
Interestingly, we observed C\,{\sc iv}~$\lambda$5811. C\,{\sc iv} 
$\lambda$5801/11 has been detected in PNe with Wolf Rayet type central 
stars, suggesting that the central star is very active. In the case of 
BoBn 1 C\,{\sc iv} lines might be be nebular origin rather than the 
central star origin, because the 2$V_{\rm exp}$ of C\,{\sc iv}~$\lambda$5811 is 
14.3 {\kms} comparable with the value in close I.P. ions such as 
[Ne\,{\sc iv}] and [F\,{\sc iv}] (see Table \ref{exp}).

\subsubsection{Nitrogen}
All of the observed N\,{\sc ii} lines are triplets. 
Since the ground level of N\,{\sc ii} is a triplet (2$p^{2}$ $^{3}P$), 
we adopted Case B assumption. 
The N\,{\sc ii} resonance line 2$p^2$ $^{3}P$$-$2$p$4$s$ 
$^{3}P_{1}$ $\lambda$508.668 {\AA} can be enhanced by the {\hei} resonance line 1
s$^{2}$ $^{1}S$$-$1$s$8$p$ $^{1}P_{0}$ $\lambda$508.697 {\AA}. The 
cascade transition from 2$p$4$s$ $^{3}P_{1}$ can enhance the line intensity
of the multiplet V3 lines. But, this transition cannot enhance the line intensity of 
3f-4d transition (multiplet V43b, V48a, V50a, and V55a) due to the lack of a direct resonance or cascade excitation
path. Comparison of N$^{2+}$ abundances derived from the 3f-4d with those from 
the V3 lines implies that the fluorescence is negligible 
in BoBn 1. 

The multiplet V1, V2, and V17 N\,{\sc iii} lines are observed. 
We adopted Case B assumption except for the V17 multiplet. For the V17 line, 
we adopted Case A assumption.
The intensity of the resonance N\,{\sc iii} line 
$\lambda$374.36 {\AA} (2$p$ $^{2}P^{0}$$-$3$d$ $^{2}D$) may be enhanced by O\,{\sc iii}
resonance at 374.11 {\AA} (2$p^{2}$ $^{3}P$$-$3$s^{3}$ $P^{0}$).
The line intensity of the multiplet V1 and V2 lines might be enhanced by the 
O\,{\sc iii} lines. The multiplet V17 line (4f-5g transition) does not appear to 
be enhanced. Therefore, we adopt the N$^{3+}$ abundance from this line.

\subsubsection{Oxygen \label{rox}}
We observed O\,{\sc ii} doublet (3d-4f) and quadruplet lines 
(multiplet V1, V4, V10, V19). Most of the V 1 lines and all of the V 2 
lines are observed. Since the ground level of O\,{\sc ii} 
is a quadruplet, we adopted Case A for the doublet lines and Case B 
for quadruplet lines. It seems that the multiplet V1 and V10 lines give 
the most reliable value. 

A number of O\,{\sc iii} lines are observed.
We consider Case B for the triplet lines (multiplet 
V2) and Case A for the singlet line (multiplet V5). There is a
possibility that the multiplet V2 lines would be excited by the 
Bowen fluorescence mechanism or by the charge exchange of O$^{3+}$ 
and H$^{0}$ instead of recombination and the multiplet V5 line 
could be excited by charge exchange. Therefore, we did 
not use O$^{3+}$ abundances in the estimation of a total oxygen abundance
from ORLs.

\subsubsection{Neon}
The observed Ne\,{\sc ii} lines are doublet (multiplets V9 and V21) 
and quartet lines (V1 and V2). We considered Case B 
for the doublet lines and Case A for the quartet lines. 
The multiplet V1 and V2 lines are 
insensitive to the case assumption and are pure
recombination lines (Grandi 1976). Therefore, we adopted 
the Ne$^{2+}$ abundance derived from multiplet V 1 and V2 lines.


\begin{table}
\centering
\footnotesize
\caption{He and C ionic abundances from ORLs.\label{rec_hec}}
\begin{tabular}{@{}l@{\hspace{5pt}}c@{\hspace{5pt}}c@{\hspace{5pt}}c@{}}
\hline\hline
Multi.&$\lambda_{\rm lab}$&$I$($\lambda_{\rm lab}$)&\\
      &(\AA)&[$I$({\hb})=100]&He$^{+}$/H$^{+}$\\
\hline
V11&5876.62&   18.1 $\pm$ 0.14&	9.93(--2) $\pm$ 1.26(--3)\\
V14&4471.47&	4.82 $\pm$ 0.04&	9.63(--2) $\pm$ 2.89(--3)\\
V46&6678.15&	4.02 $\pm$ 0.11&	9.59(--2) $\pm$ 3.80(--3)\\
V48&4921.93&	1.32 $\pm$ 0.02&	9.66(--2) $\pm$ 3.02(--3)\\
V51&4387.93&	0.59 $\pm$ 0.01&	9.54(--2) $\pm$ 3.57(--3)\\
   &{\bf Adopted}&		     &   {\bf 9.81(--2) $\pm$ 2.01(--3)}\\
\hline
&&&He$^{2+}$/H$^{+}$\\
\hline
3.4	&4685.68&	24.8 $\pm$ 0.79&	{\bf 2.03(--2) $\pm$ 6.47(--4)}\\
\hline
&&&C$^{2+}$/H$^{+}$\\
\hline
V2 & 6578.05 & 3.75(--1) $\pm$ 6.87(--3) & 7.30(--4) $\pm$ 3.50(--5) \\ 
V6 & 4267.15 & 7.90(--1) $\pm$ 4.42(--2) & 7.55(--4) $\pm$ 5.02(--5) \\ 
V16.04 & 6151.27 & 3.81(--2) $\pm$ 3.40(--3) & 8.74(--4) $\pm$ 8.26(--5) \\ 
V17.04 & 6461.95 & 7.80(--2) $\pm$ 6.77(--3) & 7.24(--4) $\pm$ 6.97(--5) \\ 
V17.06 & 5342.43 & 5.39(--2) $\pm$ 3.94(--3) & 9.74(--4) $\pm$ 8.16(--5) \\ 
 & {\bf Adopted} &  &  {\bf 7.58(--4) $\pm$ 4.92(--5)} \\  
\hline
&&&C$^{3+}$/H$^{+}$\\
\hline
V1 & 4647.42 & 4.41(--1) $\pm$ 3.82(--3) & 7.60(--4) $\pm$ 2.15(--5) \\ 
V1 & 4650.25 & 2.61(--1) $\pm$ 4.77(--3) & 7.49(--4) $\pm$ 2.44(--5) \\ 
V16 & 4067.87 & 2.49(--1) $\pm$ 8.01(--3) & 6.11(--4) $\pm$ 2.80(--5) \\ 
V16 & 4070.20 & 4.07(--1) $\pm$ 1.10(--2) & 5.54(--4) $\pm$ 2.36(--5) \\ 
V18 & 4186.90 & 3.46(--1) $\pm$ 5.45(--3) & 5.80(--4) $\pm$ 2.11(--5) \\ 
V43 & 8196.50 & 4.39(--1) $\pm$ 8.34(--3) & 5.66(--4) $\pm$ 2.17(--5) \\ 
 & {\bf Adopted} &  &   {\bf 5.74(--4) $\pm$ 2.32(--5)} \\ 
\hline
&&&C$^{4+}$/H$^{+}$\\
\hline
V8 & 4658.64 & 1.19(--1) $\pm$ 1.91(--2) & 2.69(--5) $\pm$ 4.41(--6) \\ 
V8.01 & 7725.90 & 3.11(--2) $\pm$ 1.52(--3) & 1.49(--5) $\pm$ 8.81(--7) \\ 
 & {\bf Adopted} &  &   {\bf 2.69(--5) $\pm$ 4.41(--6)} \\ 
\hline
\end{tabular}
\end{table}

\begin{table}
\footnotesize
\centering
\caption{N, O, and Ne ionic abundances from ORLs.\label{rec_none}}
\begin{tabular}{@{}l@{\hspace{5pt}}c@{\hspace{5pt}}c@{\hspace{5pt}}c@{}}
\hline\hline
Multi.&$\lambda_{\rm lab}$&$I$($\lambda_{\rm lab}$)&\\
      &(\AA)&[$I$({\hb})=100]&N$^{2+}$/H$^{+}$\\
\hline
V3  & 5710.76 & 5.15(--3) $\pm$ 4.49(--3) & 1.26(--4) $\pm$ 1.10(--4) \\ 
V3  & 5685.26 & 2.52(--2) $\pm$ 2.36(--3) & 6.62(--4) $\pm$ 6.48(--5) \\ 
V3  & 5679.56 & 1.62(--2) $\pm$ 4.31(--3) & 6.76(--5) $\pm$ 1.81(--5) \\ 
V19  & 5001.47$^{a}$ & 1.79(--2) $\pm$ 3.19(--3) & 4.75(--5) $\pm$ 8.61(--6) \\ 
V43b  & 4171.61 & 1.15(--2) $\pm$ 2.07(--3) & 1.63(--4) $\pm$ 2.99(--5) \\ 
V48a  & 4247.22 & 2.42(--2) $\pm$ 5.05(--3) & 1.14(--4) $\pm$ 2.41(--5) \\ 
V50a  & 4179.67 & 1.36(--2) $\pm$ 7.54(--3) & 3.43(--4) $\pm$ 1.91(--4) \\ 
V55a  & 4442.02 & 1.27(--2) $\pm$ 3.88(--3) & 3.62(--4) $\pm$ 1.11(--4) \\ 
      &{\bf Adopted} &   & {\bf 2.62(--4) $\pm$ 5.99(--5)} \\ 
\hline
&&&N$^{3+}$/H$^{+}$\\
\hline 
V1  & 4097.35 & 5.00(--1) $\pm$ 2.47(--2) & 1.39(--3) $\pm$ 7.97(--5) \\ 
V1  & 4103.39 & 3.14(--1) $\pm$ 3.72(--2) & 1.75(--3) $\pm$ 2.13(--4) \\ 
V2  & 4634.12 & 1.67(--1) $\pm$ 5.69(--3) & 1.31(--4) $\pm$ 5.98(--6) \\ 
V2  & 4640.64 & 3.22(--1) $\pm$ 3.62(--3) & 1.41(--4) $\pm$ 4.55(--6) \\ 
V2  & 4641.85 & 4.88(--2) $\pm$ 5.97(--3) & 1.92(--4) $\pm$ 2.42(--5) \\ 
V17  & 4379.11 & 5.97(--2) $\pm$ 5.13(--3) & 2.56(--5) $\pm$ 2.36(--6) \\    
&{\bf Adopted} &    &{\bf 2.56(--5) $\pm$ 2.36(--6)}\\
\hline
&&&O$^{2+}$/H$^{+}$\\
\hline
V1 & 4638.86 & 1.40(--2) $\pm$ 1.21(--3) & 1.27(--4) $\pm$ 9.78(--6) \\ 
V1 & 4641.81 & 3.46(--2) $\pm$ 9.54(--3) & 1.30(--4) $\pm$ 3.75(--5) \\ 
V1 & 4649.13 & 1.86(--2) $\pm$ 1.43(--3) & 3.95(--5) $\pm$ 2.38(--6) \\ 
V1 & 4650.84 & 2.82(--2) $\pm$ 2.48(--3) & 2.72(--4) $\pm$ 2.10(--5) \\ 
V1 & 4661.63 & 2.32(--2) $\pm$ 4.71(--3) & 1.85(--4) $\pm$ 3.81(--5) \\ 
V1 & 4673.73 & 1.85(--2) $\pm$ 9.08(--3) & 9.87(--4) $\pm$ 4.68(--4) \\ 
V1 & 4676.23 & 7.65(--3) $\pm$ 4.19(--3) & 8.01(--5) $\pm$ 4.40(--5) \\ 
V4 & 6721.39 & 2.99(--3) $\pm$ 3.24(--4) & 5.13(--4) $\pm$ 5.73(--5) \\ 
V10 & 4069.62 & 1.03(--2) $\pm$ 3.01(--3) & 1.06(--4) $\pm$ 3.12(--5) \\ 
V10 & 4069.88 & 1.60(--2) $\pm$ 1.60(--2) & 1.03(--4) $\pm$ 3.04(--5) \\ 
V19 & 4153.30 & 1.68(--2) $\pm$ 3.11(--3) & 2.19(--4) $\pm$ 4.11(--5) \\ 
3d-4f & 4089.29 & 1.43(--2) $\pm$ 5.64(--3) & 1.30(--4) $\pm$ 5.14(--5) \\ 
3d-4f & 4292.21$^{b}$ & 1.76(--2) $\pm$ 5.00(--3) & 6.28(--4) $\pm$ 1.79(--4) \\ 
 & {\bf Adopted} &   &   {\bf 1.45(--4)} $\pm$ {\bf 2.32(--5)} 
\\ 
\hline
&&&O$^{3+}$/H$^{+}$\\
\hline 
V2 & 3754.70 & 1.71(--1) $\pm$ 6.17(--3) & 3.31(--4) $\pm$ 1.54(--5) \\ 
V2 & 3757.21 & 8.29(--2) $\pm$ 7.35(--3) & 3.62(--4) $\pm$ 3.38(--5) \\ 
V2 & 3759.88 & 6.09(--1) $\pm$ 1.97(--2) & 6.47(--4) $\pm$ 2.82(--5) \\ 
V2 & 3791.27 & 6.71(--2) $\pm$ 5.85(--3) & 4.29(--4) $\pm$ 3.95(--5) \\ 
V5 & 5592.37 & 1.07(--2) $\pm$ 1.96(--3) & 4.42(--4) $\pm$ 8.19(--5) \\ 
\hline
&&&Ne$^{2+}$/H$^{+}$\\
\hline
V1 & 3694.21	&4.14(--2) $\pm$ 8.49(--3)&	1.28(--4) $\pm$ 2.71(--5)\\
V2 & 3334.87	&1.04(--1) $\pm$ 1.93(--2)&	1.60(--4) $\pm$ 3.08(--5)\\
V9 & 3568.50	&6.95(--2) $\pm$ 7.83(--3)&	2.24(--3) $\pm$ 2.62(--4)\\
V21 &3453.07	&1.25(--2) $\pm$ 3.82(--3)&	4.94(--4) $\pm$ 1.53(--4)\\
&{\bf Adopted}			&&{\bf 1.51(--4) $\pm$ 2.98(--5)}\\
\hline
\end{tabular}
\tablenotetext{a}{blending line ($\lambda$ 5001.12,5001.47 lines).}
\tablenotetext{b}{blending line ($\lambda$ 4291.26,4291.86,4292.21,4292.98).}
\end{table}

\subsection{Ionization Correction}
If the ionic abundances in all ionization stages are known, an elemental
abundance will be simply the sum of its ionic abundances. Actually, it is, 
however, impossible to probe all of the ionization stages of an element
using UV to mid-infrared spectra. To estimate elemental abundances, 
we must correct for unobserved ionic abundances. This correction was 
performed using ionization correction factors, ICF(X). ICFs(X) for
each element are listed in Table \ref{icf}. 

\subsubsection{Helium, Carbon, Nitrogen, Oxygen and Neon}
The He abundance is the sum of He$^{+}$ and He$^{2+}$. 

The C abundance is 
the sum of C$^{+}$, C$^{2+}$, C$^{3+}$, and C$^{4+}$. 
For the C abundance derived from ORLs, we corrected for unseen C$^{+}$ 
assuming (C$^{+}$/C)$_{\rm ORLs}$ = (N$^{+}$/N)$_{\rm
CELs}$. For the C abundance from CELs, we corrected for C$^{4+}$ 
assuming (C$^{4+}$/C)$_{\rm CELs}$ = (C$^{4+}$/C)$_{\rm ORLs}$.

The N abundance is the sum of N$^{+}$, N$^{2+}$, and N$^{3+}$.
For the ORL N abundance, we corrected for 
N$^{+}$ assuming (N$^{+}$/N)$_{\rm ORLs}$ = (N$^{+}$/N)$_{\rm CELs}$.

The O abundance is the sum of O$^{+}$,
O$^{2+}$, and O$^{3+}$. For the ORL O abundance, we used 
only O$^{2+}$ because most of the O {\sc iii} lines 
are not pure recombination lines. We assumed
(O$^{2+}$/O)$_{\rm ORLs}$ = (O$^{2+}$/O)$_{\rm CELs}$.

The Ne abundance is the sum of Ne$^{+}$, Ne$^{2+}$, Ne$^{3+}$, and Ne$^{4+}$. 
For the ORL Ne abundance, we corrected for the unseen Ne$^{+}$, Ne$^{3+}$ and
Ne$^{4+}$ assuming (Ne$^{2+}$/Ne)$_{\rm ORLs}$ = (Ne$^{2+}$/Ne)$_{\rm CELs}$.

\subsubsection{Other elements}
Assuming that the F abundance is the sum of F$^{+}$, F$^{2+}$, F$^{3+}$,
and F$^{4+}$, we corrected for unseen F$^{4+}$ using the CEL Ne abundance. 
The S abundance is the sum of S$^{+}$, S$^{2+}$, and S$^{3+}$, 
and S$^{4+}$. Unseen S$^{4+}$ was corrected for assuming
S$^{4+}$/S = (N$^{3+}$/N)$_{\rm CELs}$. We assume that the Cl 
abundance is the sum of Cl$^{+}$, Cl$^{2+}$, Cl$^{3+}$, and
Cl$^{4+}$. The unseen Cl$^{+}$ and Cl$^{4+}$ are
corrected for assuming Cl/(Cl$^{+}$+Cl$^{4+}$) = O/(O$^{+}$+O$^{3+}$)$_{\rm CELs}$. 
For Ar, its abundance 
is assumed to be the sum of Ar$^{2+}$, Ar$^{3+}$, and Ar$^{4+}$, and unseen Ar$^{4+}$ was 
corrected for assuming (Ar$^{4+}$/Ar) = (Ne$^{4+}$/Ne)$_{\rm CELs}$.
For Fe, we assume that its abundance is the sum of 
Fe$^{2+}$, Fe$^{3+}$, and Fe$^{4+}$. The unseen Fe$^{4+}$ was corrected for 
assuming (Fe$^{4+}$/Fe) = (O$^{3+}$/O)$_{\rm CELs}$.

We assume that the Kr abundance is the sum of Kr$^{2+}$, Kr$^{3+}$,
and Kr$^{4+}$, the unseen Kr$^{2+}$ was corrected for assuming
Kr$^{2+}$/Kr$^{3+}$ = Cl$^{2+}$/Cl$^{3+}$. We assume that the Xe abundance 
is the sum of Xe$^{+}$, Xe$^{2+}$, Xe$^{3+}$, and Xe$^{4+}$. The Xe ionic
abundances except Xe$^{2+}$ were corrected for assuming Xe$^{2+}$/Xe =
S$^{2+}$/S. We give the lower limit of the Ba abundance, which is 
equal to the Ba$^{+}$ abundance (IP = 5.2 eV), since we could not 
detect higher excited Ba lines with IP $>$ 13.5 eV. It should 
take care in handling the Ba abundance.

\begin{table}
\centering

\caption{Adopted ionization correction factors (ICFs).\label{icf}}
\begin{tabular}{@{}l@{\hspace{8pt}}ll@{\hspace{4pt}}l@{}}
\hline\hline
X &Line&ICF(X)&X/H\\
\hline
He &ORLs &1&He$^{+}$+He$^{2+}$\\

C  &CELs &1&C$^{+}$+C$^{2+}$+C$^{3+}$\\
   &ORLs &$\rm \left(\frac{N}{N^{2+} +
 N^{3+}}\right)^{\dagger}$&ICF(C)$\rm \left(C^{2+}+C^{3+}+C^{4+}\right)$\\
\noalign{\medskip}

N &CELs & 1 &N$^{+}$+N$^{2+}$+N$^{3+}$\\
  &ORLs &$\rm \left(\frac{N}{N^{2+}+N^{3+}}\right)^{\dagger}$&ICF(N)$\rm \left(N^{2+}+N^{3+}\right)$\\
\noalign{\medskip}

O  &CELs & 1 &O$^{+}$+O$^{2+}$+O$^{3+}$\\
   &ORLs &$\rm \left(\frac{O}{O^{2+}}\right)$$^{\dagger}$ &ICF(O)O$^{2+}$\\
\noalign{\medskip}

F &CELs &$\rm \left(\frac{Ne}{Ne^{+}+Ne^{2+}+Ne^{3+}}\right)^{\dagger}$&ICF(F)$\rm \left(F^{+}+F^{2+}+F^{3+}\right)$\\
\noalign{\medskip}

Ne &CELs & 1 &Ne$^{+}$+Ne$^{2+}$+Ne$^{3+}$+Ne$^{4+}$ \\
   &ORLs &$\rm \left(\frac{Ne}{Ne^{2+}}\right){\dagger}$ &ICF(Ne)Ne$^{2+}$\\
\noalign{\medskip}

S &CELs &$\rm \left(\frac{N}{N^{+}+N^{2+}}\right)^{\dagger}$ &ICF(S)$\rm\left(S^{+}+S^{2+}+S^{3+}\right)$\\
\noalign{\medskip}

Cl &CELs &$\rm {\left(\frac{O}{O^{2+}}\right)}^{\dagger}$ &ICF(Cl)$\rm \left(Cl^{2+}+Cl^{3+}\right)$\\
\noalign{\medskip}

Ar &CELs &$\rm \left(\frac{Ne}{Ne^{+}+Ne^{2+}+Ne^{4+}}\right)^{\dagger}$ &ICF(Ar)$\rm \left(Ar^{2+}+Ar^{3+}\right)$\\
\noalign{\medskip}

Fe &CELs &$\rm {\left(\frac{O}{O^{+}+O^{2+}}\right)}^{\dagger}$ &ICF(Fe)$\rm \left(Fe^{2+}+Fe^{3+}\right)$\\
\noalign{\medskip}

Kr &CELs &$\rm \frac{Cl^{2+}+Cl^{3+}}{Cl^{3+}}$ &ICF(Kr)Kr$^{3+}$+Kr$^{4+}$\\
\noalign{\medskip}

Xe &CELs &$\rm \frac{S}{S^{2+}}$ &ICF(Xe)Xe$^{2+}$\\
\noalign{\medskip}

Ba &CELs$^{\ddagger}$ &1&Ba$^{+}$\\
\hline
\end{tabular}
\tablenotetext{$^{\dagger}$}{Ionic and elemental abundances derived from CELs.}
\tablenotetext{$^{\ddagger}$}{The value is the lower limit (see text).}
\end{table}

\subsection{Elemental abundances}
The resultant elemental abundances are presented in Table 
\ref{ea1}. We recognized that BoBn1 is a C-, N-, and Ne-rich PN: 
the [C/O], [N/O], and [Ne/O] abundances from ORLs are +1.23, +1.12, +0.81.  
The ratios derived from CELs are +1.58, +1.10, and +1.04, respectively. 
Comparing the C, N, O, Ne ORL and CEL abundances, 
ORLs might be emitted from O-, Ne-rich region. 
The ORL C, N, O, Ne abundances are larger by 0.14-0.49 dex 
than the CEL abundances. We need to look for reasons for the abundance 
discrepancy.

\begin{table*}
\centering
\footnotesize
\caption{The elemental abundances derived from CELs and ORLs in the case
 of no temperature fluctuation.\label{ea1}}
\begin{tabular}{@{}ccccccccc@{}}
\hline\hline
X &\multicolumn{2}{c}{X/H}&&\multicolumn{2}{c}{log(X/H)+12$^{a}$}&&\multicolumn{2}{c}{[X/H]$^{b}$}\\
\cline{2-3}\cline{5-6}\cline{8-9}
&CELs&ORLs&&CELs&ORLs&&CELs&ORLs\\
\hline
He&$\cdots$&1.18(--1) $\pm$ 2.12(--3)&&$\cdots$&11.07$\pm$0.01&&$\cdots$&+0.17$\pm$0.01\\
C&1.05(--3) $\pm$ 1.93(--4)&1.44(--3) $\pm$ 4.96(--4)&&9.02$\pm$0.08&9.16$\pm$0.16&&+0.63$\pm$0.09&+0.77$\pm$0.16\\
N&1.07(--4) $\pm$ 2.50(--5)&3.06(--4) $\pm$ 1.22(--5)&&8.03$\pm$0.10&8.49$\pm$0.18&&+0.15$\pm$0.15&+0.66$\pm$0.21\\
O&5.51(--5) $\pm$ 3.84(--6)&1.68(--4) $\pm$ 3.22(--5)&&7.74$\pm$0.03&8.23$\pm$0.08&&$-$0.95$\pm$0.06&$-$0.46$\pm$0.10\\
F&7.01(--7) $\pm$ 1.38(--7)&$\cdots$&&5.85$\pm$0.09&$\cdots$&&+1.39$\pm$0.11&$\cdots$\\
Ne&9.04(--5) $\pm$ 4.42(--6)&1.64(--4) $\pm$ 3.44(--5)&&7.96$\pm$0.02&8.22$\pm$0.09&&+0.09$\pm$0.10&+0.35$\pm$0.14\\
S&2.07(--7) $\pm$ 7.53(--8)&$\cdots$&&5.32$\pm$0.17&$\cdots$&&$-$1.87$\pm$0.17&$\cdots$\\
Cl&2.47(--9) $\pm$ 4.02(--10)&$\cdots$&&3.39$\pm$0.07&$\cdots$&&$-$1.94$\pm$0.09&$\cdots$\\
Ar&2.13(--8) $\pm$ 1.75(--9)&$\cdots$&&4.33$\pm$0.04&$\cdots$&&$-$2.22$\pm$0.09&$\cdots$\\
Fe&1.21(--7) $\pm$ 3.69(--8)&$\cdots$&&5.08$\pm$0.14&$\cdots$&&$-$2.39$\pm$0.14&$\cdots$\\
Kr&$<$7.63(--10) &$\cdots$&&$<$2.88&$\cdots$&&$<$$-$0.48&$\cdots$\\
Xe&$<$9.33(--11) &$\cdots$&&$<$1.97&$\cdots$&&$<$$-$0.27&$\cdots$\\
Ba&1.98(--10) $\pm$ 6.44(--11)&$\cdots$&&2.30$\pm$0.15&$\cdots$&&+0.12$\pm$0.15&$\cdots$\\
\hline
\end{tabular}
\tablenotetext{a}{The number density of the hydrogen is 12.}
\tablenotetext{b}{Solar abundances are taken from Lodders (2003).}
\end{table*}

\section{Discussion}
First, in this section, 
we will discuss the abundance discrepancies between CELs and ORLs using
three models (Section 4.1). Second, we will compare elemental abundances
estimated by us with others (Section 4.2). Third, we will build a photo-ionization model to
derive the parameters of the central star, ionized nebular gas, and dust (Section 4.3). Next, 
the empirically derived elemental abundances will be compared with theoretical
nucleosynthesis model predictions for low- to intermediate mass stars (Section 4.4). 
Finally, we will guess the evolutionary status or provide a presumable evolutionary scenario 
for BoBn 1 (Section 4.5).

\subsection{The abundance discrepancy between CELs and ORLs}

\begin{figure}
\centering
\includegraphics[width=80mm]{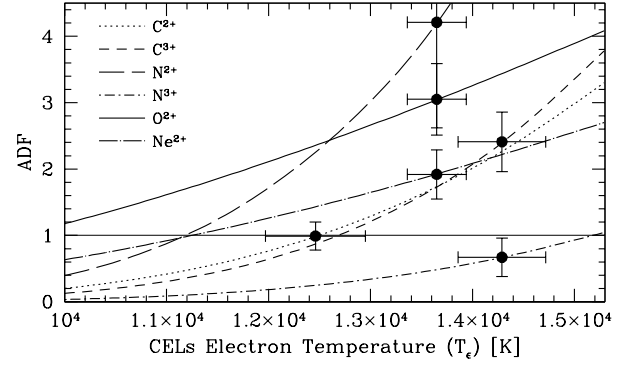}
\caption{The abundance discrepancy factor (ADF) vs. the electron
 temperature from CELs. The filled circles are estimated values when
 adopting the $T_{\epsilon}$ and $n_{\epsilon}$ values listed in Table
 \ref{temp_ne_cles}.}
\label{adf_f}
\end{figure}

We derived ionic and elemental abundances using CELs and ORLs and found somewhat 
large abundance discrepancies between them. 
So far, abundance discrepancies have been found in about 90 
Galactic disk PNe, 3 Magellanic PNe (Tsamis et al. 2003, 2004; Liu et al. 2004; 
Robertson-Tessi \& Garnett 2005; Wesson et al. 2005, etc.), and 1 Halo
PN (DdDm 1; Otsuka et al. 2009). 
We define the ionic abundance discrepancy factor ADF 
as the ratio of the ORL to the UV or optical CEL
abundances. In BoBn 1, the ADFs are 0.98$\pm$0.21 for C$^{2+}$,
2.39$\pm$0.45 for C$^{3+}$, 4.21$\pm$1.59 for N$^{2+}$, 0.67$\pm$0.29
for N$^{3+}$, 3.05$\pm$0.54 for O$^{2+}$, and 1.82$\pm$0.39 for Ne$^{2+}$,
respectively

Up to now, three models have been proposed to explain abundance
discrepancies in PNe: temperature fluctuations, high density components, 
and hydrogen-deficient cold components. We examine what can 
cause the abundance discrepancies in BoBn 1 using these models.

\subsubsection{Temperature fluctuations}
The emissivities of the CELs increase exponentially as the
electron temperature becomes higher. The electron temperature derived 
from the CELs, $T_{\epsilon}$(CELs) will be indicative of the hot region 
nearby the radiation source. If we adopt $T_{\epsilon}$(CELs) for abundance estimations using the CELs,  
the ionic abundances might be underestimated.

Peimbert (1967) considered the effect of electron temperature fluctuation in 
a nebula, which sometimes gave high electron temperature, 
on the determinations of the ionic abundances derived from CELs. For example, 
Torres-Peimbert et al. (1980) characterized the electron temperature fluctuations in term of
$t^{2}$ as the cause of the abundance discrepancy between CELs and
ORLs. Assuming the validity of the temperature fluctuation paradigm,
the comparison of the ionic abundances derived from CELs and ORLs
may provide an estimation of $t^{2}$.

The relation between ADFs and $T_{\epsilon}$(CELs) is presented in
Fig. \ref{adf_f}. We recognize that ADFs would approach to 1 if
$T_{\epsilon}$(CELs) for each zone dropped by $>$1000 K. Using the 
formulations for temperature fluctuations given by Peimbert (1967), we
have estimated the $t^{2}$ parameter and the mean electron temperatures
$T_{0}$ for each zone . The resultant $t^{2}$ and $T_{0}$ 
are listed in Table \ref{t2}. The derived $t^{2}$ = 0.027$\pm$0.011 
indicates that the temperature fluctuations are $\sim$16 $\%$ inside
nebula. 
The temperature fluctuation in BoBn 1 is low, compared with typical 
Galactic PNe, i.e. $t^{2}$$<$0.1 (cf. Zhang et al. 2004).

When we take the temperature fluctuation effect into account, the
derived ADFs become lower than the $t^{2}$=0 case; ADFs are 
0.87$\pm$0.25 for C$^{2+}$, 1.12$\pm$0.38 for C$^{3+}$, 2.95$\pm$1.32 for N$^{2+}$, 0.32$\pm$0.16
for N$^{3+}$, 2.63$\pm$0.55 for O$^{2+}$, and 1.69$\pm$0.37 for Ne$^{2+}$,
respectively. The great improvements of C$^{2+}$ and C$^{3+}$ might have been caused 
by the large temperature dependency of C\,{\sc iii}] $\lambda\lambda$1906/09 and 
C\,{\sc iv} $\lambda\lambda$1549/50; the energy difference between upper and lower level, 
$\Delta\,E$ = $k$$\Delta\,T$, where $\Delta\,T$ = 75\,380 K and 44\,820 K, 
respectively. This also implies that C$^{2+}$ and 
C$^{3+}$ abundances from ORLs are more reliable than those from CELs. Concerning 
O$^{2+}$ ADFs, large discrepancy still exists. Since  compared with the 
C\,{\sc iii}] $\lambda\lambda$1906/09 and C\,{\sc iv} $\lambda\lambda$1549/50 lines, 
the [O\,{\sc iii}] nebular lines depend more weakly on the electron temperature 
($\Delta\,T$$\sim$14\,100 K), so we realize that the temperature 
fluctuation model alone cannot improve CEL O$^{2+}$ over $>1$ dex. We need to seek other 
explanations for the large ADF(O$^{2+}$). Potentially this model can explain the discrepancies of 
N$^{2+}$ and Ne$^{2+}$ abundances, taking into account the uncertainties
of measured fluxes of the observed ORLs N\,{\sc ii} and Ne\,{\sc ii}.

In Table \ref{t2abund}, we present elemental abundances from the CELs and ORLs, 
taking into account the above temperature fluctuations. The CEL C, N, and Ne 
abundances become comparable to the ORL abundances. For the O abundance, 
there still exists a large discrepancy.

\begin{table}
\centering
\caption{$t^{2}$ and $T_{0}$ for each zone. \label{t2}}
\begin{tabular}{@{}lcc@{}}
\hline\hline
zone&$t^{2}$&$T_{0}$\\
\hline
0&0.027$\pm$0.011&8920$\pm$840\\
1,2&0.027$\pm$0.011&11\,540$\pm$400\\
3&0.027$\pm$0.011&12\,220$\pm$630\\
4&0.027$\pm$0.011&12\,950$\pm$610\\
5&0.027$\pm$0.011&12\,870$\pm$730\\
6&0.027$\pm$0.011&12\,790$\pm$840\\
\hline
\end{tabular}
\end{table}

\begin{table*}
\centering
\footnotesize
\caption{The elemental abundances derived from CELs and ORLs in the case
 of $t^{2}\neq0$.\label{t2abund}}
\begin{tabular}{@{}ccccccccc@{}}
\hline\hline
X &\multicolumn{2}{c}{X/H}&&\multicolumn{2}{c}{log(X/H)+12$^{a}$}&&\multicolumn{2}{c}{[X/H]$^{b}$}\\
\cline{2-3}\cline{5-6}\cline{8-9}
&CELs&ORLs&&CELs&ORLs&&CELs&ORLs\\
\hline
He&$\cdots$&1.18(--1) $\pm$ 2.12(--3)&&$\cdots$&11.07$\pm$0.01&&$\cdots$&+0.17$\pm$0.01\\
C&1.44(--3) $\pm$ 3.31(--4)&1.44(--3) $\pm$ 4.96(--4)&&9.16$\pm$0.10&9.16$\pm$0.16&&+0.77$\pm$0.11&+0.77$\pm$0.16\\
N&1.77(--4) $\pm$ 5.37(--5)&2.99(--4) $\pm$ 1.45(--5)&&8.25$\pm$0.14&8.48$\pm$0.23&&+0.42$\pm$0.18&+0.66$\pm$0.26\\
O&6.35(--5) $\pm$ 7.38(--6)&1.67(--4) $\pm$ 3.98(--5)&&7.80$\pm$0.05&8.22$\pm$0.11&&$-$0.89$\pm$0.07&$-$0.47$\pm$0.12\\
F&5.43(--7) $\pm$ 1.57(--7)&$\cdots$&&5.73$\pm$0.13&$\cdots$&&+1.27$\pm$0.14&$\cdots$\\
Ne&1.01(--4) $\pm$ 9.26(--6)&1.71(--4) $\pm$ 4.08(--5)&&8.00$\pm$0.04&8.23$\pm$0.11&&+0.13$\pm$0.11&+0.36$\pm$0.15\\
S&2.48(--7) $\pm$ 1.16(--7)&$\cdots$&&5.32$\pm$0.22&$\cdots$&&$-$1.80$\pm$0.23&$\cdots$\\
Cl&2.47(--9) $\pm$ 4.02(--10)&$\cdots$&&3.39$\pm$0.07&$\cdots$&&$-$1.94$\pm$0.09&$\cdots$\\
Ar&2.36(--8) $\pm$ 3.61(--9)&$\cdots$&&4.37$\pm$0.07&$\cdots$&&$-$2.18$\pm$0.10&$\cdots$\\
Fe&1.53(--7) $\pm$ 5.71(--8)&$\cdots$&&5.18$\pm$0.17&$\cdots$&&$-$2.29$\pm$0.17&$\cdots$\\
Kr&$<$8.74(--10) &$\cdots$&&$<$2.94&$\cdots$&&$<$$-$0.42&$\cdots$\\
Xe&$<$1.33(--10) &$\cdots$&&$<$2.12&$\cdots$&&$<$$-$0.09&$\cdots$\\
Ba&2.54(--10) $\pm$ 1.00(--10)&$\cdots$&&2.41 $\pm$ 0.18&$\cdots$&&+0.23$\pm$0.18&$\cdots$\\
\hline
\end{tabular}
\tablenotetext{a}{The number density of the hydrogen is 12.}
\tablenotetext{b}{Solar abundances are taken from Lodders (2003).}
\end{table*}

\subsubsection{High density components}  
It was proposed by Rubin (1989) and Viegas \& Clegg (1994). 
It assumes that small high density components 
within nebula weaken the intensity of the nebular lines 
due to their collisional de-excitation, assuming that
chemical abundances are homogeneous. In this
situation, the nebular to auroral line intensity ratios become
smaller, from which we would derive falsely high electron
temperatures. Accordingly, the CEL ionic abundances would be 
underestimated. Since the ORLs have very large critical densities, the ORL
ionic abundances are hardly affected by collisional
de-excitation. In the case of the halo PN DdDm 1, Otsuka et al. (2009) discussed the 
possibility that the O$^{2+}$ abundance discrepancy could be explained by this model. 
However, as we argued in \S 3.3.2, no such high density components in BoBn 1 gas. 
Hence this model would not provide a sound ground for the observed 
abundance discrepancy.

\subsubsection{hydrogen deficient cold components}
This model was proposed by Jacoby \& Ford (1983), Liu et al. (2000), 
P\'{e}quignot et al. (2002), Wesson et al. (2003), and others. This model 
assumes the situation as follows; the central star of a PN first
ejects an envelope at low expansion velocity (of the order 10 {\kms}) 
with "normal" heavy metal abundances, and later ejects the 
high-velocity, hydrogen deficient, cold, and rich heavy metal
components. Here, the ORLs are assumed to be emitted mainly from high-velocity 
hydrogen deficient cold components, whereas the CELs are from the hot, normal
metal gas surrounding the ORL emitters.

So far, such components are directly or indirectly observed in the some PNe, for example, 
Abell 30 (Wesson et
al. 2003), NGC 6153 and NGC 7009 (Barlow et al. 2006). In high-velocity components of 
Abell 30, Wesson et al. (2003) found that the electron temperature
derived from O\,{\sc ii} lines is 500--2500 K and the ORL oxygen abundance
is $\sim$100 times larger than from CELs. Abell 30 is a well known PN, for having a hydrogen 
deficient central star and it is suspected to have experienced a very 
late thermal pulse. If this model is the case of Abell 30, the ORL oxygen abundance
 might indicate the amount of O synthesized in the He-rich intershell.  
Barlow et al. (2006) measured the expansion velocities of
{\oiii} and O\,{\sc ii} lines in NGC 6153 and NGC 7009 and 
found that the O\,{\sc ii} expansion velocity is smaller than {\oiii}. Hence, 
they concluded that the O\,{\sc ii} and {\oiii} lines do not 
originate from material of identical physical properties.

To verify whether the large O (and O$^{2+}$) discrepancy in BoBn 1 is due to
difference physical properties between O\,{\sc ii} and {\oiii} lines or not, following Barlow et al. (2006), 
we compare the expansion and the radial velocities of O\,{\sc ii} 
and those of {\oiii} lines. The resultant values are
summarized in Table \ref{coldden}. The third and last columns are the radial velocity $V_{\rm r}$ 
and twice the expansion velocity, $2V_{\rm exp}$, respectively. We also 
estimated $V_{\rm r}$ and $2V_{\rm exp}$ of Ne\,{\sc ii} and {\neiii}. In the last one of the 
line series of each ion, we present the adopted velocities with the bold
face characters. These values are estimated from the line intensity
weighted means. The radial velocities of the O\,{\sc ii}, {\oiii}, 
Ne\,{\sc ii} and {\neiii} lines are almost consistent with the average radial
velocity of 191.6 $\pm$ 1.3 {\kms} that are derived from over 300 lines detected in the
HDS spectra. However, the $2V_{\rm exp}$ values of the O\,{\sc ii} and Ne\,{\sc ii} lines 
are$\sim$10 {\kms} smaller than those of {\oiii} and {\neiii} lines, respectively. 
These findings does not agree to the foregoing high velocity hydrogen deficient cold model. 
We can attribute the difference between the expansion velocities in ORLs and CELs  
to their thermal motions. Then, we can assume a presence of 
oxygen and neon-rich components with the same normal radial velocity, 
surrounded by hot normal-oxygen and neon gas.

Otsuka et al. (2008a) argued that the rich-neon abundance of BoBn 1 
might have been caused by a late-thermal pulse, and the 
central star might have been hydrogen-deficient. At that phase, 
hydrogen-deficient, oxygen and neon-rich cold components might be incidentally ejected 
from the central star. The ORL overabundance could be an evidence indicative of relatively 
recent yields in the central star. Georgiev et al. (2008) found that the ORLs He, C, and 
O abundances in the nebula of NGC 6543 are in good agreement with those in the stellar wind zone, while 
Morisset \& Gorgiev (2009) found that the ORL C, N, and O abundances in the nebula of IC 418 
are in good agreement with those in the stellar wind. In the halo PN K 648, 
Rauch et al. (2002) estimated the C, N, and O abundances using the stellar spectra; C=1.0(--3), 
N=1.0(--6), and O=1.0(--3). Their estimated C abundance agrees with 
the ORL C abundance of 1.8(--3) within the errors (Otsuka 2007). To verify whether 
the ORL abundances in BoBn 1 indicate recent yields in the central star or not, 
we need to estimate the stellar abundances and compare those with the nebular 
ORL abundances using high-dispersion UV spectra.

\begin{table}
\centering
\caption{Radial and twice the expansion velocities for O$^{2+}$ and Ne$^{2+}$ lines. 
\label{coldden}}
\begin{tabular}{@{}lccc@{}}
\hline\hline
Ion &$\lambda_{lab}$ &$V_{\rm r}$ &$2V_{\rm exp}$\\
     & (\AA)         &(km/s)     &(km/s)\\
\hline 
O\,{\sc ii} & 4089.29 & +190.8 $\pm$ 2.8 & 23.7 $\pm$ 6.4 \\ 
     & 4153.30 & +206.8 $\pm$ 2.6 & 33.6 $\pm$ 5.6 \\ 
     & 4292.21$^{a}$ & +190.8 $\pm$ 2.8 & 30.0 $\pm$ 4.9 \\ 
     & 4638.86 & +190.1 $\pm$ 1.3 & 31.0 $\pm$ 2.2 \\ 
     & 4641.81 & +187.2 $\pm$ 4.4 & 24.2 $\pm$ 6.2 \\ 
     & 4649.13 & +205.4 $\pm$ 0.9 & 31.2 $\pm$ 2.0 \\ 
     & 4650.84 & +205.3 $\pm$ 1.6 & 28.3 $\pm$ 1.7 \\ 
     & 4673.73 & +185.0 $\pm$ 4.8 & 55.2 $\pm$ 24.2 \\ 
     & 4676.23 & +194.2 $\pm$ 5.8 & 41.2 $\pm$ 17.3 \\ 
     & 6721.39 & +194.2 $\pm$ 5.8 & 24.5 $\pm$ 1.9 \\ 
     & {\bf Adopted}& {\bf +195.1 $\pm$ 3.0} & {\bf 31.8 $\pm$ 6.9} \\ 
     &         &                 &  \\ 
{\oiii}&4363.21  & +191.0 $\pm$ 0.2 & 40.3 $\pm$ 0.1 \\ 
       & 4931.80 & +194.2 $\pm$ 1.7 & 47.5 $\pm$ 3.9 \\ 
       & 4958.91 & +193.7 $\pm$ 0.8 & 40.5 $\pm$ 0.3 \\ 
       & 5006.84 & +193.8 $\pm$ 1.3 & 42.8 $\pm$ 0.3 \\ 
       &{\bf Adopted} &{\bf +193.7 $\pm$ 1.2} &{\bf 42.2 $\pm$ 0.3} \\ 
       &         &                 &\\ 
Ne\,{\sc ii}&3694.21  &{\bf +193.8 $\pm$ 3.2} &{\bf 34.4 $\pm$ 5.3} \\ 
            &         &                 &\\ 
{\neiii}    & 3342.42 & +199.5 $\pm$ 0.5 & 42.7 $\pm$ 0.9 \\ 
            & 3868.77 & +191.1 $\pm$ 0.7 & 41.5 $\pm$ 0.1 \\ 
            & 3967.46 & +191.9 $\pm$ 0.1 & 41.5 $\pm$ 0.2 \\ 
            &{\bf Adopted} &{\bf +191.3 $\pm$ 0.6} &{\bf  41.5 $\pm$ 0.1}\\
\hline 
\end{tabular}
\tablenotetext{a}{O\,{\sc ii}
	    $\lambda$ 4291.26,4291.86,4292.21,4292.98 are blended.}
\end{table}

\subsection{Comparison of elemental abundances from this work with others}
In Table \ref{ea2}, we compiled results for BoBn 1 from the past 30 years. 
Our estimated CEL elemental abundances except for Fe are in good 
agreement with previous works. 
The large discrepancy for Fe between us and Kniazev et al. (2008) could
be due to adopted electron temperatures for the Fe$^{2+}$ abundance
estimation. The depletion of Fe relative to the Sun is almost 
consistent with that of Ar (Table \ref{ea1}). Since Ar and Fe are not to be 
synthesized in low-mass stars, the abundances of these elements must be
roughly same, if the large amount of the dust does not 
co-exist in the nebula. In addition, the mid-IR spectra show no 
astronomical silicate or iron dust such as FeS features. Therefore, 
our estimated Fe abundance seems more reliable than Kniazev et al. (2008). 
Sneden et al. (2000) estimated [Fe/H] = --2.37$\pm$0.02 as the
metallicity of M 15 using $>$30 giants. The [Fe/H] abundance of BoBn 
1 corresponds to that of M15 within error, implying that the progenitor of
BoBn 1 might have formed at $\sim$10 Gyr ago.

BoBn 1 is a quite Ne-rich PN, and this Ne abundance is comparable 
with those of bulge and disk PNe; the averaged Ne abundance is 
8.09 (CELs) and 9.0 (ORLs) for bulge PNe (Wang \& 
Liu 2007) and 7.99 (CELs) and 9.06 for disk PNe (Tsamis et al. 
2004; Liu et al. 2004; Wesson et al. 2005). The Ne isotope, $^{20}$Ne 
is the most abundant, and it is not altered significantly by 
H- or He-burning (Karakas \& Lattanzio 2003). Therefore, 
the Ne overabundance would be due to an increase of the 
neon isotope, $^{22}$Ne. During helium burning, $^{14}$N captures 
two $\alpha$ particles, and $^{22}$Ne are produced. The Ne overabundance also implies 
that BoBn 1 might have experienced a very late thermal pulse (Otsuka et al. 2008a). 
If this is the case, Ne abundance might be an indirect evidence of He-rich intershell activity.
The Ne overabundance would be concerned with extra mixing during the RGB phase, which 
would increase N. The Ne and N enhancements of BoBn 1 might be also involved 
with the chemical environment where the progenitor formed. So far, 4 objects 
including BoBn 1 have been regarded as Sagittarius dwarf galaxy PNe, and 3 objects 
of them showed C, N, and Ne-rich ([C,N,Ne/O] $>$ 0; cf. Zjilstra et al. 2006).

The leading and trailing streams of the Sagittarius dwarf galaxy
trace several globular clusters. Terzan 8 is a member of the Sagittarius dwarf galaxy. 
Mottini et al. (2008) investigated the chemical abundances in three red giants 
in Terzan 8. The averaged [Fe/H], [O/Fe], and 
[Mg,Si,Ca,Ti/Fe] among
these objects were --2.37$\pm$0.04, +0.71$\pm$0.14, and +0.37$\pm$0.14,
respectively. Note that 
the metallicity of Terzan 8 is very close to BoBn 1. The amounts of O 
and other $\alpha$ elements such as Mg, Ne, S, and Ar 
do not significantly change during RGB phase. Therefore, 
the pattern of $\alpha$ elements derived from these RGB stars should be 
close to those in BoBn 1's progenitor. Based on 
this assumption, the initial [O/H] abundance of BoBn 1 is estimated 
to be --1.66$\pm$0.15, which is $\log$(O/H) + 12 = 7.03$\pm$0.15. [O/H]$\sim$+0.77 
would need to have been synthesized in BoBn 1 during helium burning. 
Assuming that the $^{22}$Ne($\alpha$,{\it n})$^{25}$Mg reaction is inefficient, the 
initial [Mg/H] abundance is estimated to be --2.07$\pm$0.15, which is comparable
to the observed [S,Ar/H] in BoBn 1. The observed S and Ar abundances in BoBn 1 could show 
the original abundances of the progenitor. [Ne/H] = +2.16$\pm$0.18 
could have been synthesized in BoBn 1 by $^{14}$N capturing two $\alpha$ particles.
Mottini et al. (2008) also estimated the light and heavy 
{\it s}-process enhancements; [Ba/Fe] = --0.09$\pm$0.17 and [Y/Fe] 
= --0.29$\pm$0.17. From these values, the initial {\it s}-process elemental 
abundances would be --2.46$\pm$0.18 for heavy {\it
s}-process elements such as Xe and Ba and --2.66$\pm$0.18 for 
light {\it s}-process elements such as Kr. 
The progenitor of BoBn 1 could have synthesized [{\it s}/H] $\sim$+2 
during He-burning phase.

\begin{table*}
\centering
\footnotesize
\caption{Elemental abundances derived by previous works and by this work. \label{ea2}}
\begin{tabular}{@{}llccccccccccccc@{}}
\hline\hline
&&\multicolumn{13}{c}{Abundances ($\log$(X/H) + 12)}\\
\cline{3-15}\\
Nebula&Ref.& He& C& N& O& F&Ne& S& Cl& Ar&Fe&Kr&Xe&Ba\\
\hline
BoBn 1 &(1)&$\cdots$ &{\bf 9.02} &{\bf 8.03} &{\bf 7.74}&{\bf 5.85}&{\bf 7.96} &{\bf 5.32}&{\bf 3.39}&{\bf 4.33}&{\bf 5.08}&{\bf $<$2.88}&{\bf $<$1.97}&{\bf 2.30}\\
&(2)&{\bf 11.07} &{\bf 9.16} &{\bf 8.49} &{\bf 8.23}&$\cdots$&{\bf 8.22}  &$\cdots$ &$\cdots$ &$\cdots$&$\cdots$&$\cdots$ &$\cdots$&$\cdots$\\
&(3)&$\cdots$ &{\bf 9.16} &{\bf 8.25} &{\bf 7.80}&{\bf 5.73}&{\bf 8.00} &{\bf 5.32}&{\bf 3.39}&{\bf 4.37}&{\bf 5.18}&{\bf $<$2.94}&{\bf $<$2.12}&{\bf 2.41}\\
&(4)&{\bf 11.07} &{\bf 9.16} &{\bf 8.48} &{\bf 8.22}&$\cdots$&{\bf 8.23}  &$\cdots$ &$\cdots$ &$\cdots$&$\cdots$&$\cdots$ &$\cdots$&$\cdots$\\
&(5)$^{a}$&{\bf 11.11} &{\bf 8.63} &{\bf 7.96} &{\bf 7.70}&{\bf 5.85}&{\bf 7.90} &{\bf 5.01}&{\bf 3.22}&{\bf 4.29}&{\bf 5.05}&$\cdots$&$\cdots$&$\cdots$\\
&(6)$^{a}$       &11.02 &9.20 &7.90 &7.70&$\cdots$ &7.80 &5.80&$\cdots$&4.70&$\cdots$&$\cdots$ &$\cdots$&$\cdots$\\
&(7)&11.00 &9.39$^{b}$    &8.08 &8.03&$\cdots$ &7.94 &$\cdots$&$\cdots$&$\cdots$&$\cdots$&$\cdots$ &$\cdots$&$\cdots$\\
&(8)$^{a}$    &11.05 &8.95 &8.00 &7.83&$\cdots$ &7.72&5.50&$\cdots$&4.50&$\cdots$&$\cdots$ &$\cdots$&$\cdots$\\
&(9)&10.95 &$\cdots$    &7.70 &7.89&$\cdots$ &8.10 &4.89&$\cdots$&4.19&$\cdots$&$\cdots$ &$\cdots$&$\cdots$\\
&(10)&$\cdots$ &$\cdots$&$\cdots$&$\cdots$&$\cdots$  &$\cdots$&$<$5.45&$\cdots$&$\cdots$&$\cdots$&$\cdots$ &$\cdots$&$\cdots$\\
&(11)&$\cdots$ &$\cdots$&$\cdots$&$\cdots$&$\cdots$
 &$\cdots$&5.67&$\cdots$&$\cdots$&$\cdots$&$\cdots$ &$\cdots$&$\cdots$\\
&(12)&10.98     &9.09 &8.34&7.90&$\cdots$ &8.00  &$\cdots$ &$\cdots$ &4.59&$\cdots$ &$\cdots$ &$\cdots$&$\cdots$\\
&(13)&$\cdots$ &$\cdots$&$\cdots$&$\cdots$&$\cdots$
 &$\cdots$&$\cdots$&$\cdots$&4.59&$\cdots$&$\cdots$ &$\cdots$&$\cdots$\\
&(14)$^{a}$ &10.98 &8.48 &6.94 &7.70&$\cdots$ &7.62 &6.48&$\cdots$&$\cdots$&$\cdots$&$\cdots$ &$\cdots$&$\cdots$\\
&(15)&11.06 &$\cdots$ &8.52
 &7.89&$\cdots$ &7.72&$\cdots$&$\cdots$&$\cdots$&$\cdots$&$\cdots$ &$\cdots$&$\cdots$\\
&(16)&10.99 &$\cdots$ &$\cdots$ &7.44&$\cdots$ &7.76
 &$\cdots$&$\cdots$&$\cdots$&$\cdots$&$\cdots$ &$\cdots$&$\cdots$\\
&(17)     &11.00 &9.20$^{b}$&7.64 &7.81&$\cdots$ &7.91 &5.16&3.14&4.57&5.72&$\cdots$ &$\cdots$&$\cdots$\\
\hline
K 648 &(18)&10.86&9.25&6.36&7.78&$\cdots$&6.87&5.10&$\cdots$&4.50&$\cdots$ &$\cdots$&$\cdots$&$\cdots$\\
\hline
\end{tabular}
\tablenotetext{a}{derived from photo-ionization modeling.} 
\tablenotetext{b}{derived from C\,{\sc ii} $\lambda$4267.}
\tablerefs{
(1) This work (CELs) with $t^{2}=0$. (2) This
	    work (ORLs) with $t^{2}=0$. (3) This work (CELs) with $t^{2}\neq0$. (4) This
	    work (ORLs) with $t^{2}\neq0$ . (5) This work (P-I model. see text).
	   (6) Pe\~{n}a et al. (1991). (7)
	    Kwitter \& Henry (1996). (8) Howard et al. (1997). (9) Henry
	    et al. (2004). (10) Garnett \& Lacy (1993). (11) Barker
	    (1983). (12) Torres-Peimbert et al. (1981). (13) Barker
	    (1980). (14) Aldrovandi (1980). (15) Hawley \& Miller
	    (1978). (16) Boeschaar \& Bond (1977). (17) Kniazev et
	   al. (2008). (18) Otsuka (2007,  $t^{2}$=0).
}
\label{sam.abun}
\end{table*}

\subsection{Comparison of observations and photo-ionization models}

\begin{table*}
\centering
\footnotesize
\caption{Comparison between the PI-model and the observations.}
\begin{tabular}{@{}lcccclcccc@{}}
\hline\hline
Ion/Band &$\lambda_{\rm lab}$&type&$I({\rm Cloudy})$&$I({\rm
 Obs.})$&Ion/Band &$\lambda_{\rm lab}$&type&$I({\rm Cloudy})$&$I({\rm Obs.})$\\
         &({\AA}/$\mu$m)&&[$I$({\hb})=100]&[$I$({\hb})=100]& &({\AA}/$\mu$m)&&[$I$({\hb})=100]&[$I$({\hb})=100]\\
\hline
He\,{\sc i} & 4471 & ORL & 5.40 & 4.82 &  O\,{\sc ii} & 4094 & ORL & 1.35(--2) & 1.43(--2) \\ 
He\,{\sc i} & 4922 & ORL & 1.33 & 1.32 &  O\,{\sc ii} & 4152 & ORL & 7.18(--3) & 1.68(--2) \\ 
He\,{\sc i} & 5876 & ORL & 1.65(+1) & 1.81(+1) &  O\,{\sc ii} & 4294 & ORL & 6.05(--3) & 1.76(--2) \\ 
He\,{\sc i} & 6678 & ORL & 3.64 & 4.02 &  O\,{\sc ii} & 4651 & ORL & 5.61(--2) & 4.68(--2) \\ 
He\,{\sc ii} & 4686 & ORL & 2.79(+1) & 2.48(+1) &  $[$Ne\,{\sc ii}$]$ &  12.81 & CEL & 9.33(--1) & 2.49 \\ 
$[$C\,{\sc i}$]$ & 8727 & CEL & 6.61(--2) & 7.93(--2) &  $[$Ne\,{\sc iii}$]$ & 3343 & CEL & 1.42 & 8.47(--1) \\ 
$[$C\,{\sc ii}$]$ & 2326 & CEL & 7.56(+1) & 3.51(+1) &  $[$Ne\,{\sc iii}$]$ & 3869 & CEL & 2.74(+2) & 2.17(+2) \\ 
 C\,{\sc iii}$]$ & 1907 & CEL & 9.26(+2) & 8.39(+2) &  $[$Ne\,{\sc iii}$]$ & 3968 & CEL & 8.25(+1) & 6.39(+1) \\ 
 C\,{\sc iii}$]$ & 1910 & CEL & 6.55(+2) & 6.03(+2) &  $[$Ne\,{\sc iii}$]$ &  15.55 & CEL & 1.30(+2) & 1.61(+2) \\ 
 C\,{\sc iv}$]$ & 1548 & CEL & 3.75(+2) & 1.05(+3) &  $[$Ne\,{\sc iii}$]$ &  36.01 & CEL & 1.11(+1) & 1.33(+1) \\ 
 C\,{\sc iv}$]$ & 1551 & CEL & 1.90(+2) & 5.19(+2) &  $[$Ne\,{\sc iv}$]$ & 2424 & CEL & 2.54(+1) & 1.71(+1) \\ 
 C\,{\sc ii} & 4267 & ORL & 3.34(--1) & 7.90(--1) &  $[$Ne\,{\sc iv}$]$ & 4725 & CEL & 1.32(--1) & 9.60(--2) \\ 
 C\,{\sc ii} & 6580 & ORL & 4.84(--2) & 3.75(--1) &  $[$Ne\,{\sc v}$]$ & 3346 & CEL & 1.65(--1) & 3.22(--1) \\ 
 C\,{\sc iii} & 4069 & ORL & 1.57(--1) & 2.49(--1) &  $[$Ne\,{\sc v}$]$ & 3426 & CEL & 4.51(--1) & 8.71(--1) \\ 
 C\,{\sc iii} & 4187 & ORL & 5.48(--2) & 3.46(--1) &  $[$S\,{\sc ii}$]$ & 4070 & CEL & 5.21(--2) & 3.95(--1) \\ 
 C\,{\sc iii} & 4649 & ORL & 1.48(--1) & 2.61(--1) &   $[$S\,{\sc ii}$]$ & 4078 & CEL & 1.68(--2) & 2.45(--2) \\ 
 C\,{\sc iii} & 8197 & ORL & 5.60(--2) & 4.39(--1) & $[$S\,{\sc ii}$]$ & 6716 & CEL & 1.01(--1) & 1.23(--1) \\ 
 C\,{\sc iv} & 4659 & ORL & 7.25(--3) & 1.19(--1) &  $[$S\,{\sc ii}$]$ & 6731 & CEL & 1.53(--1) & 2.16(--1) \\ 
 C\,{\sc iv} & 7726 & ORL & 3.30(--3) & 3.11(--2) &   $[$S\,{\sc iii}$]$ & 6312 & CEL & 9.57(--2) & 4.80(--2) \\ 
$[$N\,{\sc i}$]$ & 5198 & CEL & 3.28(--2) & 2.73(--1) & $[$S\,{\sc iii}$]$ & 9069 & CEL & 7.13(--1) & 3.78(--1) \\ 
$[$N\,{\sc i}$]$ & 5200 & CEL & 1.26(--2) & 1.91(--1) &  $[$S\,{\sc iii}$]$ &  18.67 & CEL & 6.38(--1) & 6.92(--1) \\ 
$[$N\,{\sc ii}$]$ & 5755 & CEL & 1.38 & 1.23 &   $[$S\,{\sc iv}$]$ &  10.51 & CEL & 1.95 & 1.92 \\ 
$[$N\,{\sc ii}$]$ & 6548 & CEL & 1.46(+1) & 1.56(+1) &  $[$Cl\,{\sc iii}$]$ & 5518 & CEL & 1.81(--2) & 1.80(--2) \\ 
$[$N\,{\sc ii}$]$ & 6584 & CEL & 4.31(+1) & 5.04(+1) &  $[$Cl\,{\sc iv}$]$ & 8047 & CEL & 2.08(--2) & 2.10(--2) \\ 
 N\,{\sc iii}$]$ & 1750 & CEL & 7.43(+1) & 4.81(+1) & $[$Ar\,{\sc iii}$]$ & 5192 & CEL & 5.55(--3) & 4.00(--3) \\ 
N\,{\sc iv}$]$ & 1486 & CEL & 2.91(+1) & 4.58(+1) &   $[$Ar\,{\sc iii}$]$ & 7135 & CEL & 3.18(--1) & 2.73(--1) \\ 
 N\,{\sc ii} & 4176 & ORL & 3.26(--3) & 1.36(--2) & $[$Ar\,{\sc iii}$]$ & 7751 & CEL & 7.67(--2) & 6.10(--2) \\ 
 N\,{\sc ii} & 4239 & ORL & 1.56(--2) & 2.42(--2) &  $[$Ar\,{\sc iv}$]$ & 4711 & CEL & 6.40(--2) & 9.40(--2) \\ 
 N\,{\sc ii} & 4435 & ORL & 9.53(--3) & 1.27(--2) &   $[$Ar\,{\sc iv}$]$ & 4740 & CEL & 6.82(--2) & 9.00(--2) \\ 
 N\,{\sc ii} & 5005 & ORL & 4.34(--2) & 1.79(--2) &  $[$Ar\,{\sc iv}$]$ & 7171 & CEL & 2.34(--3) & 6.53(--3) \\ 
 N\,{\sc ii} & 5679 & ORL & 2.52(--2) & 1.62(--2) & $[$Ar\,{\sc iv}$]$ & 7263 & CEL & 1.97(--3) & 4.58(--3) \\ 
 N\,{\sc iii} & 4110 & ORL & 1.01(--2) & 1.40(--2) &  $[$Fe\,{\sc iii}$]$ & 5271 & CEL & 3.89(--2) & 2.20(--2) \\ 
 N\,{\sc iii} & 4379 & ORL & 4.44(--2) & 5.97(--2) &  $[$Fe\,{\sc iii}$]$ & 4755 & CEL & 1.37(--2) & 2.10(--2) \\ 
 N\,{\sc iii} & 4641 & ORL & 4.91(--4) & 3.46(--2) &  $[$Fe\,{\sc iii}$]$ & 4881 & CEL & 2.13(--2) & 2.10(--2) \\ 
$[$O\,{\sc i}$]$ & 5577 & CEL & 1.48(--3) & 1.70(--2) &  2MASS $J$ & 1.24 &  & 4.98(+1) & 6.10(+1) \\ 
$[$O\,{\sc i}$]$ & 6300 & CEL & 4.02(--2) & 8.72(--1) & 2MASS $H$ & 1.66 &  & 3.09(+1) & 5.50(+1) \\ 
$[$O\,{\sc i}$]$ & 6363 & CEL & 1.28(--2) & 2.91(--1) &   2MASS $K$ & 2.16 &  & 2.26(+1) & 3.27(+1) \\ 
$[$O\,{\sc ii}$]$ & 3726 & CEL & 1.05(+1) & 1.09(+1) & PAH & 6.40 &  & 5.28(+1) & 7.32(+1) \\ 
$[$O\,{\sc ii}$]$ & 3729 & CEL & 5.96 & 6.61 &  PAH & 7.90 &  & 1.32(+2) & 9.47(+1) \\ 
$[$O\,{\sc ii}$]$ & 7323 & CEL & 1.03 & 1.02 &  IRS B & 20.00 &  & 6.08(+1) & 6.48(+1) \\ 
$[$O\,{\sc ii}$]$ & 7332 & CEL & 8.21(--1) & 7.81(--1) &   IRS C & 30.00 &  & 3.13(+1) & 2.97(+1) \\ 
$[$O\,{\sc iii}$]$ & 4363 & CEL & 6.86 & 5.57 & &  &  &  &  \\ 
$[$O\,{\sc iii}$]$ & 4931 & CEL & 5.03(--2) & 4.36(--2) &  &  &  &  &  \\ 
$[$O\,{\sc iii}$]$ & 4959 & CEL & 1.23(+2) & 1.22(+2) &  &  &  &  &  \\ 
$[$O\,{\sc iii}$]$ & 5007 & CEL & 3.70(+2) & 3.51(+2) &  &  &  &  &  \\ 
$[$O\,{\sc iv}$]$ &  25.88 & CEL & 1.21(+1) & 1.25(+1) &  &  &  &  &  \\ 
\hline
\end{tabular}
\label{cloudy}
\end{table*}

\begin{table*}
\centering
\footnotesize
\caption{The derived properties of the PN central star, ionized nebula, and dust by
 the P-I model.}
\begin{tabular}{@{}llllll@{}}
\hline\hline
\multicolumn{2}{c}{central star}&\multicolumn{2}{c}{nebula}&\multicolumn{2}{c}{dust}\\
\hline
$d$ (kpc)&24.8&composition&He:11.11,C:8.63,N:7.96,O:7.70,&grains&amC \& PAHs\\
$L_{\star}$ ($L_{\odot}$)    &1180     &(conti.)&F:5.85,Ne:7.90,S:5.01,Cl:3.22,&$M_{\rm dust}$ ($M_{\odot}$)&5.78(--6)\\ 
$T_{\star}$ (K)&125\,260    &(conti.)   &Ar:4.29,Fe:5.05,others:[X]=--2.13&$T_{\rm dust}$ (K)&80-180\\ 
$\log\,g$ (cm$^{2}$ s$^{-1}$)&6.5                            &$R_{\rm in}$/$R_{\rm out}$ ($''$)&0.43/0.60&$M_{\rm dust}$/$M_{\rm gas}$&5.84(--5)\\
composition&[X,Y]=0, [Z]=--1                           &$N_{\rm
 H}$($R_{\rm in}$) (cm$^{-3}$)&3890&$\dot{M}$$_{\rm dust}$ ($M_{\odot}$ yr$^{-1}$)&$\sim$3(--9)\\
$M_{\rm core}$ ($M_{\odot}$)&$\sim$0.62                           &geometry&spherical&\\
&&$\log$$F$({\hb}) &--12.44\\    
&&$M_{\rm gas}$/$M_{\rm atom}$ ($M_{\odot}$)&0.09/0.04\\                    
\hline
\end{tabular}
\label{cloudypara}
\end{table*}

\begin{figure*}[t]
\centering
\includegraphics[width=130mm]{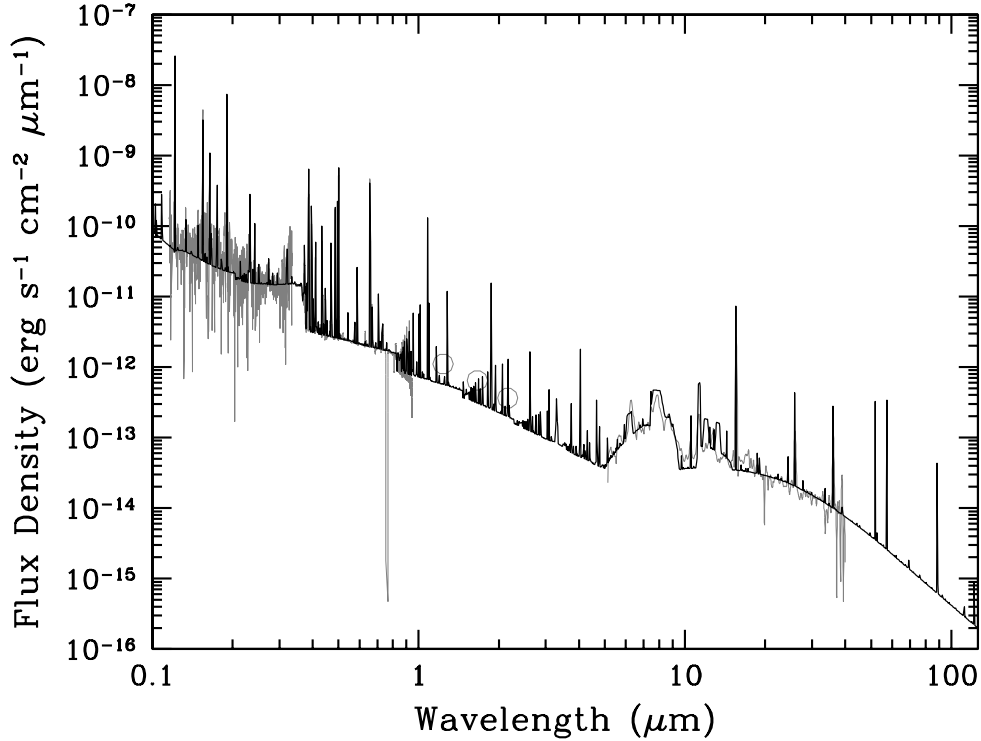}
\caption{The predicted SED from the P-I modeling (black line) and the
 observed spectrum of BoBn 1 from the $IUE$, Subaru/HDS, VLT/UVES, and
 $Spitzer$/IRS (gray lines). The circles are 2MASS data.}
\label{sed}
\end{figure*}

\begin{figure}
\centering
\includegraphics[width=80mm]{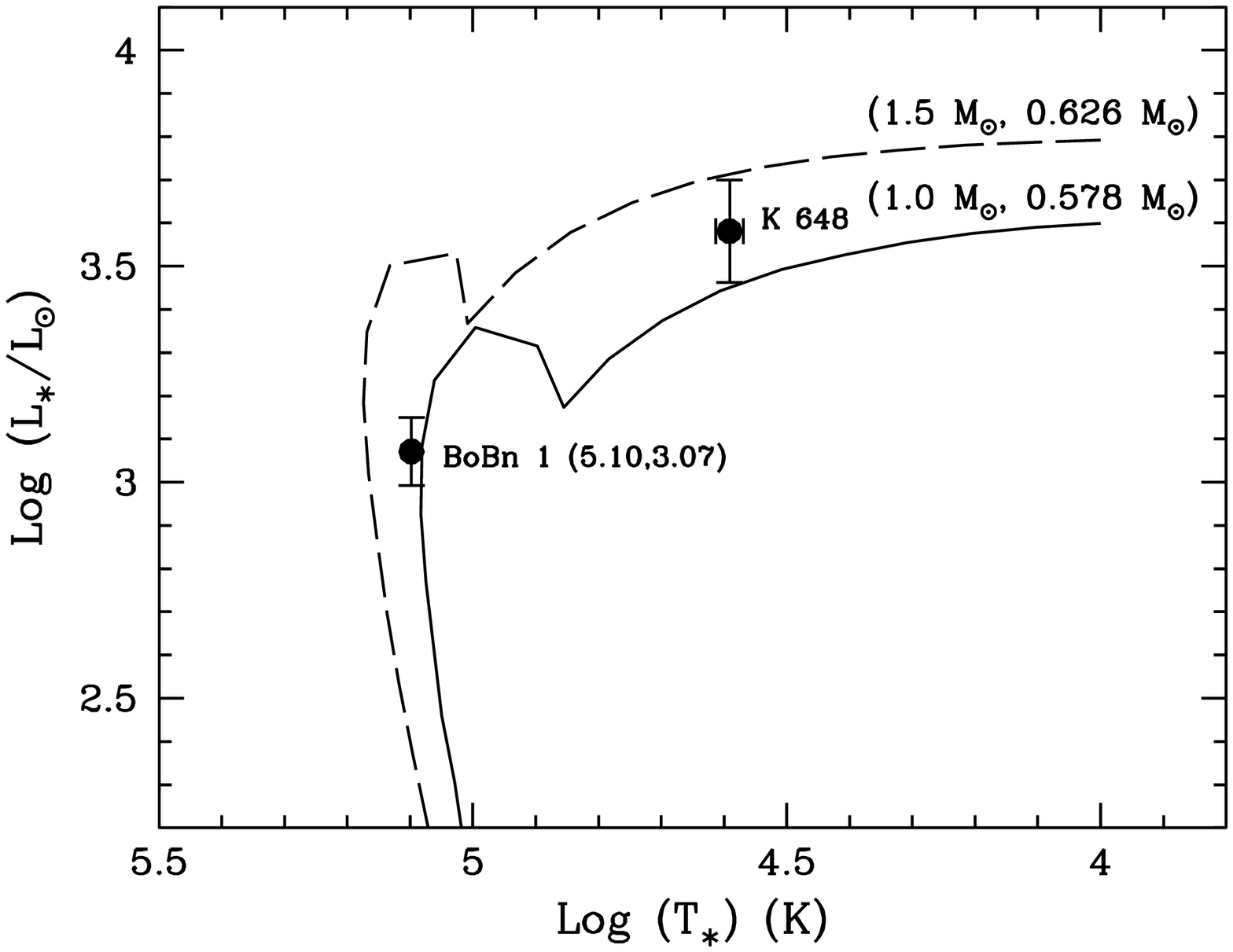}
\caption{The location of BoBn 1 and K 648 on the HR-diagram. The values of K
 648 are from Rauch et al. (2002). The solid and broken lines represent
 the post-AGB He-burning evolution tracks of Vassiliadis 
\& Wood (1994) for a metallicity of $\sim$
 0.5 $Z_{\odot}$.}
\label{hr}
\end{figure}

To investigate the properties of the ionized gas, dust, 
and the PN central star in a self-consistent 
way, we have constructed a theoretical photo-ionization (P-I) model which aims to 
match the observed flux of emission lines and the spectral energy 
distribution (SED) between UV and mid-IR wavelength, using 
{\sc Cloudy} c08.00 (Ferland 2004).

First, a rough value of the distance to BoBn 1 is necessary in 
fitting the observed fluxes. The distance to this object is estimated to be in the 
range between 16.5 and 29 kpc (see Table 1). Based on the assumption
that BoBn 1 is a member of the Sagittarius dwarf galaxy, we fix the distance to be 
24.8 kpc (Kunder \& Chaboyer 2009). P-I model construction needs information about the incident SED from the
central star and the elemental abundances, geometry, density
distribution, and size of the nebula. Once one gets a proper prediction 
of line intensities from a proper modeling procedure, the central
stellar physical properties employed in the P-I model can give 
us a hint of the nebular evolutionary history 
or that of its progenitor star. Especially, the central star temperature 
$T_{\star}$ and the SED of the PN central star are an important physical
parameter in constructing the correct P-I model.

We estimated $T_{\star}$ of 125\,930 $\pm$ 6100 K using the energy balance methods 
proposed by Gurzadyan (1997). Being guided by this $T_{\star}$, we used 
theoretical model atmosphere for a series of values of $T_{\rm eff}$ to
supply the SED from the central star. We used Thomas Rauch's non-LTE 
theoretical model atmospheres\footnote{See
http://astro.uni-tuebingen.de/$\sim$rauch/} for halo stars ($[X, Y]$ = 0 and $[Z]$ =
--1) with the surface gravity $\log$\,{\it g} = 6.0, 6.125, 6.25, 6.375,
6.5, and 6.625. 
We varied $T_{\rm eff}$ and the luminosity $L_{\star}$ to match the
observations.

For the elemental abundances X/H, we used the values from the case of
$t^{2}$ = 0 as a starting point. For the C, N, O, and Ne abundances, we used
the CELs abundances. We assumed no
high-density cold clumps. Using the HDS slit-viewer image (Fig.
1), we measured the radius of the outer nebular shell
$R_{\rm out}$ of $\sim$0.6$''$ (=0.072 pc) and fixed this value. 
We assumed the hydrogen density $N_{\rm H}$ to be a $R^{-2}$ smooth
distribution, i.e., $N_{\rm H}$ = 
$N_{\rm H}$($R_{\rm in}$) $\times$ ($R_{\rm in}/R$)$^{2}$. In the
models, within a small range we varied X/H, $N_{\rm H}$($R_{\rm in}$), and $R_{\rm
in}$ to match the observed line fluxes from UV to
mid-infrared wavelength, including 2MASS $JHK$ bands, and our 
mid-infrared  bands, i.e., between 17 and 23 $\mu$m (IRS B) and between 27 and 33
$\mu$m (IRS C).

Since the IRS spectra show that the dust grains co-exist in the nebula of
BoBn 1, we need information about the dust composition. 
Here we considered amorphous carbon and PAH grains. The optical constants were 
taken from Rouleau \& Martin (1991) for amorphous carbon and from
Desert et al. (1990), Schutte et al. (1993), Geballe (1989), and Bregman et al. (1989) 
for PAHs. 
The observed emission-line and base line continuum 
fluxes between 5.9 and 6.9 $\mu$m (PAH 6.4 $\mu$m) and between 
7.4 and 8.4 $\mu$m (PAH 7.9 $\mu$m) were used to determine the PAH abundance.
We assumed that the gas and dust co-exist in the same sized
ionized nebula. We adopted a standard MRN $a^{-3.5}$ 
distribution (Mathis, Rumpl \& Nordsieck 1977) with $a_{\rm min}$=0.001 
$\mu$m and $a_{\rm max}$=0.25 $\mu$m for amorphous carbon. For PAHs
we adopted an $a^{-4}$ size distribution with $a_{\rm min}$=0.00043 $\mu$m and $a_{\rm
max}$=0.0011 $\mu$m.  We adopted an $R^{-2}$ smooth dust density
distribution. High-density clumped dust grains were not considered.

In Table \ref{cloudy}, we compare the 
predicted with observed relative fluxes where $I$(H$\beta$) = 100. 
For most CELs and He lines and the wide band fluxes, as well, 
the agreement between the P-I model and 
the observation is within 30$\%$. The poor fit of {\Ni}, {\oi}, and {\sii} would be 
due to the assumed density profile. The relation between $n_{\epsilon}$ and I.P. 
(Fig. \ref{ip_te_ne} lower panel) shows that the electron density jumps 
from $\sim$2000 to $\sim$6000 cm$^{-3}$ around {\sii} emitting region. In our model, 
such a density jump was not considered. For 
important lines such as He\,{\sc i, ii}, and C\,{\sc iii}$]$, {\nii}, {\oii},
{\oiii}, {\neiii}, fairly good
agreements  in calculating ionic abundances are achieved. However, most of the C and O ORLs fit
to the observations poorly. In most cases, the P-I models underestimate
their line fluxes. As we discussed above, the O ORLs, and likely the 
Ne ORLs too, might be emitted from cold and metal-enhanced clumps.

In Table \ref{cloudypara}, we list the derived parameters of the PN central star,
ionized nebula, and dust, and in Fig. \ref{sed} we present 
the predicted SED (black line). The predicted SED 
matches the UV to mid-infrared region well. Through the P-I modeling, we found 
the PN central star's parameters: $T_{\rm eff}$ = 125\,260 $\pm$ 200 K, $L_{\star}$
= 1180 $\pm$ 240 $L_{\odot}$, $\log$\,{\it g} = 6.5, and a 
core mass of $\sim$0.62 $M_{\odot}$. The estimated 0.62 
core mass is comparable to that of 
K 648 (0.62 $M_{\odot}$, Bianchi et al. 2001; 0.57 $M_{\odot}$ Rauch et al. 2002) 
and the high-excitation halo PN NGC 4361 
(0.59 $M_{\odot}$: Traulsen et al. 2005). The ionized mass of BoBn 1 is 0.09 $M_{\odot}$,
which is comparable to K 648 (0.07 $M_{\odot}$; Bianchi et al. 1995 ).
In Fig. \ref{hr} we plot 
the locations of BoBn 1 and K 648 (Rauch et al. 2002) 
and the post-AGB He-burning evolutionary tracks for 
LMC metallicity ($Z$$\sim$0.5$Z_{\odot}$) by Vassiliadis 
\& Wood (1994). These evolutionary tracks would suggest the possibility that 
the progenitors of BoBn 1 and K 648 were single 1-1.5 $M_{\odot}$ stars
which would end their lives as white dwarfs with a core mass of 
$\sim$0.6 $M_{\odot}$. Alternatively, these halo PNe 
might have evolved from binaries composed of a 0.8 $M_{\odot}$ (= a typical halo star mass) 
secondary and a more massive primary, and gained mass ($\sim$0.1 $M_{\odot}$) from the primary through mass 
transfer or coalescence.

For K 648, Alves et al. (2000) support a binary evolution scenario. 
From F enhancement and similarity to carbon-enhanced metal poor stars (CEMP), 
Otsuka et al. (2008a) argued that BoBn 1 might have evolved from a binary composed of 
a 0.8 $M_{\odot}$ secondary and a $>$ 2 $M_{\odot}$ primary star. 
We should consider two possibilities: these halo PNe have evolved from single stars or from 
binaries.

The P-I model indicated elemental abundances except for C and S are in excellent
agreement with those estimated by the empirical method using the ICFs. 
The discrepancy for C is due to the underprediction of C\,{\sc
iv}  lines. The model predicted C$^{+}$ = 2.0(--5) and C$^{2+}$ = 
3.4(--4), which are comparable to the observations. However, 
it predicted lower line fluxes of C\,{\sc iv} $\lambda\lambda$1548/51 
than the observations, and accordingly an underestimated C$^{3+}$ as 6.95(--5). 
These might suggest that the origin of C\,{\sc iv} lines is not 
the ionized nebula but the stellar wind zone. The discrepancy for S could 
be due to the low ionic the S$^{+}$ abundance, which depends strongly 
on the assumed radial density profile.

For BoBn 1, we have for the first time estimated a dust 
mass of 5.78(--6) $M_{\odot}$ and the temperature of 80--180 K. 
The dust in BoBn 1 is carbon rich. The dust composition suggests 
that BoBn 1 had experienced the TDU 
during the latest thermal pulsing AGB phase (TP-AGB). Since 
the TDU efficiently takes place 
in $>$1 $M_{\odot}$ stars, the progenitor of BoBn 1 might 
be 1--3.5 $M_{\odot}$ from the aspect of elemental
abundances and dust composition.

The dust-to-gas mass ratio $\psi$ of 5.84(--5) is much lower than the typical value in PNe
($<$$\sim$10$^{-3}$; Pottash 1984). For AGB stars, Lagadec et al. (2009) argued that 
$\psi$ scales linearly with the metallicity. They assume the ratio
is 

\begin{equation}
\psi = \psi_{\odot}\times10^{[{\rm Fe/H}]}
\end{equation}

\noindent
where $\psi_{\odot}$ is 0.005. When we adopt $[$Fe/H$]$ = --2.22 for
BoBn 1, we obtain $\psi$ = 3.01(--5).

Assuming that the inner and outer shells were expanding with
$\sim$10 {\kms} and most of the dust was formed during the ending period of the 
thermal pulse AGB phase, we estimated the dust mass-loss rate $\dot{M}$$_{\rm dust}$ of 
$\sim$3(--9) $M_{\odot}$ yr$^{-1}$. Lagadec et al. (2009) estimated
$\dot{M}$$_{\rm dust}$ in metal-poor ([Fe/H]$\sim$--1) carbon stars IRAS16339-0317 and
18120+4530 in the Galactic halo and IRAS12560+1656 in the Sgr
stream. They estimated 4--18(--9) $M_{\odot}$ yr$^{-1}$. For
IRAS12560+1656, Groenewegen et al. (1997) estimated $\dot{M}$$_{\rm
dust}$ of 1.9(--9) $M_{\odot}$ yr$^{-1}$. These $\dot{M}$$_{\rm dust}$ values
are comparable to that of BoBn 1.

\subsection{Comparison of observations and theoretical models}

Through P-I modeling, we found two possibilities: BoBn 1 might have
evolved from (a) a 1-1.5 $M_{\odot}$ single star or (b) a binary
composing of $\sim$0.8 $M_{\odot}$ secondary and a more massive primary. In this
section, we explore these possibilities by comparing the observed and
predicted elemental abundances employing theoretical
nucleosynthesis models for low- to intermediate mass stars.

In Table \ref{theo}, we present observed elemental abundances 
and predicted values from the theoretical models of Karakas \& Lugaro (2009) 
for 1.0, 1.5, and 2.0 $M_{\odot}$ stars with 
$Z$=10$^{-4}$ ([Fe/H]$\sim$--2.3). The abundances from the
models are the values at the end of the AGB
phase. For these models, Karakas \& Lugaro (2009) chose an initial 
$\alpha$-enhanced abundance pattern, i.e., [$\alpha$/Fe]=+0.4. 
For {\it s}-process elements, they chose scaled solar abundances, 
i.e., [X/Fe]=0. These [$\alpha$/Fe] and [X/Fe] ratios are consistent
with the values for RGB stars in Terzan 8, therefore the 
assumption of initial abundances seems very
reasonable for BoBn 1. The accuracy of the predicted abundances by the models 
is within 0.3 dex. For the observed abundances, 
we adopted the $t^{2}$=0 CEL abundances except for C.
The adopted C abundance was from ORLs.

On the possibility (a) that BoBn 1 has evolved from a single star 
and has survived in the Galactic halo, 
the abundances of BoBn 1 except for N can be properly explained by the 
1.5 $M_{\odot}$ star model including a partial 
mixing zone of 0.004 $M_{\odot}$, which produces a $^{13}$C 
pocket during the interpulse period and releases free neutrons 
through $^{13}$C($\alpha$,{\it n})$^{16}$O. This model assumes that the
stars end as white dwarfs with a core mass of $\sim$0.7 $M_{\odot}$, 
which is comparable to our estimated core mass.

The $^{13}$C($\alpha$,{\it n})$^{16}$O reaction proceeds in the 
upper surface layer of the He-burning shell. The F and {\it s}-process elements 
are synthesized by capturing these neutrons in the He-intershell. 
The observed F, probably Kr and Xe abundances are systematically larger ($\sim$+0.5 dex) 
than 1.5 $M_{\odot}$ star + partial mixing model. This 
suggests that hydrogen mixing mass is likely to be $>$ 0.004 $M_{\odot}$ or that 
BoBn1 had experienced helium-flash driven deep mixing (He-FDDM, Fujimoto et
al. 2000; Suda et al. 2004) and obtained the extra neutrons. 
This process can occur in stars with
[Fe/H] $<$--2.5 at the bottom of the He-burning shell because the entropy 
barrier between the H- and He-shell becomes low. The lower limit to [Fe/H] for BoBn1
is $-$2.46. This process would also produce $^{14}$N through the 
$^{13}$C($p$,$\gamma$)$^{14}$N reaction, by mixing protons into the 
He-burning shell.

We note that BoBn 1 is similar to K 648, for the latter 
is also known as an extremely metal poor, C- and N-rich halo PN.
On the assumption that K 648 has been evolved from a single
star, the abundances of K 648 except for Ne can be explained by a 1.5 
$M_{\odot}$ model. K 648 would not have experienced He-FDDM for
abnormally increasing N.

However, can halo single stars with an initial mass of $\sim$1.5 $M_{\odot}$ and
$Z$ = 10$^{-4}$ survive up to now? Such stars would end as white dwarfs in 
$\sim$2-3 Gyr. Their lifetime is much shorter than the age of
Terzan 8; Forbes et al. (2004) estimated the age of Terzan 8 to be 13 $\pm$
1.5 Gyr from an age-metallicity relation for the Sagittarius dwarf globular cluster. 
If the progenitor of BoBn 1 was a $\sim$0.8 $M_{\odot}$ single star, then
it can have survived up to now, however it cannot evolve into a visible PN
and cannot become extremely C-rich. To circumvent the evolutionary time scale
problem, we should consider the other evolutionary scenario for BoBn 1.

\begin{table}
\centering
\caption{Comparison of observations and the theoretical models for single and binary stars 
with $Z$=10$^{-4}$.}
\begin{tabular}{lcccccccc}
\hline\hline
&\multicolumn{8}{c}{Abundances ($\log$(X/H) + 12)}\\
\cline{2-9}\\
Model & C &  N & O & F & Ne & Fe & Kr & Xe \\ 
\hline
1.00 $M_{\odot}$ & 8.13 & 6.52 & 6.98 & 3.79 & 6.45 & 5.18 & 2.00 & 0.97 \\ 
1.50 $M_{\odot}$ & 9.21 & 6.81 & 7.76 & 5.28 & 8.11 & 5.21 & 2.43 & 1.46 \\ 
+ partial mixing & 9.17 & 6.78 & 7.85 & 5.30 & 8.31 & 5.21 & 2.47 & 1.50 \\
2.00 $M_{\odot}$ & 9.55 & 6.87 & 7.92 & 5.93 & 8.66 & 5.24 & 2.37 & 1.53 \\ 
\hline
0.75 $M_{\odot}$ + 1.50 $M_{\odot}$ & 9.51 & 7.09 & 7.67 &5.04  &7.47  & 5.22 &2.11  &1.30  \\ 
0.75 $M_{\odot}$ + 1.80 $M_{\odot}$ & 9.31 & 6.72 & 7.50 &4.74  &7.17  & 5.20 &1.88  &1.07 \\ 
0.75 $M_{\odot}$ + 2.10 $M_{\odot}$ & 9.23 & 6.58 & 7.41 &4.69  &7.11  & 5.20 &1.73  &0.91  \\ 
\hline
BoBn 1 ($t^{2}$=0) & 9.16 & 8.03 & 7.74 & 5.85 & 7.96 & 5.08 & $<$2.88 & $<$1.97 \\ 
K 648 ($t^{2}$=0) & 9.25 & 6.36 & 7.78 &$\dots$ & 6.87 &$\dots$ &$\dots$ & $\dots$ \\ 
\hline
\end{tabular}
\label{theo}
\end{table}

\subsection{The Origin of BoBn 1}
As mentioned earlier, BoBn 1 is likely to be a binary origin 
because the [C,N,F/Fe] abundances of BoBn 1 are comparable to
those of the carbon-enhanced metal poor star (CEMP) HE 1305+0132
(Schuler et al. 2007) and other CEMP stars.

\begin{figure}
\centering
\includegraphics[width=80mm]{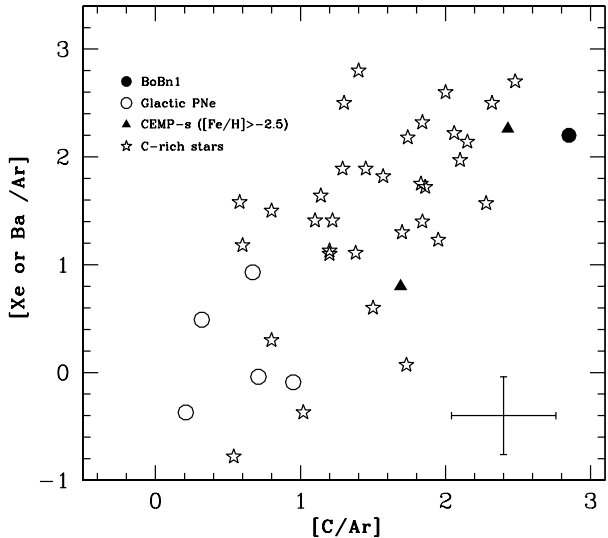}
\caption{[Xe or Ba/Ar]-[C/Ar] diagram. The [Xe/Ar] value of BoBn 1 is upper
 limit. 
For PNe, we adopt Ar as a metallicity reference. For CEMP-{\it s} and 
C-rich AGB stars, we use Fe as a metallicity reference.}
\label{cxe}
\end{figure}

Most CEMP stars  
show large enhancements of C and N 
abundances. Some evolutionary models for CEMP stars have demonstrated that the C 
and N overabundances would be reproduced by binary interactions. 
Schuler et al. (2007) concluded that HE 1305+0132 might have experienced 
mass transfer and that {\it s}-process elements should be enhanced. 
Lugaro et al. (2008) concluded that HE 1305+0132 consisted of 
$\sim$2 M$_{\odot}$ (primary) and $\sim$0.8 M$_{\odot}$ (secondary)
stars with {\it Z} = 10$^{-4}$ and that the enhanced C and F could be explained by 
binary mass transfer from the primary star via Roche lobe overflow and/or wind
accretion. In Fig. \ref{cxe}, we present the diagram of
[Xe,Ba/Ar]-[C/Ar]. The [Xe/Ar] for BoBn 1 is an upper limit. 
The data for Galactic PNe except BoBn 1 are taken from Sharpee et al. (2007). The data for 
{\it s}-process elements 
enhanced CEMP (CEMP-{\it s}) with [Fe/H] $>$ --2.5 and C-rich AGB are from the SAGA
data base (Suda et al. 2008). For PNe, we use Xe as a heavy 
{\it s}-process element and adopt Ar as a metallicity reference. For CEMP-{\it s} 
and C-rich AGB stars, we use Ba and Fe as a metallicity reference. 
The diagram indicates that C and {\it
s}-process elements are certainly synthesized in the same layer and brought up to the
stellar surface by the TDU. We note that the enhancement of heavy {\it
s}-process elements in BoBn1 is comparable to CEMP-{\it s} stars, in particular
CS22948-027 (Aoki et al. 2007;[Fe/H] = --2.21, [Ba/Fe] = +2.31, 
[C/Fe] = +2.12, [N/Fe] = +2.48). The chemical similarities between BoBn 1 and CEMP-{\it s} 
stars suggest that this PN shares a similar origin and evolutionary 
history.

BoBn 1 is similar 
to K 648 in elemental abundances and nebular shape (see Fig.
\ref{bb1.image} and Table \ref{sam.abun}). K 648 has been for a
long time suspected to have experienced binary evolution. Rauch et
al. (2002) and Bianchi et al. (2001) analyzed the spectrum of the 
central star and estimated a core mass $\sim$0.6 $M_{\odot}$. 
The mass $\sim$0.6 $M_{\odot}$ corresponds to the initial mass of 1-1.5 
$M_{\odot}$ from the HR-diagram as shown in Fig. \ref{hr}. 
The initial mass of 1-1.5 $M_{\odot}$ suggests 
that K 648 might have evolved from a binary and accreted a part of the 
ejected mass by a massive primary or coalescence 
during the evolution. Alves et al. (2000) argued that 
K 648 has experienced mass augmentation in a close-binary merger 
and evolved as a higher mass star to become a PN. Such a 
high mass star would be a blue straggler. Ferraro et al. (2009) 
observed stars in the globular cluster M 30 using the $HST$/WFPC2 
and concluded that blue stragglers are results of coalescence 
or binary mass-transfer.

In view of the internal kinematics 
and chemical abundances, the progenitor of 
K 648 seems to be a binary. K 648 has a bipolar outflow (Tajitsu \& Otsuka 
2006) and bipolar nebula (Alves et al. 2000). 
The statistical study of PN morphology shows that bipolar PNe have evolved from 
massive stars with initial mass $\ga$ 2.4 $M_{\odot}$ or binaries. 
Otsuka et al. (2008b) showed that the [C/Fe] and [N/Fe] abundances of 
K 648 are compatible with CEMP stars.  So far, there have been 
no reports on the detection for any binary signatures in both objects. 
The contradiction to the evolutionary time scale of this object can be avoided 
if BoBn 1 has indeed evolved from a binary. 
Similar to K 648, BoBn 1 could have evolved from a binary 
and undergone coalescence to become a PN.

We explore the possibility of binary evolution of BoBn 1 using 
binary nucleosynthesis models by Izzard et al. (2004, 2009). We assume a binary
system composed initially of a 0.75 $M_{\odot}$ secondary and a 1.5/1.8/2.1
$M_{\odot}$ primary with $Z$=10$^{-4}$, separation = 219/340/468
$R_{\odot}$, respectively. We set eccentricity $e$ = 0, common envelope efficiency $\alpha$ = 0.5 and
structure parameter $\lambda_{\rm CE}$ = 0.5. He-FDDM is not
considered. We set a $^{13}$C pocket mass of 7.4$\times$10$^{-4}$ 
$M_{\odot}$. The $^{13}$C pocket contains 4.1$\times$10$^{-6}$ 
$M_{\odot}$ $^{13}$C and 1.3$\times$10$^{-7}$ $M_{\odot}$ $^{14}$N. We set 
the $^{13}$C pocket efficiency = 2 and adopt wind mass-loss rates from Reimers formula on
the RGB and by Vassiliadis \& Wood (1993) on the TP AGB.

When we choose 
these initial primary and secondary masses and $\alpha$, the binary system will 
experience Roche lobe overflow; and it will merge into a 1.2-1.4 $M_{\odot}$
single star at the TP-AGB phase and end its life as a white 
dwarf with a core mass of 0.62-0.68 
$M_{\odot}$, 10.4-12.5 Gyr after the progenitor was born. We
present the results in Table \ref{theo}. The binary models might seem 
to explain the elemental abundances of BoBn 1 and 
K 648. Concerning K 648, the 0.75 $M_{\odot}$ + 1.5 $M_{\odot}$/1.8 $M_{\odot}$ 
models fairly well match the prediction to the observed abundances. Among the models, 
the 0.75 $M_{\odot}$ + 1.5 $M_{\odot}$ model seems to be the best fit
to BoBn 1 for the moment because this model can explain not 
only the abundance patterns but also the observed core mass (the predicted core mass
$\sim$0.64 $M_{\odot}$). 
If the progenitor has experienced extra mixing in the RGB phase and increased N, 
the N overabundance can be accommodated by this model. The issues with the 
evolutionary time scale and C and N enhancements might be resolved 
simultaneously if BoBn 1 has evolved from such a binary. At the present, 
we conclude consider that binary evolution scenario is more plausible for BoBn 1.

To discuss further the evolution of BoBn 1 and K 648, we need 
to increase detection cases of {\it s}-process elements and 
to investigate the isotope ratios of $^{12}$C/$^{13}$C, 
$^{14}$N/$^{15}$N, $^{16}$O/$^{17}$O, and $^{16}$O/$^{18}$O, which would be 
useful to investigate nucleosynthesis in the progenitors. It would 
be also necessary to completely trace mass-loss history to improve mass-loss rate. 
So far, mass-loss history of evolved stars has been revealing by 
investigating spatial distribution of dust grains and molecular gas using 
far-infrared to millimeter wavelength data. The Atacama Large Millimeter Array (ALMA) 
and the thirty meter telescope (TMT) could open new windows to study the 
evolution of metal-poor stars such as halo PNe.

\section{Conclusion}
We have performed a comprehensive chemical abundance analysis of BoBn 1
 using $IUE$ archive data, Subaru/HDS spectra, VLT/UVES archive data,
 and $Spitzer$/IRS spectra. We calculated the ionic and elemental abundances of
 13 elements using ORLs and CELs. The estimations of C, N, O, and Ne 
abundances from the ORLs and Kr, Xe, and Ba from the CELs are done 
for the first time. The C, N, O,
 and Ne ORL abundances are systematically larger than those from CELs. 
We investigated the cause of the abundance discrepancies.
The discrepancies except for O could be explained by a temperature fluctuation model, and
 that of O might be due to hydrogen deficient cold components.

In the optical high-dispersion spectra, we detected emission-lines of 
fluorine and {\it s}-process elements such as rubidium, krypton, xenon, and barium. The values 
of [F/H], [Kr/H], and [Xe/H] suggest that BoBn 1 is the most F-rich
 among F detected PNe and is a heavy {\it s}-process element rich PN. The
 enhancement of C, N, and heavy {\it s}-process is comparable to
 CEMP-{\it s} stars with [Fe/H] $>$--2.5. This suggests that BoBn 1 
shares a similar origin and evolutionary history with CEMP-{\it s} stars.

We built photo-ionization model using non-LTE theoretical stellar
atmosphere models to check consistency between elemental
abundances derived by empirical methods and from the model and 
to investigate the properties of the central star,
 ionized nebula, and dust in a self-consistent way to fit the IR wavelength region. 
In the modeling, we considered the presence
 of dust. We compared the observed elemental abundances with theoretical nucleosynthesis
 model predictions for single stars and binaries with $Z$ = 10$^{-4}$. 
The observed elemental abundances except for N could be explained either by a 1.5 $M_{\odot}$ single star model 
or a binary model composed of 0.75 $M_{\odot}$ + 1.5 $M_{\odot}$ stars. 
Through the modeling, we estimated the 
luminosity and effective temperature and surface gravity of the central
star and the total mass of ionized gas and dust and even for the SED of
BoBn 1. Using theoretical
evolutionary tracks for post-AGB stars, we found that the progenitor of the central star 
was perhaps a 1-1.5 $M_{\odot}$ star and evolved into a system of a white dwarf with a core mass of
 $\sim$0.62 $M_{\odot}$ and an $\sim$0.09 $M_{\odot}$ ionized
 nebula. We estimated a total mass of 5.8$\times$10$^{-6}$ $M_{\odot}$ 
in the nebula, which composes of amorphous carbon and PAHs. The presence of carbon dust
 indicates that BoBn 1 has experienced the third dredge-up during the
 thermal pulse AGB phase.

The progenitor
 might have been initially quite N-rich. The He-flash driven deep mixing might be responsible for 
the overabundance of N. From careful  consideration of observational results and a comparison between BoBn 1
 and K 648 in M 15, we propose that the progenitor was a 0.75 $M_{\odot}$ + 1.5 $M_{\odot}$
 binary with, e.g. an initial separation of 219 $R_{\odot}$ and had experienced 
coalescence during its evolution to become a C- and N-rich PN. The similar evolutionary 
scenario would be also applicable to K 648.

\section*{Acknowledgments}
The authors express their thanks to Amanda Karakas, Mike Barlow, and 
Roger Wesson for fruitful discussion and a critical reading of the manuscript. 
They wish to thank the anonymous referee for valuable comments.
S.H. acknowledges the support by Basic Science Research Program through 
the National Research Foundation of Korea funded by the Ministry of Education, 
Science and Technology (NRF-2010-0011454). 
This work is mainly based on data 
collected at the Subaru Telescope, which is operated by 
the National Astronomical Observatory of Japan (NAOJ).
This work is in part based on ESO archive data obtained by ESO
Telescopes at the Paranal Observatory. 
This work is in part based on archival data obtained 
with the Spitzer Space Telescope, which is operated by the 
Jet Propulsion Laboratory, California Institute of 
Technology under a contract with NASA. Support for this 
work was provided by an award issued by JPL/Caltech. This work in in
part based on IUE archive data downloaded from the MAST.

\appendix
\section{Observed line list (on-line table). \label{line_list}}

\begin{table}
\centering
\caption{Observed and reddening corrected line ratios [$I(\rm H\beta)$=100] and identifications of BoBn 1. The few lines from the Table A1.}
\begin{tabular}{@{}ccccccccl@{}}
\hline\hline
{$\lambda_{\rm obs}$}& 
{Ion} & 
{$\lambda_{\rm lab}$} &
{Comp.}  & 
{$f(\lambda)$} &  
{$I(\lambda)$} &
{$\delta$$I(\lambda)$} &
{Source}&
{Note}\\
{({\AA})}& 
{}& 
{({\AA})} &  
{}& 
{}&   
{}& 
{}&
{}&
{}\\
\hline
3301.48 & O~{\sc iii} & 3299.39 & 1 & 0.423 & 0.632 & 0.031 & UVES1 &  \\ 
3314.43 & O~{\sc iii} & 3312.33 & 1 & 0.418 & 1.644 & 0.038 & UVES1 &  \\ 
3336.76 & Ne~{\sc ii} & 3334.87 & 1 & 0.411 & 0.104 & 0.019 & UVES1 &  \\ 
$\cdots$\\
4717.13 & [Ne~{\sc iv}] & 4714.25 & 1 & 0.041 & 0.055 & 0.003 & HDS &  \\ 
4718.61 & [Ne~{\sc iv}] & 4715.80 & 1 & 0.041 & 0.019 & 0.002 & HDS &  \\ 
4727.07 & [Ne~{\sc iv}] & 4724.15 & 1 & 0.039 & 0.051 & 0.002 & HDS &  \\ 
4728.51 & [Ne~{\sc iv}] & 4725.62 & 1 & 0.038 & 0.045 & 0.003 & HDS &  \\ 
4743.20 & [Ar~{\sc iv}] & 4740.17 & 1 & 0.034 & 0.090 & 0.004 & HDS &  \\ 
4757.92 & [Fe~{\sc iii}] & 4754.69 & 1 & 0.030 & 0.021 & 0.005 & HDS &  \\ 
4788.04 & C~{\sc iv}? & 4785.90 & 1 & 0.021 & 0.044 & 0.005 & HDS &  \\ 
4790.11 & N~{\sc iv} & 4786.92 & 1 & 0.020 & 0.013 & 0.005 & UVES1 &  \\ 
4792.87 & [F~{\sc ii}] & 4789.45 & 1 & 0.020 & 0.056 & 0.005 & HDS &  \\ 
4801.07 & [Fe~{\sc ii}] & 4798.27 & 1 & 0.017 & 0.008 & 0.005 & HDS &  \\ 
4805.62 & Ne~{\sc ii} & 4802.58 & 1 & 0.016 & 0.034 & 0.006 & HDS &  \\ 
4855.14 & He~{\sc ii} & 4852.00 & 1.00 & 0.003 & 0.007 & 0.004 & HDS & Raman line? \\ 
4856.93 & [Fe~{\sc iii}] & 4853.70 & 1 & 0.002 & 0.004 & 0.003 & HDS &  \\ 
4862.36 & He~{\sc ii} & 4859.32 & 1 & 0.001 & 0.869 & 0.130 & HDS &  \\ 
4864.46 & H4 & 4861.33 & 1 & 0.000 & 100.000 & 0.202 & HDS &  \\ 
4872.75 & [F~{\sc ii}] & 4868.99 & 1 & --0.002 & 0.013 & 0.003 & HDS &  \\ 
4879.17 & [Co~{\sc vi}] & 4876.26 & 1 & --0.004 & 0.009 & 0.004 & HDS &  \\ 
4884.34 & [Fe~{\sc iii}] & 4881.00 & 1 & --0.005 & 0.021 & 0.005 & HDS &  \\ 
4924.71 & He~{\sc i} & 4921.93 & 1 & --0.016 & 0.277 & 0.011 & UVES1 &  \\ 
4925.21 & He~{\sc i} & 4921.93 & 2 & --0.016 & 1.043 & 0.013 & UVES1 &  \\ 
 &  &  & Tot. &  & 1.320 & 0.017 & UVES1 &  \\ 
4934.42 & [O~{\sc iii}] & 4931.23 & 1 & --0.019 & 0.044 & 0.004 & HDS &  \\ 
4937.32 & Ba~{\sc ii} & 4934.08 & 1 & --0.019 & 0.006 & 0.002 & UVES1 &  \\ 
4961.87 & [O~{\sc iii}] & 4958.91 & 1 & --0.026 & 29.603 & 2.389 & HDS &  \\ 
4962.03 & [O~{\sc iii}] & 4958.91 & 2 & --0.026 & 32.311 & 1.991 & HDS &  \\ 
4962.28 & [O~{\sc iii}] & 4958.91 & 3 & --0.026 & 60.490 & 5.079 & HDS &  \\ 
 &  &  & Tot. &  & 122.404 & 5.956 & HDS &  \\ 
4967.84 & C~{\sc ii} & 4964.73 & 1 & --0.027 & 0.027 & 0.004 & HDS &  \\ 
5004.63 & N~{\sc ii} & 5001.47 & 1 & --0.036 & 0.018 & 0.003 & HDS &  \\ 
5009.85 & [O~{\sc iii}] & 5006.84 & 1 & --0.038 & 55.674 & 13.550 & HDS &  \\ 
5009.97 & [O~{\sc iii}] & 5006.84 & 2 & --0.038 & 137.072 & 3.874 & HDS &  \\ 
5010.25 & [O~{\sc iii}] & 5006.84 & 3 & --0.038 & 158.090 & 16.602 & HDS &  \\ 
 &  &  & Tot. &  & 350.836 & 21.776 & HDS &  \\ 
5018.53 & He~{\sc i} & 5015.68 & 1 & --0.040 & 0.291 & 0.012 & HDS &  \\ 
5019.01 & He~{\sc i} & 5015.68 & 2 & --0.040 & 1.481 & 0.014 & HDS &  \\ 
 &  &  & Tot. &  & 1.772 & 0.019 & HDS &  \\ 
5033.51 & [Fe~{\sc iv}] & 5030.33 & 1 & --0.043 & 0.089 & 0.008 & HDS &  \\ 
5035.14 & C~{\sc ii} & 5032.07 & 1 & --0.044 & 0.053 & 0.005 & HDS &  \\ 
5038.93 & [Fe~{\sc ii}] & 5035.48 & 1 & --0.045 & 0.011 & 0.004 & HDS &  \\ 
5039.16 & C~{\sc ii} & 5035.94 & 1 & --0.045 & 0.018 & 0.003 & UVES1 &  \\ 
5050.58 & He~{\sc i} & 5047.74 & 1 & --0.048 & 0.044 & 0.009 & HDS &  \\ 
5051.11 & He~{\sc i} & 5047.74 & 2 & --0.048 & 0.211 & 0.011 & HDS &  \\ 
 &  &  & Tot. &  & 0.254 & 0.014 &  &  \\ 
5059.25 & Si~{\sc ii} & 5055.98 & 1 & --0.050 & 0.007 & 0.004 & UVES2 &  \\ 
5069.55 & N~{\sc ii} & 5066.46 & 1 & --0.052 & 0.042 & 0.009 & UVES2 &  \\ 
5117.37 & O~{\sc v} & 5114.06 & 1 & --0.063 & 0.013 & 0.030 & HDS &  \\ 
5122.44 & [Fe~{\sc iii}] & 5119.97 & 1 & --0.065 & 0.014 & 0.004 & HDS &  \\ 
5125.06 & C~{\sc ii} & 5121.83 & 1 & --0.065 & 0.029 & 0.002 & HDS &  \\ 
5130.66 & Fe~{\sc iii} & 5127.39 & 1 & --0.066 & 0.007 & 0.002 & HDS &  \\ 
$\cdots$\\
9074.84 & [S~{\sc iii}] & 9068.60 & 1 & --0.594 & 0.378 & 0.010 & UVES2 &  \\ 
9101.95 & N~{\sc ii} & 9096.16 & 1 & --0.596 & 0.011 & 0.004 & UVES2 &  \\ 
9114.25 & He~{\sc ii} & 9108.54 & 1 & --0.597 & 0.032 & 0.003 & UVES2 &  \\ 
9205.28 & Fe~{\sc i} & 9199.45 & 1 & --0.603 & 0.012 & 0.002 & UVES2 &  \\ 
9216.31 & He~{\sc i} & 9210.34 & 1 & --0.604 & 0.093 & 0.006 & UVES2 &  \\ 
9219.44 & He~{\sc i} & 9213.24 & 1 & --0.604 & 0.025 & 0.004 & UVES2 &  \\ 
9223.29 & Mg~{\sc ii}? & 9218.25 & 1 & --0.604 & 0.008 & 0.003 & UVES2 &  \\ 
9230.87 & He~{\sc ii} & 9225.23 & 1 & --0.605 & 0.032 & 0.006 & UVES2 &  \\ 
9234.92 & H~{\sc i} & 9229.01 & 1 & --0.605 & 2.545 & 0.043 & UVES2 &  \\ 
9350.69 & He~{\sc ii} & 9344.94 & 1 & --0.613 & 0.572 & 0.011 & UVES2 &  \\ 
\hline
\end{tabular}
\end{table}

\end{document}